%% file: main.tex
\documentclass[fleqn,usenatbib]{mnras}

\usepackage{newtxtext,newtxmath}
\usepackage[T1]{fontenc}
\usepackage{graphicx}
\usepackage{amsmath}	

\DeclareRobustCommand{\VAN}[3]{#2}
\let\VANthebibliography\thebibliography
\def\thebibliography{\DeclareRobustCommand{\VAN}[3]{##3}\VANthebibliography}

\newcommand{\Msun}{\mathrm{M}_{\odot}}
\newcommand{\Hz}{\mathrm{Hz}}
\newcommand{\Msmbh}{M_{\rm SMBH}}
\newcommand{\rS}{r_{\rm S}}

\renewcommand{\theenumi}{(\arabic{enumi}}
\renewcommand{\labelenumi}

 \input{Title_Abstract/Title.tex}

\begin{document}
 
 \label{firstpage}
 
 \pagerange{\pageref{firstpage}--\pageref{lastpage}}
 
 \maketitle

\input{Title_Abstract/Abstract.tex}

 \input{Sections/Introduction.tex}

\input{Sections/Magnif_Delay_Dist.tex}

\input{Sections/Results}

\input{Sections/Summary_Conclusions.tex}

\input{Sections/Acknowledgments.tex}

\input{Sections/Data_Availability.tex}

 \bibliographystyle{mnras}
 \bibliography{refs}

 \appendix

\input{Appendices/Properties_AmplDist.tex}

\input{Appendices/SignalTimeDuration}

\input{Appendices/Additional_Effects}

\input{Appendices/Regimes.tex}

\bsp
\label{lastpage}
\end{document}

%% file: Title_Abstract/Title.tex
\title[Astrophysical GW Echoes from Galactic Nuclei]{Astrophysical Gravitational-Wave Echoes from Galactic Nuclei}

\author[L. Gond\'an \& B. Kocsis]{
L\'aszl\'o Gond\'an,$^{1}$\thanks{E-mail: laszlo.gondan@ttk.elte.hu}
Bence Kocsis$^{2}$\thanks{E-mail: bkocsis.@gmail.com}
\\
$^{1}$ ELTE E\"otv\"os Lor\'and University, P\'azm\'any P. s. 1/A, Budapest 1117, HU\\
$^{2}$ Rudolf Peierls Centre for Theoretical Physics, Clarendon Laboratory, Parks Road, Oxford OX1 3PU, UK
}

\date{Accepted XXX. Received YYY; in original form ZZZ}

\pubyear{2022}

%% file: Title_Abstract/Abstract.tex
\begin{abstract}
 Galactic nuclei (GNs) are dense stellar environments abundant in gravitational-wave (GW) sources for LIGO, VIRGO, and KAGRA. The GWs may be generated by stellar-mass black hole (BH) or neutron star mergers following gravitational bremsstrahlung, dynamical scattering encounters, Kozai-Lidov type oscillations driven by the central supermassive black hole (SMBH), or gas-assisted mergers if present. In this paper, we examine a smoking gun signature to identify sources in GNs: the GWs scattered by the central SMBH. This produces a secondary signal, an astrophysical GW echo, which has a very similar time-frequency evolution as the primary signal but arrives after a time delay. We determine the amplitude and time-delay distribution of the GW echo as a function of source distance from the SMBH. Between $\sim10\%-90\%$ of the detectable echoes arrive within $\sim(1-100)M_6\,\mathrm{sec}$ after the primary GW for sources between $10-10^4$ Schwarzschild radius, where \mbox{$M_6=M_{\rm SMBH,z}/(10^6\,\Msun)$}, and $M_{\rm SMBH,z}$ is the observer-frame SMBH mass. The echo arrival times are systematically longer for high signal-to-noise ratio (SNR) primary GWs, where the GW echo rays are scattered at large deflection angles. In particular, $\sim10\%-90\%$ of the distribution is shifted to $\sim(5-1800)M_6\,\mathrm{sec}$ for sources, where the lower limit of echo detection is $0.02$ of the primary signal amplitude. We find that $\sim5\%-30\%$ ($\sim1\%-7\%$) of GW sources have an echo amplitude larger than $0.2-0.05$ times the amplitude of the primary signal if the source distance from the SMBH is $50$ ($200$) Schwarzschild radius. Non-detections can rule out that a GW source is near an SMBH.
\end{abstract}


%% file: Sections/Introduction.tex
\section{Introduction} 
\label{sec:Intro}
 
 The detection of gravitational waves (GWs) from stellar-mass BH and neutron star mergers by the Advanced Laser Interferometer Gravitational-Wave Observatory\footnote{\url{http://www.ligo.caltech.edu/}} (aLIGO; \citealt{Aasietal2015}), Advanced VIRGO\footnote{\url{http://www.ego-gw.it/}} (AdV; \citealt{Acerneseetal2015}), and KAGRA\footnote{\url{https://gwcenter.icrr.u-tokyo.ac.jp/en/}} \citep{KagraColletal2018} has opened the field of GW astronomy \citep{Abbottetal2016,Abbottetal2017,Abbottetal2019a,Abottetal2020a}. In the past years, a significant effort has been invested in explaining how these compact objects may form binaries and merge as frequently as observed to constrain the possible astrophysical origin of the detected mergers (e.g. \citealt{Baracketal2019,Abottetal2020c,Abottetal2020b}, and references therein).

 GWs of compact binaries encode masses, mass ratios, spins (e.g. \citealt{CutlerFlanagan1994,PoissonWill1995}), and in some cases, orbital eccentricity (e.g. \citealt{Gondanetal2018a,Loweretal2018,GondanKocsis2019,RomeroShaw2019}). The GW signal can also carry information on the formation environment through external effects affecting the GW signal. Gas dynamics affect the waveform by slowing down or accelerating the inspiral (e.g. \citealt{BarausseRezzolla2008,Kocsisetal2011,Yunesetal2011,Barausseetal2014,DOrazioLoeb2018,Derdzinskietal2019,CardosoMaselli2020,Toubiana2020,Derdzinski2021}). Furthermore, there are multiple situations in which the merger environment can affect the GW signal. An intervening mass distribution (e.g. \citealt{Kocsis2006,Bonvinetal2017,RandallXianyu2019}) or the relative motion of the GW source due to peculiar velocities of GW sources and the orbital motion around a companion (e.g.  \citealt{Inayoshietal2017,Meironetal2017,Robsonetal2018,Chamberlainetal2019,Wongetal2019,Tamaninietal2020,TorresOrjuelaetal2020}) can result in Doppler boosting, gravitational lensing of the rest-frame GWs, time-dependent gravitational redshift, GW phase shift, or the Shapiro delay.

 High-frequency GWs emitted by stellar/intermediate-mass compact binaries are scattered by the gravitational field of an SMBH (e.g. \citealt{Futterman1988,Dolan2008}) producing second and higher-order scattered GWs analogous to the relativistic images of electromagnetic (EM) waves (e.g. \citealt{DolanOttewill2011,ZenginogluGalley2012}).  These second and higher-order scattered GW signals have the same waveform as the unperturbed signal in the geometrical optics limit, including the inspiral, merger, and ringdown waveforms. They only differ in amplitude, arrival time, overall phase, and polarization-angle \citep{Daietal2017,Ezquiagaetal2021}.\footnote{The so-called saddle point images may experience nontrivial waveform changes if the GWs exhibit modes higher than (2,2) modes or precession \citep{Daietal2017,Ezquiagaetal2021}. In addition, the polarization of the GW signal may also change by a small amount during the scattering by a Kerr-BH with a small impact parameter \citep{Dolan2008} and due to the intrinsic anisotropy of the emission pattern, Doppler effect, and de Sitter precession \citep{Torres-Orjuela2019,Gong2021,Yu2021}. Note that we do not take into account these effects in this paper.} Similar to EM waves, the amplitudes of the scattered GW signals decay asymptotically exponentially with the order of the scattered waves (e.g. \citealt{ZenginogluGalley2012,Kocsis2013}), indicating that the detections of scattered secondary GWs are typically the most prominent besides the primary GWs. Note that the scattering of GWs and second and higher-order scattered GWs are referred to as lensing and multiple images, respectively, in the GW lensing community.\footnote{Multiple images are produced when the stationary phase approximation holds, and the diffraction integral can be solved in terms of its stationary points (e.g. \citealt{TakahashiNakamura2003}).}

 In this study, we examine GW lensing and multiple images due to a single SMBH in the vicinity of the stellar-mass compact binary sources generating a secondary GW signal (hereafter referring to it as an \textit{astrophysical GW echo}). We investigate the main characteristics of the GW echo signal, including its amplitude and time delay relative to the primary GW signal. The detection of an astrophysical GW echo will be a smoking-gun signature of GW sources in close proximity to SMBHs in GN host environments. Note that sequences of distinct GW pulses also referred to as GW echoes are expected from exotic compact objects (e.g. \citealt{CardosoPani2017,Maggioetal2017,Marketal2017,Buenoetal2018}) and when going beyond the description of BHs by classical general relativity (e.g. \citealt{Abedietal2017,Barceloetal2017,DongStojkovic2021}) as well.

 Several source populations are expected to emit GWs in the vicinity of SMBHs in the inner regions of GNs for ground-based GW detectors.
\begin{enumerate}
  \item Due to the extremely high stellar number densities in GNs reaching up to $10^{10}\,{\rm pc}^{-3}$, the probability of very close dynamical encounters among single objects is non-negligible, where the binary forms and merges due to GW emission \citep{OLearyetal2009}. The heavy BHs sink to the densest central regions due to mass segregation, where the velocity dispersion is so high that only very close encounters result in captures. This population of mergers represents the highest eccentricity GW sources in the Universe, which typically form in the aLIGO/AdV/KAGRA frequency band \citep{KocsisLevin2012,Gondanetal2018b,GondanKocsis2021}.
  
  \item The long-term gravitational perturbations due to the central SMBH drive variations in the binary's orbital eccentricity, the so-called Kozai-Lidov (KL) effect \citep{Kozai1962,Lidov1962,LidovZiglin1976}. In some cases, this effect may lead to close encounters between the binary components, after which the binary merges due to GW emission \citep{AntoniniPerets2012,PetrovichAntonini2017,Hamersetal2018,Hoangetal2018,RandallXianyu2018a,RandallXianyu2018b,Fragioneetal2019,LiuLai2020}.
  
  \item In active galactic nuclei (AGNs), the gaseous accretion disk around the SMBH may facilitate binary formation and mergers of stellar-mass compact objects \citep{McKernanetal2014,Bellovaryetal2016,Bartosetal2017,Stone17,Tagawaetal2020a}. In this scenario, BHs form in situ in the vicinity of a GN and sink to the inner region due to mass segregation or they are delivered to these regions by infalling globular clusters \citep{Morris1993,MiraldaEscudeGould2000,Freitagetal2006,HopmanAlexander2006,OLearyetal2009,Antonini2014}, then get captured in the disk by hydrodynamic drag as they cross the disk (e.g. \citealt{Goldreichetal2002,Bartosetal2017,Yangetal2019b,Tagawaetal2020a}). Alternatively, some BHs may have formed in the disk itself  \citep{Levinetal2007,Stone17}. Once in the disk, BHs get transported to the inner regions by exchanging angular momentum with the surrounding gas \citep{GoldreichTremaine1979}. In certain regions, the BHs open an annular gap in the accretion disk and accumulate in a narrow range of radii, the so-called migration traps \citep{Bellovaryetal2016,Secundaetal2019,Secundaetal2020a,Secundaetal2020b}. \citet{Bellovaryetal2016} argue that migration traps may be expected to be close to the SMBH from $\sim 20$ to $\sim 300$ Schwarzschild radii ($\rS = 2 G \Msmbh / c^2$) from the central SMBH of mass $\Msmbh$.\footnote{Note that \citet{PanYang2021} have recently found that migration traps cannot exist in standard AGN discs and BHs do not open gaps as long as their mass is limited to $10 \,\Msun$; see however \citet{Kocsisetal2011}.} However, they may exist in slim disks near the innermost stable circular orbit \citep{PengChen2021} or near the boundary of a gap region if a gap opens due to a heavy stellar-mass BH or an intermediate-mass BH \citep{McKernanetal2014}. Dynamical encounters frequently happen in migration traps leading to the formation and subsequent merger of BBHs on short time scales \citep{Secundaetal2019,Yangetal2019,Secundaetal2020a}, where the binary separation is efficiently reduced by gas dynamical friction \citep{Escalaetal2004,KimKim2007,Baruteauetal2011} to the point where GW emission drives the binaries together. Alternatively, BBHs may also form and merge in the disk outside migration traps \citep{Tagawaetal2020a}. Due to the deep potential barrier of the SMBH, the merger remnant BH remains near the migration trap and may undergo subsequent mergers with additional BHs, which leads to high BH masses, characteristic spin properties, and possibly nonzero eccentricity identifiable via GW observations \citep{Yangetal2019,Secundaetal2020a,Samsing2020,Tagawaetal2020b,Tagawa2021a,Tagawa2021b}.
\end{enumerate}

 Astrophysical GW echos are produced in GNs due to strong gravitational lensing by the SMBH. The lensing of GWs has been investigated in the literature in the wave-optics and the geometrical optics regimes (e.g. \citealt{TakahashiNakamura2003}) on very different scales, including lensing by the large-scale structure of the Universe  \citep{Hilbertetal2007,Hilbertetal2008,Robertsonetal2020}, galaxy clusters and galaxies (e.g. \citealt{Wangetal1996,TakahashiNakamura2003,Daietal2017,Takahashi2017,Broadhurstetal2018,Lietal2018,Smithetal2018}), intermediate-mass BH in globular clusters (e.g. \citealt{Kainsetal2016,Laietal2018,Tatekawaetal2020}), and populations of stellar-mass microlenses in galaxies (e.g. \citealt{,Wangetal1996,Cheungetal2020,Mishraetal2021,Meenaetal2022}). In addition, lensing by SMBHs has been investigated in the geometrical optics limit for the GWs of binary BHs (BBHs) detectable by aLIGO/AdV/KAGRA \citep{Kocsis2013} and the Laser Interferometer Space Antenna\footnote{\url{https://www.elisascience.org/}} (LISA; \citealt{AmaroSeoaneetal2017}) \citep{DOrazioLoeb2020,Yu2021}. Several searches have been carried out for signatures of strongly lensed and multiply imaged GW signals in the LIGO and VIRGO data. So far, some potential candidates have been proposed that may be interpreted as signals strongly lensed by an intervening galaxy or a galaxy cluster \citep{Broadhurstetal2019,Daietal2020,Diegoetal2021,Liuetal2021}. However, no compelling evidence has been found for lensing yet \citep{Hannukselaetal2019,McIsaacetal2020,Pangetal2020,Abottetal2021a,Kimetal2022}.

 If the SMBH lens happens to lie close to the line-of-sight behind the GW source, such that the deflection angle $\alpha$ is close to the Einstein radius, the primary GW signals and astrophysical GW echoes may both be strongly magnified relative to the unlensed GW signal \citep{Einstein1936}. However, unlike for EM waves\footnote{Since intensity scales with the square of the amplitude, it is more suppressed.}, the GW echo may be remarkably significant for GW sources in the proximity of the SMBH not just around the Einstein radius but also for larger deflection angles. For large deflection angles, the GW amplitude of the echo relative to the unlensed GW signal scales as $(r/\Msmbh)^{-1} \alpha^{-1}$ (e.g. \citealt{Kocsis2013}), and the anisotropic emission of the GW source profile significantly broadens the distribution of the relative echo amplitude (see Figure \ref{Fig:MagnifPrim_MagnifEcho} below). This indicates that GW echoes with large deflection angles may be typically detected for GW sources that are close to the SMBH, although this range has not yet been determined quantitatively in the literature to our knowledge.

 In this paper, we examine the possibility of observing astrophysical GW echos with aLIGO/AdV/KAGRA from the mergers of stellar-mass BBHs in the vicinity of SMBHs as a function of the distance from the SMBH. We assume a generic lensing configuration for calculating the magnification and time delay, i.e. \textit{not} limited to the case of small deflection angles, but neglect possible interference effects between the scattered GW signals. We resort to the geometrical optics limit (i.e. the short-wavelength limit) and assume that the BBH is fixed with respect to the SMBH (i.e. BBHs are in the stationary-lens regime). We account for the anisotropy of the GW emission pattern but neglect the deflection-angle-dependence of the Doppler effect. We find that these are typically valid approximations to leading order for stellar-mass BBH mergers around SMBHs in LIGO-Virgo-KAGRA searches for SMBHs with masses above $10^5 \, \Msun$ and due to the short time interval the GW signal spends in the aLIGO/AdV/KAGRA band. Since lensing does not alter the waveform in the geometrical optics limit, it is scaled with a constant amplitude and shifted with a time delay (hence the name astrophysical GW echo). This allows us to make population-model and detector-independent predictions on the detected SNR distribution and detectability of the GW echo compared to that of the primary GW. We determine the probability of detecting GWs that undergo an arbitrary large-angle deflection due to the SMBH. To this end, we generate random Monte Carlo (MC) samples of BBH mergers in the vicinity of SMBHs to determine the magnification distribution of GW echoes relative to the primary GW signals and the relative amplitude of the GW echo as a function of the radial distance of merging BBHs from the SMBH and the inclination of the plane of merging BBHs. The inclination dependence may be relevant for BBH mergers in AGN disks \citep{Bartosetal2017,Corleyetal2019,Tagawaetal2020a} or BH disks \citep{Szolgyen_Kocsis2018,Gruzinovetal2020,Matheetal2022} in GNs. We also calculate the fraction of detectable GW echoes in mock binary catalogues by accounting for observational bias and the corresponding time-delay distribution as a function of SMBH mass and binary distance from the SMBH.

 In Section \ref{sec:MagnifGWs}, we derive the deflection and magnification of primary GWs and astrophysical GW echoes for a generic lensing configuration for an arbitrary binary position relative to the SMBH and the line-of-sight to the observer. Next, we describe the methodologies with which we carry out our investigations and present the main results of the paper in Section \ref{sec:Results}. Finally, we summarize the results of the paper and conclude in Section \ref{sec:Conclusions}. Several details of our analysis are included in the Appendix. First, we discuss the properties of magnification distribution for primary GWs and GW echoes in Appendix \ref{Sec:Anistropy_Properties}, then estimate the time duration the GW signals of merging BBHs around SMBHs spend in the aLIGO/AdV/KAGRA band in Appendix \ref{sec:SignalTimeDuration}. We estimate the impact of additional effects caused by the SMBH on the properties of the GW source in addition to the deflection of GW rays in Appendix \ref{Sec:Additional_Effects}. Finally, we investigate the validity of our assumptions on the geometrical optics limit and the stationary-lens regime for BBH mergers in the vicinity of SMBHs in Appendix \ref{sec:LensingRegimes}.

 We use geometric units ($G = 1 = c$) throughout the paper, where mass $M$ and distance $r$ have units of time: $GM/c^3$ and $r/c$.

%% file: Sections/Magnif_Delay_Dist.tex
\section{Deflection and magnification of gravitational waves} 
\label{sec:MagnifGWs}
 
 We introduce the geometric conventions we use to describe the deflection of primary GWs and astrophysical GW echoes around an SMBH in Section \ref{subsec:Lensing_Geom}. Section \ref{subsec:GeometricConvects} introduces the geometric conventions with which GW sources are generated in our models in disks and spherical nuclear star clusters around SMBHs. We determine the deflection angle and time delay in Section \ref{subsec:Lens_Delay}, and the amplitude magnification of a GW echo relative to the primary GW is given in Section \ref{subsec:Magnif_GWs}.

\subsection{Lensing geometry}
\label{subsec:Lensing_Geom}

\begin{figure}
    \centering
    \includegraphics[width=86mm]{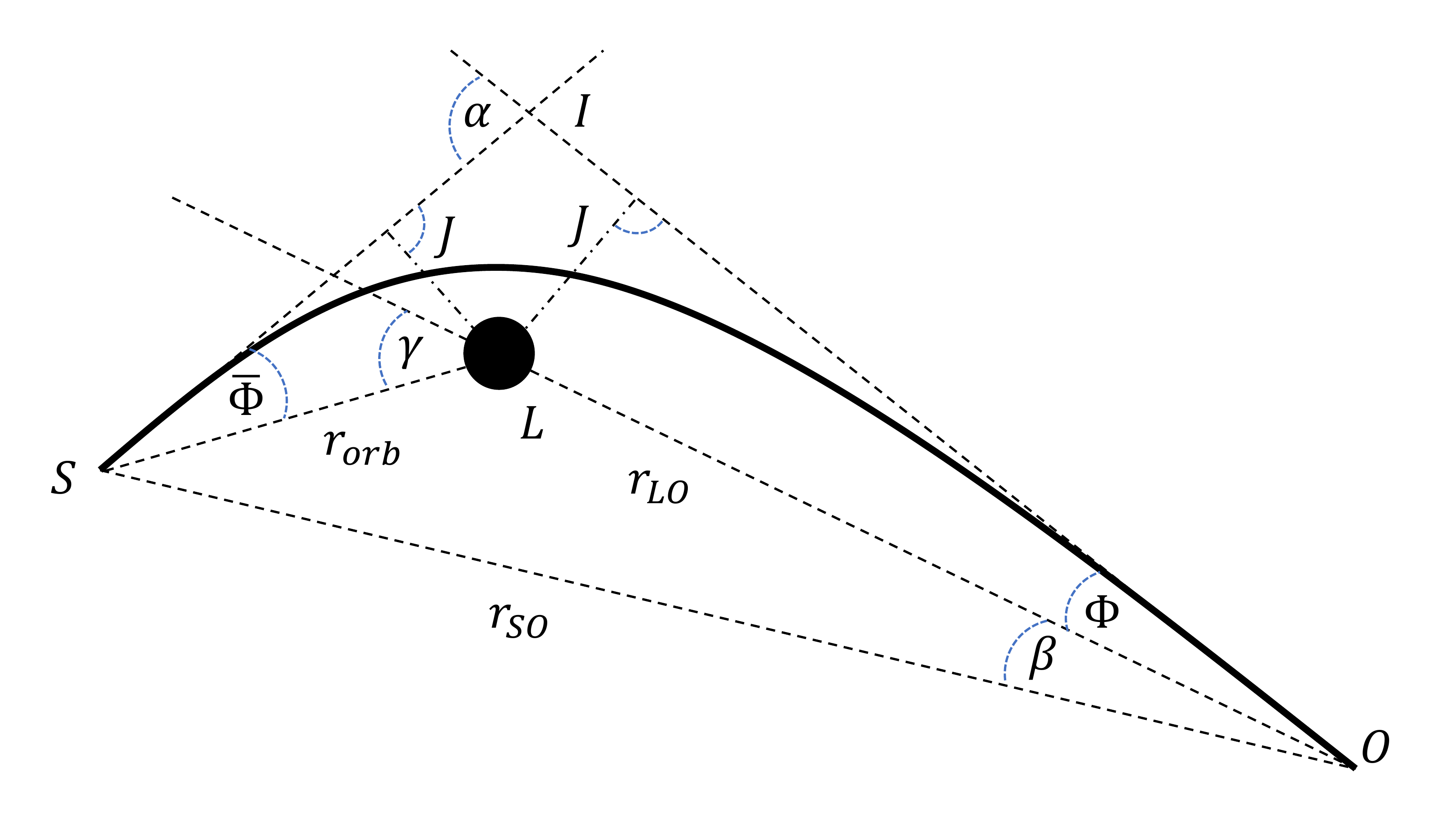}
    \\
    \includegraphics[width=90mm]{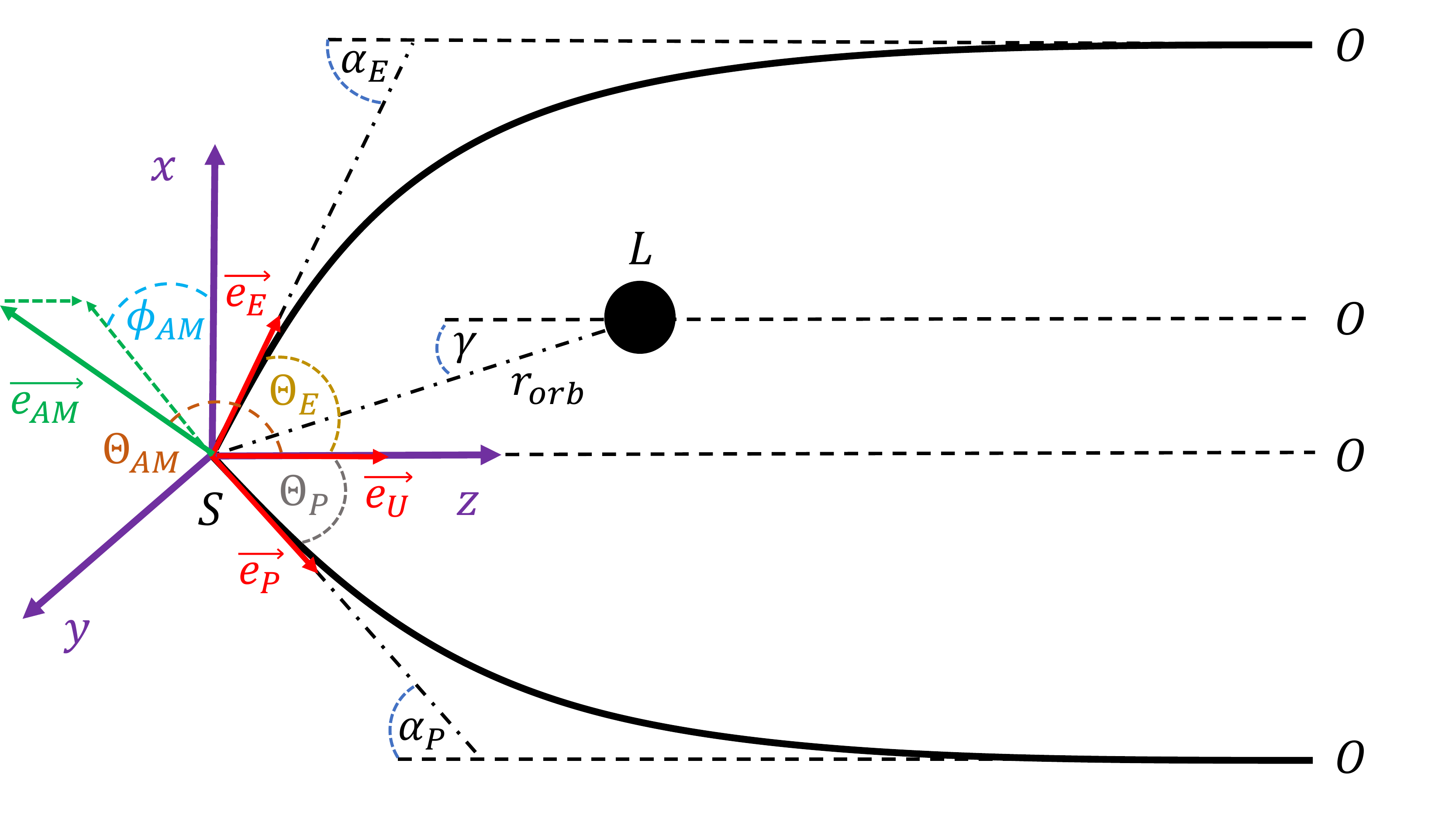}
\caption{\textit{Top panel:} Generic lensing geometry by a static spherically symmetric SMBH lens (L). The source is in S, the observer in O, and the image as seen by the observer in I. Illustrated quantities are defined in Section \ref{subsec:Lensing_Geom}, and the bending of primary GWs and GW echoes around the SMBH lens are described in Section \ref{subsec:Lens_Delay}. \textit{Bottom panel:} Schematic representation of the emission of unlensed GW (U), primary GW (P), and GW echo (E) with respect to the merging binary (i.e. the source) assuming a distant observer. See definitions in Section \ref{subsec:Lensing_Geom}. 
\label{Fig:GeomConvMagnif} } 
\end{figure}

 The top panel of Figure \ref{Fig:GeomConvMagnif} illustrates the geometric configuration. We assume that the SMBH lens is a Schwarzschild BH for which the bending of rays happens in a plane defined by the lens (L), the observer (O), and the source (S) (e.g. \citealt{Weinberg1972}). We define all distances and angles of the GW trajectory in the asymptotic flat (Minkowski) spacetime.

 The line joining the observer and the lens is the optical axis, $\overline{OL}$. $r_{\rm LO}$ and $r_{\rm SO}$ represent the radial lens-observer and source-observer distances, respectively, while the radial lens-source distance is $r_{\rm orb}$. We introduce $D_{\rm SL}$ and $D_{\rm SO}$ such that they measure the component of the source-lens and source-observer distances along the optical axis, and the corresponding lens-observer distance is \mbox{$r_{\rm LO}$}. The angular position of the source in the observer's sky with respect to the position of the SMBH is $\beta = SOL\measuredangle$. Furthermore, $\Phi$ and $\overline{\Phi}$ are respectively the angles subtended by the tangent of the ray at the observer and the source relative to $\overline{OL}$ and $\overline{SL}$. $\gamma$ is the angle between the source-lens direction ($\overline{SL}$) and the optical axis. The impact parameters of incoming and outgoing rays are the same $J$ in the asymptotic approximation. In this lensing configuration, the primary GW signal has $-\pi \leqslant \gamma < 0$ where the GW source and its scattered GW are on the same side of the lens, and the GW echo has $0 < \gamma \leqslant \pi$. The high-magnification strong lensing scenario, where $\theta$ is at the Einstein radius, corresponds to $\gamma \sim 0$  \citep{VirbhadraEllis2000}, and $\gamma \sim \pm \pi$ describes retro-lensing scenarios (e.g. \citealt{HolzWheeler2002,EiroaTorres2004}).

 Given that $r_{\rm LO} \gg r_{\rm orb}$, generally $\beta \ll 1$ and $\Phi \ll 1$, while $\{ \alpha, \gamma,\overline{\Phi} \}$ may be of order unity. Furthermore, we will use the following simple approximate geometrical relations:
\begin{align}  \label{eq:FristSet}
  & D_{\rm SL} = r_{\rm orb} \cos{\gamma} \, , \quad  D_{\rm SO} = r_{\rm LO}  \, , \quad r_{\rm SO} = r_{\rm LO}\, , 
  \\
  & \Phi = \frac{J}{r_{\rm LO}} \, , \quad \overline{\Phi} = \arcsin \left(\frac{J}{r_{\rm orb}}\right)\, , \quad \beta = \frac{r_{\rm orb}}{r_{\rm LO}} \sin{\gamma} \, . \label{eq:SecondSet}
\end{align}

 The amplitude of GWs emitted by a merging binary is anisotropic as they are described by spin-2 weighted spherical harmonics, whose energy flux changes with the $\Psi$ polar angle with respect to the binary orbital axis (i.e. the angular momentum vector) as
\begin{equation}  \label{eq:Y(phi)}
    Y(\Psi) \propto \sin^8 \left(\frac{\Psi}{2}\right)  
    + \cos^8 \left(\frac{\Psi}{2}\right) \, .
\end{equation}
 Thus, the GW energy flux emitted in the direction perpendicular to  binary plane is a factor of $Y(0)/Y(\pi/2) = 8$ times higher than in the plane of the binary and a factor of $5/2$ higher than average. This shows that a 90-degree scattering may change the GW energy flux by a factor between $1/8$ and $8$ due to the anisotropic GW emission pattern. We take into account the anisotropic emission of GWs of the merging binaries according to Equation \eqref{eq:Y(phi)}. We describe the emission of the unlensed GWs, primary GWs, and GW echoes next; see the bottom panel of Figure \ref{Fig:GeomConvMagnif} for an illustration.

\subsection{Geometric conventions} 
\label{subsec:GeometricConvects}
 
 To account for the anisotropic GW emission pattern, we adopt a Cartesian coordinate system centred at the binary (i.e. the source) as shown in Figure \ref{Fig:GeomConvMagnif}. We define orthonormal basis vectors $\{ \mathbf{x}, \mathbf{y}, \mathbf{z} \}$ such that (i) the binary, the lens, and the observer lie in the $x - z$ plane, (ii) $\mathbf{z}$ points from the binary to the observer, (iii) and the SMBH (lens) lies in the half-space with positive $x$. We also define the tangent unit vectors along with the rays of the GWs along with the primary signal and the echo at the source respectively with $\mathbf{e}_{\rm P}$, $\mathbf{e}_{\rm E}$, and let $\mathbf{e}_{\rm U} = \mathbf{z}$ label the tangent vector along with rays to the observer in case the lens is neglected. The angular momentum unit vector of the binary is denoted by $\mathbf{e}_{\rm AM}$ and defined using spherical coordinates as
\begin{equation}  \label{eq:e_AM}
  \mathbf{e}_{\rm AM} = \left( \sin{\Theta_{\rm AM}} \cos{\phi_{\rm AM}}, \sin{\Theta_{\rm AM}} \sin{\phi_{\rm AM}} , \cos{\Theta_{\rm AM}}\right) \, ,
\end{equation}
 where the azimuthal angle $\phi_{\rm AM}$ ranges between $[0, 2 \pi]$, while the polar angle $\Theta_{\rm AM}$ takes values between $[0, \pi]$. Since the rays of the GW echo lie in the $x-z$ plane, $\mathbf{e}_{\rm E}$ is given as
\begin{equation}  \label{eq:e_E}
  \mathbf{e}_{\rm E} = \left( \sin{\Theta_{\rm E}}, 0, \cos{\Theta_{\rm E}} \right) \, ,
\end{equation}
 where for distant observers the polar angle $\Theta_{\rm E}$ equals the deflection angle of the GW echo $\alpha_{\rm E}$.\footnote{For clarity, we note that $\alpha_{\rm E}$ refers to the angle corresponding to the GW echo, which is generally \textit{not} the Einstein angle.} Similarly, $\mathbf{e}_{\rm P}$ can be given as
\begin{equation}  \label{eq:e_P}
  \mathbf{e}_{\rm P} = \left( - \sin{\Theta_{\rm P}}, 0, \cos{\Theta_{\rm P}} \right) \, ,
\end{equation}
 where the polar angle $\Theta_{\rm P}$ equals $\alpha_{\rm P}$. Finally, the $\Psi$ polar angle with respect to $\mathbf{e}_{\rm AM}$ can be given separately for the unlensed GW, the primary GW, and the GW echo as
\begin{equation}  \label{eq:Psi_angles}
  \cos{\Psi_{\rm U}} = \mathbf{e}_{\rm AM} \cdot \mathbf{e}_{\rm U} \, , \quad \cos{\Psi_{\rm P}} = \mathbf{e}_{\rm AM} \cdot \mathbf{e}_{\rm P} \, , \quad \cos{\Psi_{\rm E}} = \mathbf{e}_{\rm AM} \cdot \mathbf{e}_{\rm E} \, ,
\end{equation}
 where $\cdot$ denotes the scalar product.

 To describe GW sources in AGN/BH disks, we adopt another Cartesian coordinate system centred at the SMBH (lens) with a set of orthonormal basis vectors $\{ \mathbf{X}, \mathbf{Y}, \mathbf{Z} \}$ aligned with the AGN/BH disk. In particular, the basis vectors are oriented such that $\mathbf{X}$ and $\mathbf{Y}$ lie in the disk plane and \mbox{$\mathbf{Z} = \mathbf{X} \times \mathbf{Y}$ is the rotation axis of the disk}. The unit vector $\mathbf{e}_{\rm O}$ points from the SMBH (lens) to the observer\footnote{Note that since the observer is at infinity, $\mathbf{e}_{\rm O} = \mathbf{e}_{\rm U} = \mathbf{z}$.} where \mbox{$\mathbf{e}_{\rm O} = (\cos{\varphi_{\rm O}} \cos{i}, \sin{\varphi_{\rm O}} \cos{i}, \sin{i})$}. Here, the inclination angle of the disk $i$ is defined as the latitude angle, and $\varphi_{\rm O}$ denotes the longitude angle of the observer in the disk.

 Here, $i$ ranges between $[-\pi/2, \pi/2]$, where $i = \{ -\pi/2, \pi/2 \}$ and $i = 0$ cases describe face-on and edge-on disks, respectively, as viewed by the observer. Furthermore, $\varphi_\mathrm{O}$ ranges between $[0, 2 \pi]$. In order to describe GW sources (e.g. merging BBHs) at radius $r_{\rm orb}$ from the SMBH lens, we introduce the source position vector $\mathbf{r}_{\rm LS}$, which is defined in polar coordinates in the plane of the disk as $\mathbf{r}_{\rm LS} = r_{\rm orb}(  \cos{\varphi_S}, \sin{\varphi_S}, 0)$. Finally, $\gamma$ is calculated as
\begin{equation}  \label{eq:gammaCalc}
  \gamma = \arccos{(\mathbf{e}_{\rm O} \cdot \mathbf{r}_{\rm LS} / r_{\rm orb})}=\arccos{[\cos(\varphi_S-\varphi_{\rm O})\cos i]} \, . 
\end{equation}
 Without loss of generality, we restrict the range of $i$ to $[0, \pi/2]$ due to symmetry for $i\rightarrow -i$.

 We decompose the source population into uniform rings around the SMBH. Note that for an inclined uniform ring, the prior distribution of $\gamma$ for fixed $r_{\rm orb}$, $i$, and $\varphi_{\rm O}$ follows from $dN/d\varphi_{\rm S} = \rm const$ and $\varphi_{\rm S} = \varphi_{\rm O} + \arccos[(\cos{\gamma})/(\cos{i})]$ as
 \begin{equation}  \label{eq:gammaDist}
  \frac{dN}{d\gamma} = \frac{dN}{d\varphi_{\rm S}}\frac{ d\varphi_{\rm S} }{ d\gamma } \propto \frac{ \sin{\gamma} }{\sqrt{ \cos^2 i - \cos^2 \gamma}} 
  = \frac{ \sin{\gamma} }{ \sqrt{ \sin^2 \gamma - \sin^2 i} }  
\end{equation}
 for $|\sin \gamma| > \sin i$, and $dN/d\gamma = 0$ otherwise if $|\sin \gamma| \leqslant \sin i$. It is uniformly distributed between $[0, \pi]$ for $i = 0$ (edge-on disk) and it shrinks onto the single value $\gamma = \pi/2$ for $i = \pi/2$ (face-on disk). For $0 < i < \pi/2$, the $\gamma$ range $[\gamma_{\rm min}, \gamma_{\rm max}] = [i, \pi - i]$ shrinks gradually from $[0, \pi]$ to the point $\{ \pi/2 \}$ with increasing $i$. Note that the $\gamma$ distribution is symmetric around $\gamma = \pi/2$ with peaks at the boundaries $\gamma_{\min}=i$ and \mbox{$\gamma_{\max} = \pi-i$}.

 We use the same $(X,Y,Z)$ coordinates to describe an isotropic population of GW sources around a single SMBH (e.g. merging BBHs in either an isotropic GN or an orientation-averaged AGN/BH disk) at radius $r_{\rm orb}$ but in this case \mbox{$\mathbf{r}_{\rm LS} = r_{\rm orb}(  \cos{\varphi_r} \cos{l},  \sin{\varphi_r} \cos{l}, \sin{l} )$}, where the latitude angle $l$ plays the role of $i$ in Equations \eqref{eq:gammaCalc} and \eqref{eq:gammaDist}, but while $i$ was set to a single value for a single disk, $l$ ranges between $[-\pi/2, \pi/2]$. This implies a uniform distribution of $-1 \leqslant \cos{\gamma} \leqslant 1$. Alternatively, this is equivalent to a distribution of source disks with different $\sin{i}$ sampled uniformly in the range $-1 \leqslant \sin{i} \leqslant 1$.\footnote{For an isotropic distribution, the polar angle $p$, defined by $p = \pi/2 - i$, has a uniform distribution between $-1 \leqslant \cos{p} \leqslant 1$.}

\subsection{Deflection of primary GWs and GW echoes and the time delay of echoes}
\label{subsec:Lens_Delay}
 
 We use the Ohanian lens equation in the asymptotic flat region of spacetime \citep{Ohanian1987} 
\begin{equation}  \label{eq:OhanianLens}
  \gamma = \alpha - \Phi - \overline{\Phi} \, 
\end{equation}
 to determine the large-angle bending of GW rays around a Schwarzschild BH. This gives accurate results for the bending of rays compared to the exact general relativistic treatment even when $r_{\rm orb}$ is relatively small ($r_{\rm orb} \sim 10 \rS$), and it provides the most accurate estimate for light bending among other approximate lens equations \citep{Bozza2008}. Here, for a Schwarzschild BH, $\alpha$ is given as a function of the closest distance of approach $r_0$ as\footnote{While the integral can be evaluated by integrating between \mbox{$r_0 \leqslant r \leqslant r_{\rm orb}$} and $r_{0} \leqslant r \leqslant r_{\rm obs}$ (e.g. Equation 6 in \citealt{Bozza2008}), where $r_{\rm obs}$ is the radial distance of the source to the observer, but here we resort to the approximation of $r_{\rm orb} \rightarrow \infty$ and $r_{\rm obs} \rightarrow \infty$. This is justified as long as $r_0 \ll r_{\rm orb}$, so that $\alpha$ is not very close to 0 and $J(r_0) \leqslant r_{\rm orb}$.}
\begin{align}  \label{eq:alpha_r0}
  \alpha(r_0) + \pi = 2 \int_{r_0} ^{\infty} \frac{ dr }{ r \sqrt{  \left( \dfrac{r}{r_0} \right)^2 \left( 1 - \dfrac{r_S}{r_0} \right) - \left( 1 - \dfrac{r_S}{r} \right) } } 
  =\frac{4\,F(\vartheta,k)}{\sqrt{u_1 u_3}} 
  \, , 
\end{align}
 where $F(\vartheta,k)$ is the elliptic integral of the first kind\footnote{We use the trigonometric form defined by
\begin{equation}  \label{eq:Fvarphi}
  F(\vartheta,k) = \int_0^{\vartheta} \frac{d\chi}{\sqrt{1-k^2 \sin^2 \chi}}\,.
\end{equation}
}, and we introduce the following functions of $r_0$
\begin{align}
 & u_0  = \frac{r_S}{r_0} \, , \quad u_1 = \sqrt{1-u_0} \, , \quad u_3 = \sqrt{1-3 u_0} \, ,
 \\
 & \vartheta = \frac{1}{2}\cos^{-1}\left(\frac{-3u_0}{1+u_1u_3}\right) \, , \quad k^2 = 2 u_0 \dfrac{(u_1)^{-1}+3(u_3)^{-1}}{u_1+u_3} \, .
\end{align}
 The impact parameter $J$ for a given $r_0$ can be expressed as
\begin{equation}  \label{eq:J_r0}
  J = r_0 \left( 1 - \frac{ r_S }{ r_0 } \right)^{- \frac{1}{2}}
\end{equation}
 \citep{Weinberg1972}. As seen, Equations \eqref{eq:alpha_r0} and \eqref{eq:J_r0} relate $J$ to $\alpha$. Since we are only interested in distant sources, $\Phi$ can be safely dropped from Equation \eqref{eq:OhanianLens}.

 Next, we express Equation \eqref{eq:OhanianLens} as a function of the polar coordinates of the source position $(r_{\rm orb},\gamma)$ as
\begin{equation}  \label{eq:solve_lens_eq}
  \gamma = \alpha (r_0) - \arcsin {\left( \frac{J(r_0)}{ r_{\rm orb} } \right)} \, .
\end{equation}
 This equation leads to a collection of solutions for both $-\pi \leqslant \gamma < 0$ and $0 < \gamma \leqslant \pi$ as GWs can go around the lens multiple times, where a higher multiplicity corresponds to a lower $r_0$.\footnote{Note that GWs can go around a lens if $\pi \leqslant \alpha < \infty$, which translates to $1.5 \, \rS < r_0 \lesssim 1.76 \, \rS$ in case of a Schwarzschild BH due to the one-to-one correspondence between $\alpha$ and $r_0$ given by Equation \eqref{eq:alpha_r0}. Such condition can be given numerically in terms of $(\gamma, r_{\rm orb})$ as well by substituting the appropriate $(\alpha, r_0)$ pairs back into Equation \eqref{eq:solve_lens_eq} and investigating which $(\gamma, r_{\rm orb})$ pairs satisfy this equation.} For a given source position $-\pi \leqslant \gamma < 0$ with respect to the observer, the largest $r_0$ corresponds to the primary signal, while the other solutions describe higher odd-order scattered GWs. Similarly, the largest $r_0$ corresponds to the echo and further $r_0$ solutions describe higher even-order scattered GWs for a given $0 < \gamma \leqslant \pi$. Here and throughout the paper ``primary'' signal and ``echo'' refer to the solutions of Equation \eqref{eq:solve_lens_eq}. Finally, substituting the resulting $r_0$ back into Equation \eqref{eq:alpha_r0}, shows that the primary GWs and GW echoes satisfy $0 < \alpha < \alpha_{\rm Einstein}$ and $\alpha_{\rm Einstein} < \alpha \lesssim \pi$, respectively, where \mbox{$\alpha_{\rm Einstein} = 2 \left( 2 r_{\rm orb} / \rS \right)^{-1/2}$} is the deflection angle at the Einstein radius.

 In Schwarzschild coordinates $(t,r,\varphi,\theta)$ in the equatorial plane (\mbox{$\theta = \pi/2$}), GWs follow null geodesics that satisfy
\begin{align}
  \label{eq:drdvarphi_Schw}
  \frac{dr}{d\varphi} &=\pm r \sqrt{\frac{r^2}{J^2} - \left( 1 - \frac{\rS}{r} \right) } \, , 
  \\
  \label{eq:dvarphidt_Schw}
  \frac{d\varphi}{dt} &= \frac{J}{r^2} \left( 1 - \frac{\rS}{r} \right)
\end{align}
 Here, $J$ is a constant of motion, the impact parameter, which is expressed in terms of the closest distance of approach $r_0$ in Equation (\ref{eq:J_r0}). Solving Equations \eqref{eq:drdvarphi_Schw} and \eqref{eq:dvarphidt_Schw} for the coordinate time $t$ between two radial coordinates $r_{\rm A}$ and $r_{\rm B}$, we get the following expressions
\begin{equation}  \label{eq:t_AB}
  t(r_B,r_A,J) = \frac{1}{J} \int_{r_A}^{r_B} \frac{r\,dr}{ \left( 1 - \frac{\rS}{r} \right) \sqrt{\frac{r^2}{J^2} - \left( 1 - \frac{\rS}{r} \right) }} \, .
\end{equation}
 For GWs or EM waves passing near an SMBH with $r_0$, the total change in the coordinate time from the source $r_{\rm orb}$ to the observer $r_{\rm obs}$ can be given as
\begin{equation}  \label{eq:deltat_AB}
   \Delta t \left( r_{\rm obs}, r_{\rm orb},J \right)  = t \left( r_{\rm orb},r_0,J \right) + t \left( r_{\rm obs},r_0,J \right) \, .
\end{equation}
 We define the time delay $\tau_{\rm d}$ between the primary GW and its GW echo as
\begin{equation}  \label{eq:t_delay_gen}
  \tau_{\rm d} = \lim_{r_{\rm obs} \to \infty } \Big[ \Delta t \left( r_{\rm obs}, r_{\rm orb},J_{\rm E} \right) - \Delta t \left( r_{\rm obs}, r_{\rm orb},J_{\rm P} \right) \Big] \, ,
\end{equation}
 where $J_{\rm E}$ and $J_{\rm P}$ are calculated respectively for the primary GW and the GW echo emitted by a merging BBH at a relative position $(r_{\rm orb}, \gamma)$ to the SMBH and the line-of-sight. We consider BBH mergers with a maximum radial distance from the SMBH out to $r_{\rm orb} = 10^4 \, \rS$ throughout the paper for which setting $r_{\rm obs} = 10^8 \, \rS$ yields an excellent approximation for practical purposes so that
\begin{equation}  \label{eq:t_delay_num}
  \tau_{\rm d} \approx \Delta t \left(10^8 \, \rS, r_{\rm orb},J_{\rm E} \right) - \Delta t \left( 10^8 \, \rS, r_{\rm orb},J_{\rm P} \right) \, .
\end{equation}
 Note that $\tau_{\rm d} \propto \rS$ (Equation \ref{eq:t_AB}), and it applies to SMBHs in the local Universe. For SMBHs at cosmological redshift $z$, the cosmological redshift effect has to be considered by multiplying $\tau_{\rm d}$ by a factor of $1+z$ or equivalently by assuming redshifted SMBH mass $M_{\rm SMBH,z} = (1+z) M_{\rm SMBH}$. Finally, we note that the additional observation direction-dependent variation in the time delay caused by the Doppler shift is much smaller in comparison, which is neglected here (Appendix \ref{Sec:Additional_Effects}).

\subsection{Magnification of primary GWs and GW echoes}
\label{subsec:Magnif_GWs}
 
 For an isotropic source with an arbitrary position relative to the SMBH and the line-of-sight to the observer, the magnification of the energy flux relative to the unlensed signal can be calculated in the short-wavelength limit as
\begin{equation}  \label{eq:mu_general}
  \mu_{\rm iso}  = \left( \frac{ r_{\rm SO} }{ r_{\rm orb} } \right)^2 \frac{ \sin{\Phi} }{ \sin{\gamma} } \left\arrowvert \frac{ d \Phi }{ d \gamma } \right\arrowvert 
\end{equation}
 \citep{BozzaMancini2004}. Using Equations \eqref{eq:FristSet} and \eqref{eq:SecondSet}, this may be expressed as
\begin{equation}  \label{eq:Magnif_GWs}
  \mu_{\rm iso} = \frac{ 1 }{ r_{\rm orb} } \frac{ J }{ \left| \sin{\gamma} \right| } \left\arrowvert \frac{ d J }{ d \gamma } \right\arrowvert  
  = \frac{ 1 }{ 2r_{\rm orb} } \left| \frac{ d (J^2)/dr_0 }{ d (\cos \gamma)/dr_0 } \right|  \, ,
\end{equation}
 where $J(r_0)$ and $\gamma(r_0)$ are given analytically in Equations \eqref{eq:alpha_r0},  \eqref{eq:J_r0}, and \eqref{eq:solve_lens_eq}. Here Equation \eqref{eq:mu_general} describes the magnification of the energy flux for point sources. Finally, we note that hereafter Equations \eqref{eq:mu_general} and \eqref{eq:Magnif_GWs} refer to the primary GW signal ($\mu_{\rm iso,P}$) and GW echo ($\mu_{\rm iso,E}$) if $-\pi \leqslant \gamma < 0$ and $0 < \gamma \leqslant \pi$, respectively, or equivalently when $0 < \alpha < \alpha_{\rm Einstein}$ and $\alpha_{\rm Einstein} < \alpha \lesssim \pi$.

 Equation \eqref{eq:Magnif_GWs} gives the amplitude magnification perceived by an observer relative to the unlensed isotropic source. To take into account the effect of anisotropic GW emission pattern for a merging binary with given $\mathbf{e}_{\rm AM}$ at a relative position $(r_{\rm orb},\gamma)$ with respect to the SMBH and the line-of-sight, we multiply $ \mu_{\rm iso}$ by the energy flux of lensed GW signal relative to that of the unlensed GW signal $Y(\Psi)/ Y(\Psi_{\rm U})$; see Equation \eqref{eq:Y(phi)}. Accordingly, the observed magnification of the energy flux of the primary GW and the GW echo relative to the unlensed GW are given as
\begin{equation}  \label{eq:Magnif}
  \mu_{\rm P} = \frac{Y(\Psi_{\rm P})}{Y(\Psi_{\rm U})} \mu_{\rm iso,P} \, , \quad   \mu_{\rm E} = \frac{Y(\Psi_{\rm E})}{Y(\Psi_{\rm U})} \mu_{\rm iso,E} \, ,
\end{equation}
 respectively, where $\Psi_{\rm P}$, $\Psi_{\rm U}$, and $\Psi_{\rm E}$ are determined by \mbox{Equation \eqref{eq:Psi_angles}}.

\begin{figure}
    \centering
    \includegraphics[width=77mm]{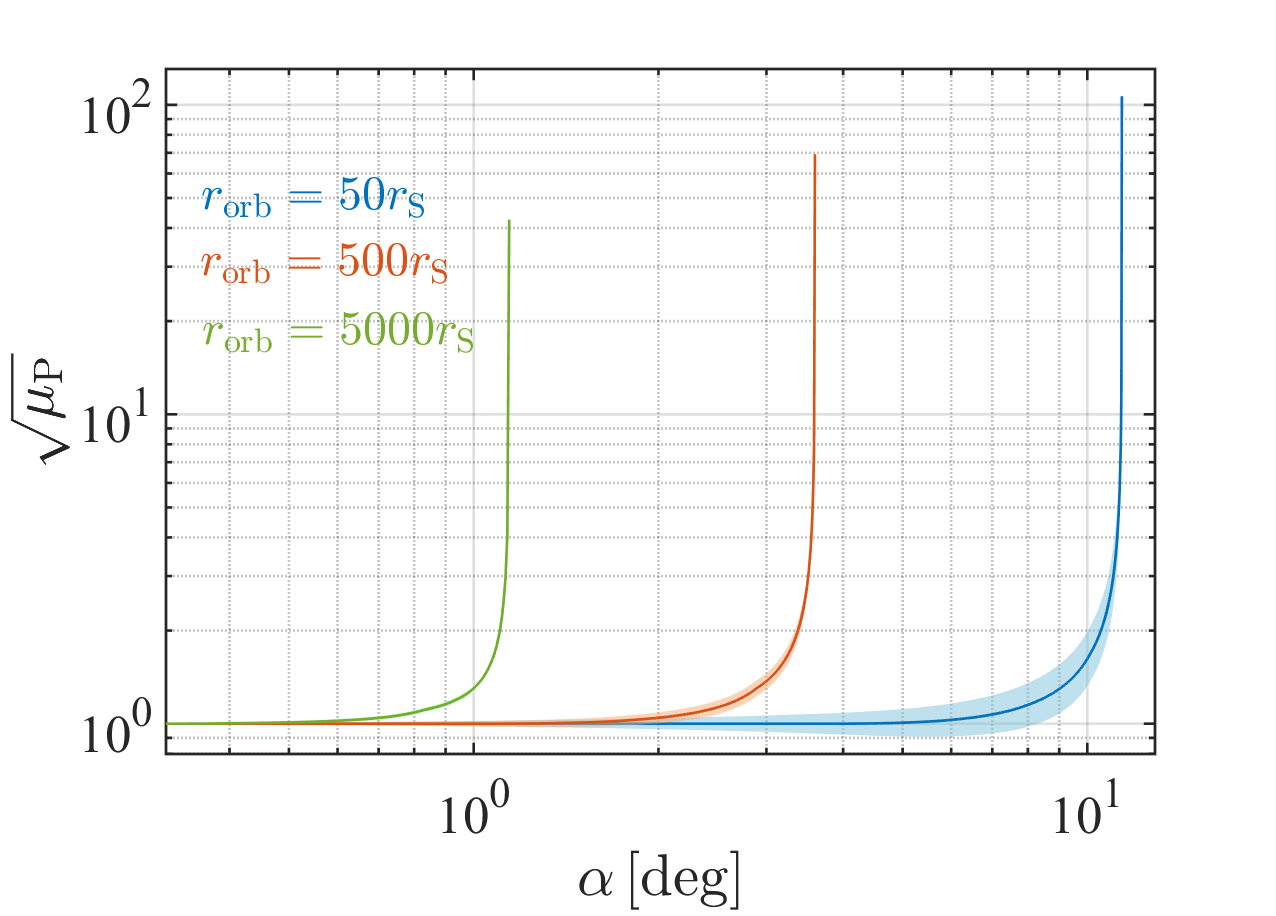}
    \\
    \includegraphics[width=77mm]{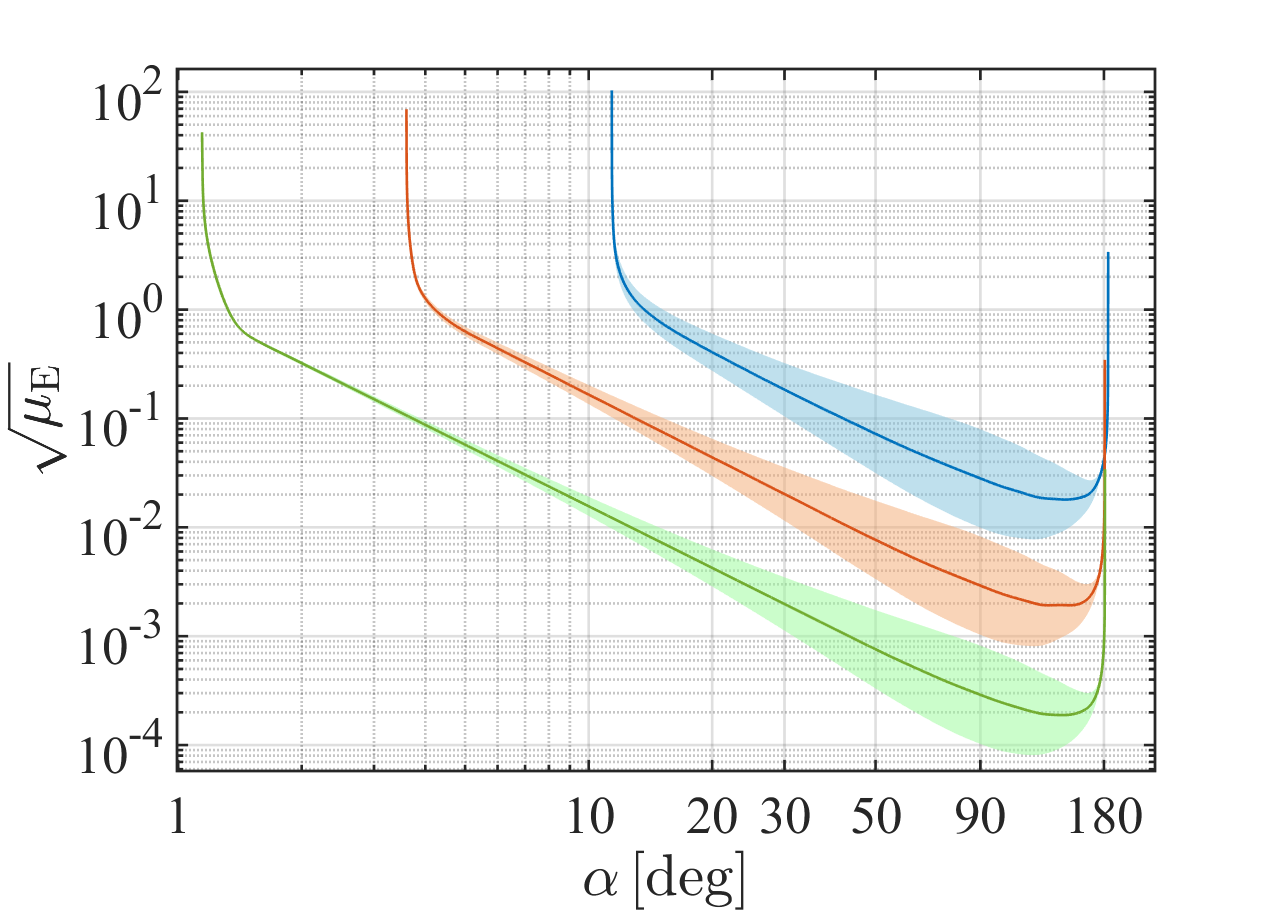}
    \\
    \includegraphics[width=77mm]{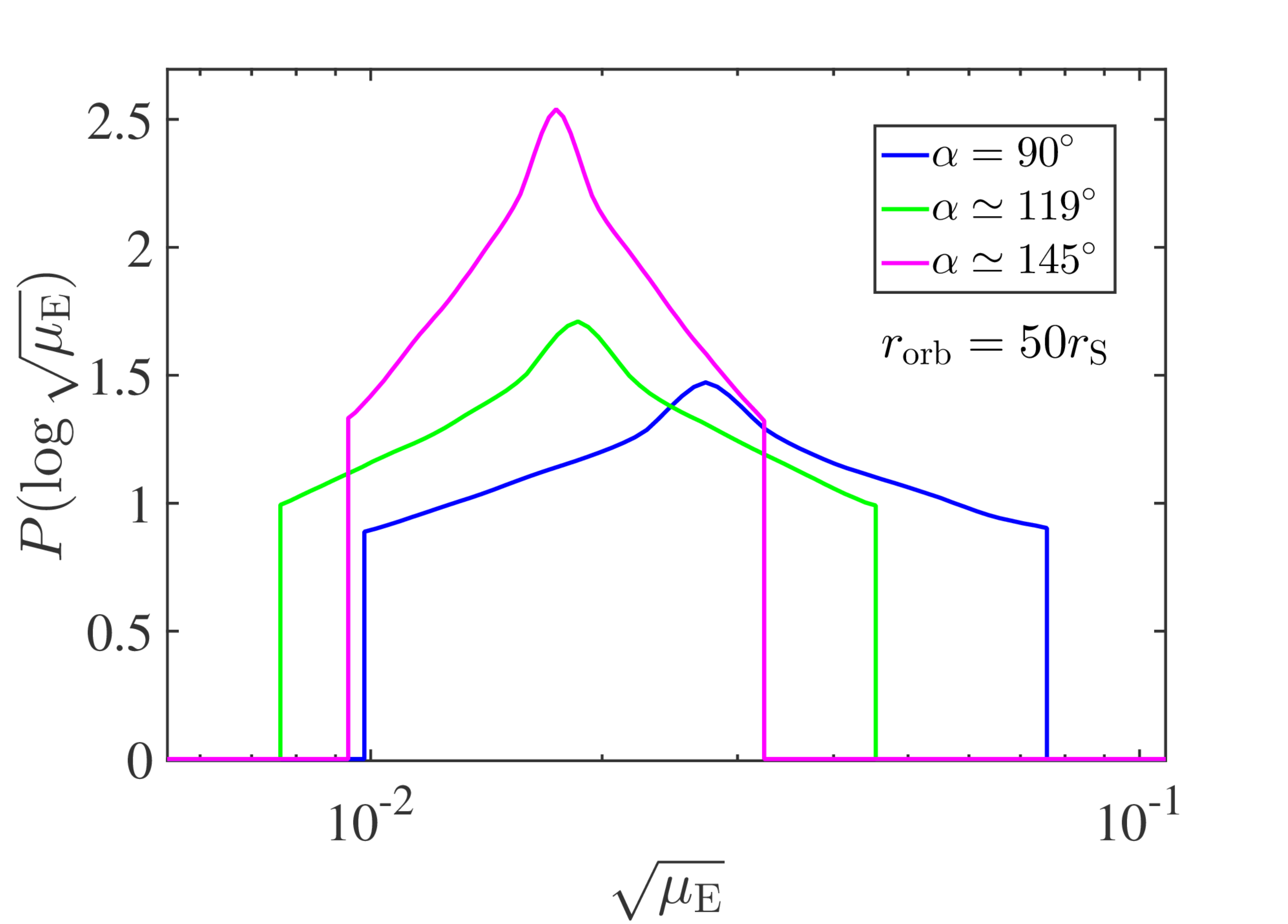}
\caption{\textit{Top and middle panels:} The GW amplification of the emitted primary GWs (top, $\sqrt{\mu_{\rm P}}$) and astrophysical GW echoes (middle, $\sqrt{\mu_{\rm E}}$) of a merging binary due to an SMBH lens as a function of deflection angle $\alpha$ for different source distances $r_{\rm orb}$ from the SMBH. We consider binaries with isotropic angular momentum distributions that are stationary with respect to the SMBH lens, relevant for aLIGO/AdV/KAGRA sources. Solid lines show the median of the amplification distribution, and shaded areas show the range of possible amplifications resulting from the anisotropic GW emission pattern of merging binaries. The strong amplification peak in the top and middle panels at small $\alpha$ corresponds to strong lensing at the Einstein radius, $\alpha_{\rm Einstein}$, while the second peak in the middle panel at $\sim \pi$ corresponds to the glory phenomenon in geometrical optics. Primary GWs and GW echos generally have $0 < \alpha < \alpha_{\rm Einstein}$ and $\alpha_{\rm Einstein} < \alpha \lesssim \pi$, respectively. \textit{Bottom panel:} The PDF of the logarithmic GW amplification for GW echoes, $P(\log \sqrt{\mu})$, at fixed $r_{\rm orb} = 50 \, \rS$ for three specific $\alpha$ values as labelled. We show the case with the broadest distribution among sources with fixed deflection angle ($\alpha = 90^{\circ}$), the case with the minimum $\sqrt{\mu_{\rm E}}$ ($\alpha \sim 119^{\circ}$, the minimum of the blue shaded area in the middle panel), and the case where the median of the $\sqrt{\mu_{\rm E}}$ distribution has a minimum ($\alpha \sim 145^{\circ}$, the minimum of the solid blue line in the middle panel).
\label{Fig:MagnifPrim_MagnifEcho} }
\end{figure}

 Since the GW amplitude scales with the square root of the energy flux (e.g. \citealt{Wangetal1996,TakahashiNakamura2003}), $\sqrt{\mu}$ represents the \textit{amplification}, i.e. the magnification of the GW amplitude relative to the unlensed GW due to lensing. Accordingly, $\sqrt{\mu_{\rm P}}$ and $\sqrt{\mu_{\rm E}}$ are the amplification of the primary GW and the GW echo, respectively. While the amplification is always larger than unity for the primary GW, it may be larger or lower than unity for the GW echo. In particular, for large-angle scattering, the echo amplitude is typically fainter than the unperturbed signal, as shown in Figure \ref{Fig:MagnifPrim_MagnifEcho}. Note that additional effects such as gravitational redshift and Doppler beaming are less important and are neglected in this study; see Appendix \ref{Sec:Additional_Effects} for details.

 Next, assuming an isotropic distribution of $\mathbf{e}_{\rm AM}$\footnote{Here and throughout the paper, we assume an isotropic distribution of the angular momentum unit vector $\mathbf{e}_{\rm AM}$ for merger events in isotropic GNs since there is no preferred direction for $\mathbf{e}_{\rm AM}$ in this environment. We make the same assumption for merger events in AGN/BH disks, as binary-single interactions are expected to randomize the direction of the binary orbital plane (e.g. \citealt{Samsing2020,Tagawaetal2020a,Tagawa2021a}).}, we investigate the main characteristics of the distribution of amplitude magnification $P(\sqrt{\mu})$ as a function of $r_{\rm orb}$ and $\alpha$ jointly for primary GWs and GW echoes. Accordingly, $P(\sqrt{\mu})$ is given as
\begin{align} \label{eq:AmplMafnifDist}
  P(\sqrt{\mu}) &= 
    \left\{
  \begin {array}{ll}
   P(\sqrt{\mu_{\rm P}})    &  \textrm{ if $0 < \alpha < \alpha_{\rm Einstein}$}  \\
   P(\sqrt{\mu_{\rm E}})   &  \textrm{ if $\alpha_{\rm Einstein} < \alpha \lesssim \pi$ } \\
   \end {array} \right.
\end{align}
 at any fixed $r_{\rm orb}$. We find that $P(\sqrt{\mu})$ peaks at $\sqrt{ \mu_{\rm iso}}$ at any fixed $\alpha$ and $r_{\rm orb}$, which is in addition to the median of the distributions. We also find that the width of the distributions is largest at $\alpha = \pi/2$, and in this case ${\rm max}(\sqrt{\mu}) / {\rm med}(\sqrt{\mu})={\rm med}(\sqrt{\mu}) / {\rm min}(\sqrt{\mu})\sim \sqrt{8}$ as expected from Equation \eqref{eq:Y(phi)} since a face-on binary is $Y(0)/Y(\pi/2) = 8$ times brighter in energy flux than an edge-on binary. We list some more properties of $P(\sqrt{\mu})$ in Appendix \ref{Sec:Anistropy_Properties}.

 We show examples for $\sqrt{\mu_{\rm P}}$ and $\sqrt{\mu_{\rm E}}$ as a function of $\alpha$ for various $r_{\rm orb}$ separately as incoming GWs are measured independently. We also depict examples for the probability density function (PDF) of the logarithmic GW amplification for GW echoes for some specific $\alpha$ values at a fixed $r_{\rm orb}$.\footnote{Note that similar distributions characterise the logarithmic GW amplification for primary GWs at any fixed $\alpha$ and $r_{\rm orb}$ values.} The top panel of Figure \ref{Fig:MagnifPrim_MagnifEcho} shows $\sqrt{\mu_{\rm E}}$ as a function of $\alpha$ for various $r_{\rm orb}$, where $0 < \alpha < \alpha_{\rm Einstein}$ (Section \ref{subsec:Lens_Delay}). \footnote{For each $r_{\rm orb}$, the presented logarithmic $\alpha$ range was divided into 8000 equal parts, and for each, we sampled $10^6$ unit vectors $\mathbf{e}_{\rm AM}$ from an isotropic distribution to determine the lower and upper limits of $P(\sqrt{\mu_{\rm P}})$.} Solid lines show the median of the $P(\sqrt{\mu_{\rm P}})$ distribution as a function of $\alpha$, which corresponds to the amplitude magnification of an isotropic-equivalent GW source. Furthermore, the shaded areas show the allowed range of amplification $\sqrt{\mu_{\rm P}}$ broadened by the anisotropic GW emission pattern of merging binaries. The peak corresponds to strong lensing at the Einstein radius $\alpha_{\rm Einstein}$. Similarly, the middle panel of Figure \ref{Fig:MagnifPrim_MagnifEcho} shows $\sqrt{\mu_{\rm E}}$ for the same $r_{\rm orb}$ values as in the top panel. In this case, $\alpha_{\rm Einstein} < \alpha \lesssim \pi$ (Section \ref{subsec:Lens_Delay}). The peak at small $\alpha$ corresponds to strong lensing at $\alpha_{\rm Einstein}$, and the second peak at $\sim \pi$ corresponds to the glory phenomenon in geometrical optics \citep{Futterman1988}. The bottom panel of Figure \ref{Fig:MagnifPrim_MagnifEcho} shows the $\log \sqrt{\mu_{\rm E}}$ distribution, $P(\log{\sqrt{\mu_{\rm E}}}) =\frac{1}{N} \frac{dN}{d\log\sqrt{\mu_{\rm E}}}$, where $N$ is the number of sources, and we also use a similar logarithmic representation for other parameters throughout the paper when plotting PDFs. The figure shows three cases with large deflection angles for fixed $r_{\rm orb} = 50 \, \rS$: (i) when the support of $P(\sqrt{\mu_{\rm E}})$ is the widest ($\alpha = 90^{\circ}$), (ii) when $\sqrt{\mu_{\rm E}}$ has a minimum ($\alpha \sim 119^{\circ}$), and (iii) when the median of $P(\sqrt{\mu_{\rm E}})$ has a minimum ($\alpha \sim 145^{\circ}$). Clearly, $P(\sqrt{\mu_{\rm E}})$ is approximately symmetric on a logarithmic $\sqrt{\mu_{\rm E}}$ scale, and so it peaks near the median.

 We define the fraction of the GW echo amplitude relative to the primary GW as
\begin{equation}  \label{eq:muEP}
   \sqrt{\mu_{\rm EP}} = \frac{ \sqrt{\mu_{\rm E}} }{ \sqrt{\mu_{\rm P}} } \, .
\end{equation}
 The echo amplitude is typically smaller than the primary amplitude, i.e. $\sqrt{\mu_{\rm EP}} < 1$. However, $\sqrt{\mu_{\rm EP}} \gtrsim 1$ is also possible for (i) $\gamma \sim \pi$ because in this case $\sqrt{\mu_{\rm P}} \sim 1$ and $\sqrt{\mu_{\rm E}}$ peaks due to the retro-lensing glory phenomenon, and for (ii) $\gamma \sim 0$ because in this case $\sqrt{\mu_{\rm P}} \simeq \sqrt{\mu_{\rm E}}$; see Figure \ref{Fig:MagnifPrim_MagnifEcho} for examples, and for (iii) cases where the energy flux at the source in the direction along the echo rays is stronger than along the primary signal rays.

 Importantly, for any $r_{\rm orb}$, the minimum of both $\sqrt{\mu_{\rm P}}$ and $\sqrt{\mu_{\rm E}}$ are finite non-zero. Accordingly, we determine the minimum value of the relative echo amplitude $\sqrt{\mu_{\rm EP}}$ as a function of $r_{\rm orb}$ in Appendix \ref{Sec:Anistropy_Properties}, which sets a strict lower bound on the allowed range of $r_{\rm orb}$ for high SNR sources, where an astrophysical GW echo is not detected.  In these investigations, we assume that the SNR of both the primary GW and the GW echo are above the SNR detection threshold.

%% file: Sections/Results.tex
\section{Results} 
\label{sec:Results}
 
 First, we study the characteristics of the $\sqrt{\mu_{\rm EP}}$ distribution, $P(\sqrt{\mu_{\rm EP}})$, for single AGN and BH disks with a fixed inclination and isotropic distributions of BBH mergers around SMBHs in Sections \ref{subsec:SingleDisk_IsotropDist}. Next, in Section \ref{subsec:Disk_IsotropDist_ObsBias}, we determine the $\sqrt{\mu_{\rm EP}}$ distribution in a mock SNR-limited survey of the primary GWs accounting for observational bias, i.e. a larger detection volume for BBH mergers with larger primary GW amplitudes. Then, in Section \ref{subsec:RelPos_TimeDelay}, we determine the distribution of angular position relative to the SMBH and the line-of-sight, the distribution of deflection angle for the echo, and the time-delay distribution between the primary GW and the GW echo for BBH mergers in isotropic populations around SMBHs in a mock SNR-limited survey of both GW echoes and primary GWs.

\begin{figure*}
    \centering
    \includegraphics[width=80mm]{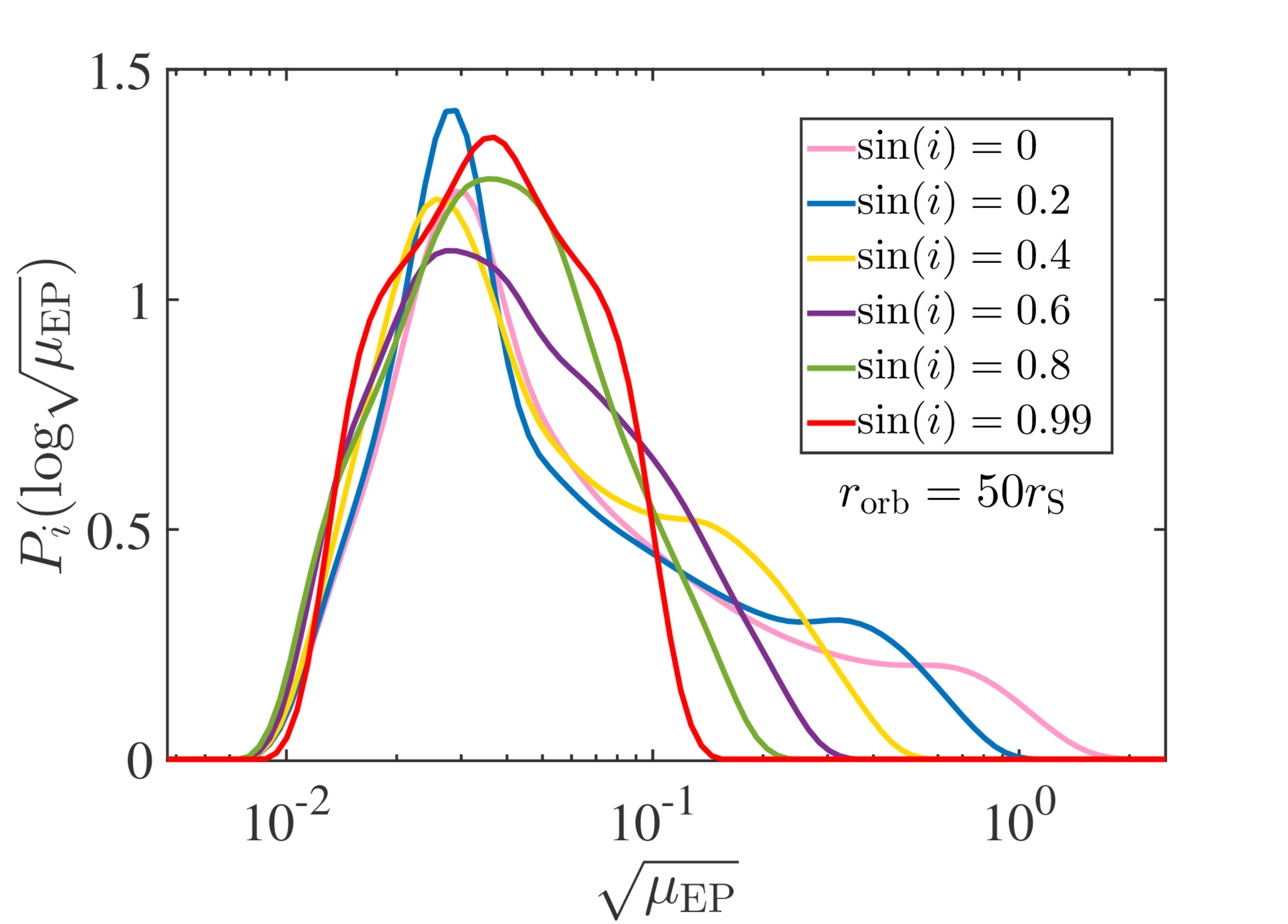}
    \includegraphics[width=80mm]{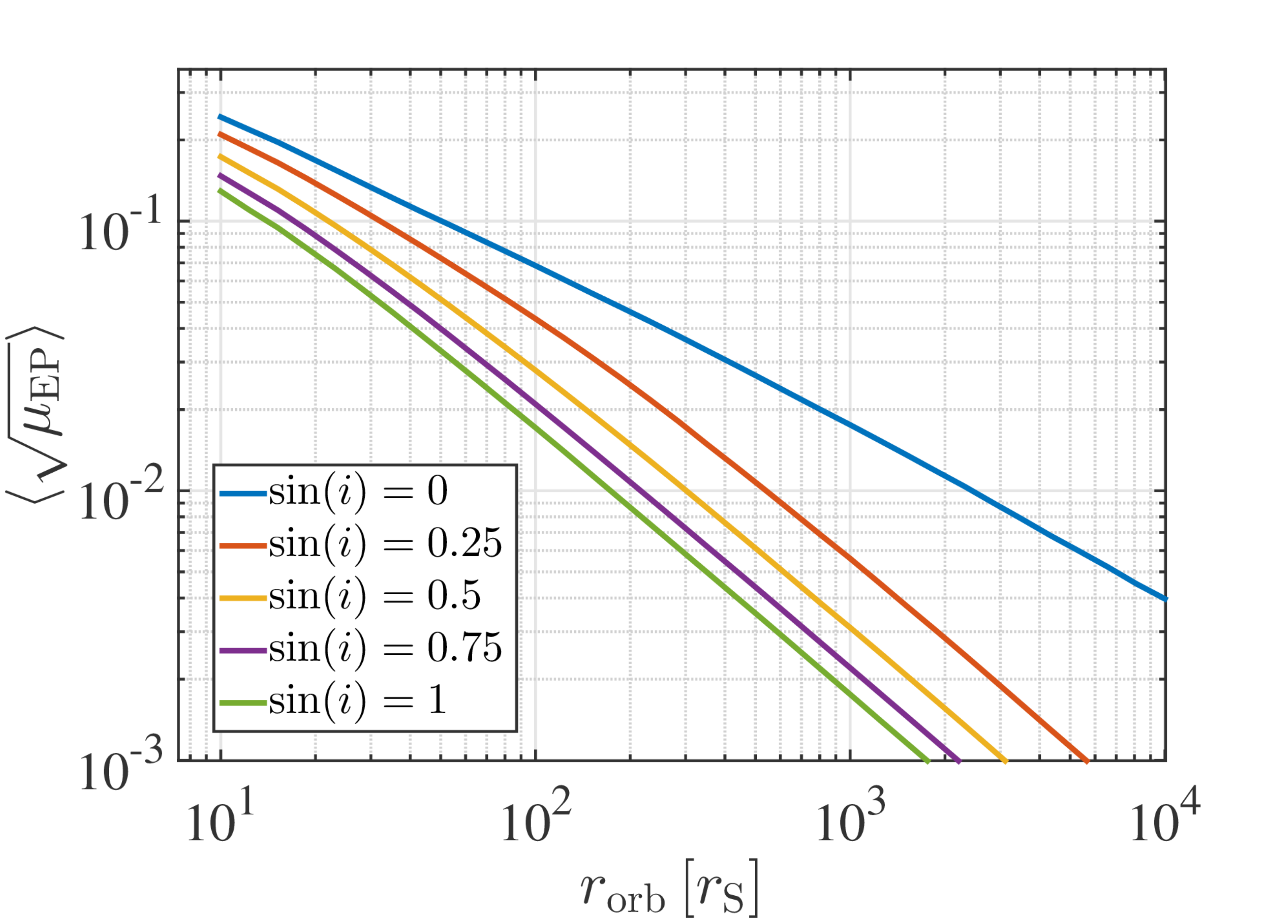}
    \\
    \includegraphics[width=80mm]{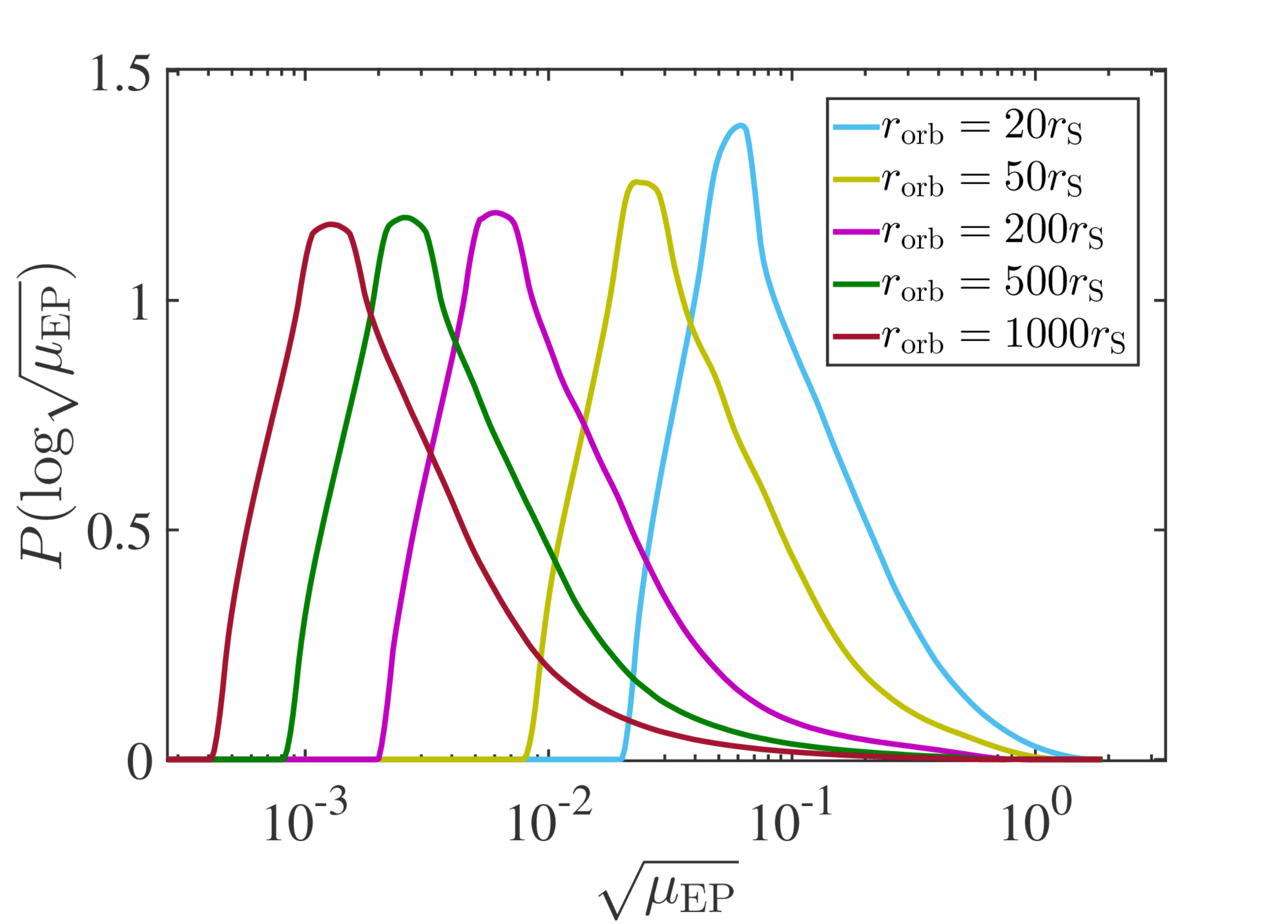}
    \includegraphics[width=80mm]{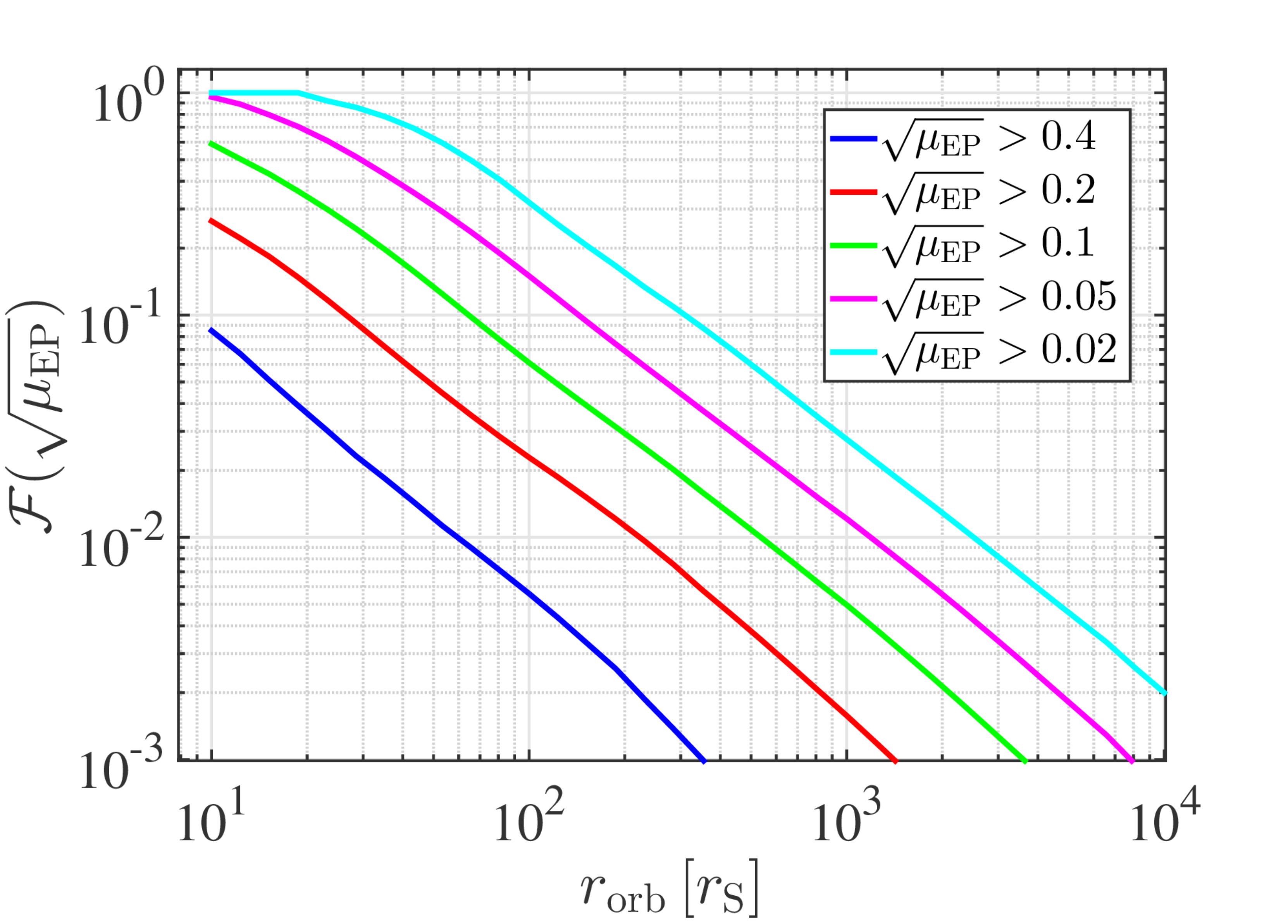}
\caption{\textit{Top left panel:} The distribution of the logarithmic GW echo amplitude relative to the primary GW signal for BBH mergers in a single AGN or a BH disk with fixed inclination $i$, $P_i(\log \sqrt{\mu_{\rm EP}})$. Examples are shown for various inclination angles as labelled in the legend, and the merger distance from the SMBH is fixed to $r_{\rm orb} = 50 \, \rS$. \textit{Top right panel:} The mean relative amplitude of the GW echo compared to the primary GW signal, $\langle \sqrt{\mu_{\rm EP}} \rangle$, as a function of $r_{\rm orb}$ for BBH mergers in an AGN/BH disk with a fixed inclination as shown in the legend. \textit{Bottom left panel:} Same as the top left panel, but for an isotropic distribution of BBH mergers around a single SMBH or around SMBHs in a mock complete volume-limited survey (e.g. either isotropic GNs or an ensemble of AGN/BH disks in both cases), $P(\log \sqrt{\mu_{\rm EP}})$, for various $r_{\rm orb}$ as labelled in the legend. \textit{Bottom right panel:} The cumulative fraction of BBH mergers among an isotropically distributed sample of mergers at a distance $r_{\rm orb}$ from a single SMBH or in a population of galactic nuclei in a mock complete volume-limited survey $\mathcal{F}(\sqrt{\mu_{\rm EP}})$ for which $\sqrt{\mu_{\rm EP}}$ is higher than the values given in the legend. For example, the bottom curve shows the fraction of mergers in a single GN where the GW echo amplitude relative to the primary GW is larger than $\sqrt{\mu_{\rm EP}} = 0.4$, which may be detectable even for weak marginally detected primary GWs. For very bright primary GWs, the fraction of mergers with a detectable GW echo with a relative amplitude of $\sqrt{\mu_{\rm EP}} > 0.02$ reaches $100\%$ for $r_{\rm orb} \lesssim 20 \, \rS$ (see Figure \ref{Fig:RelmagnifMin_vs_rorb} for the critical radius for a guaranteed GW echo detection for arbitrary $\sqrt{\mu_{\rm EP}}$).
  \label{Fig:Single_Disks_GNs} } 
\end{figure*}

\subsection{GW echos from single AGN/BH disks and isotropic nuclear star clusters around SMBHs}
\label{subsec:SingleDisk_IsotropDist}
 
 Let us first investigate how the distribution of GW echo amplitude relative to the primary GW signal,  $P_i(\sqrt{\mu_{\rm EP}})$, and its mean, $\langle \sqrt{\mu_{\rm EP}} \rangle$, depend on the inclination of the GW source's position plane around the SMBH (i.e. the AGN/BH disk plane) relative to the line-of-sight and the SMBH-binary distance. Second, we determine $P(\sqrt{\mu_{\rm EP}})$, the relative GW amplitude for an isotropic distribution of BBH mergers around a single SMBH or around SMBHs in a mock complete volume-limited survey (e.g. either isotropic GNs or an ensemble of AGN/BH disks in both cases) in which the detection of primary GWs is complete within some maximum detection distance.

 To determine $P_i(\sqrt{\mu_{\rm EP}})$ for BBH mergers in an AGN/BH disk with fixed inclination $i$ and fixed SMBH-binary distance $r_{\rm orb}$, we first sample the longitude angles of both the source $\varphi_{\rm S}$ and observer $\varphi_{\rm O}$ in the disk plane uniformly between $[0, 2 \pi]$ and compute the corresponding $\gamma$ (Equation \ref{eq:gammaCalc}). Next, we sample angular momentum unit vectors $\mathbf{e}_{\rm AM}$ from an isotropic distribution at each $\gamma$, i.e. we sample $\phi_{\rm AM}$ uniformly between $[0, 2 \pi]$ and $\cos{\Theta_{\rm AM}}$ uniformly between $[-1,1]$ (Section \ref{subsec:Lensing_Geom}), then compute $\sqrt{\mu_{\rm EP}}$ for each $(\gamma, \phi_{\rm AM}, \Theta_{\rm AM})$ in the MC sample as discussed in Section \ref{subsec:Magnif_GWs}. Finally, using the resulting MC sample, we generate the PDF of the relative echo amplitude, i.e. $P_i(\sqrt{\mu_{\rm EP}})$.\footnote{We used an MC sample of $10^6$ mergers to generate $P_i(\sqrt{\mu_{\rm EP}})$.} Throughout the paper, we verify the convergence of generated distributions as a function of sample size by evaluating the Kolmogorov–Smirnov test with respect to the final distributions.

 The top left panel of Figure \ref{Fig:Single_Disks_GNs} displays $P_i(\log \sqrt{\mu_{\rm EP}})$ for various $i$ values at fixed $r_{\rm orb} = 50 \, \rS$. The obtained distributions are bimodal, where the peak at low $\sqrt{\mu_{\rm EP}}$ corresponds to BBH mergers with $\gamma \sim \pi/2$, while the second peak is produced by mergers close to either $\gamma \sim \gamma_{\rm min} = i$ or \mbox{$\gamma_{\rm max} = \pi - i$} (Section \ref{subsec:GeometricConvects}). $P_i(\sqrt{\mu_{\rm EP}})$ is bounded between lower and upper relative amplitudes corresponding to BBH mergers respectively near \mbox{$\gamma \sim \gamma_{\rm magn,min} \sim 113^{\circ}$} and near either $\gamma \sim \gamma_{\rm min}$ or $\gamma_{\rm max}$.\footnote{Here, $\gamma_{\rm magn,min} \sim 113^{\circ}$ is the angle between the SMBH-binary direction and the optical axis where $\sqrt{\mu_{\rm EP}}$ attains its minimum (Appendix \ref{Sec:Anistropy_Properties}). Note that the upper bound of $\sqrt{\mu_{\rm EP}}$ decreases systematically with $i$, while the lower bound is the smallest at $i \sim 67^{\circ}$ ($\sin{i} \sim 0.92$)}. Furthermore, the range of $\sqrt{\mu_{\rm EP}}$ for fixed $r_{\rm orb}$ approaches \mbox{$\sqrt{\mu_{\rm EP}}(\gamma = \pi/2)$} with decreasing $i$ (Section \ref{subsec:GeometricConvects}) where $P_i(\sqrt{\mu_{\rm EP}})$ forms a sharp peak in the $i \rightarrow \pi/2$ limit. Note that $P_i(\sqrt{\mu_{\rm EP}})$ shifts toward lower values as $r_{\rm orb}$ increases because $\sqrt{\mu_{\rm E}}$ decreases systematically with increasing $r_{\rm orb}$ while $\sqrt{\mu_{\rm P}}$ remains $\sqrt{\mu_{\rm P}} \sim 1$ for the vast majority of mergers as only a small portion of them are located near $\alpha_{\rm Einstein}$.

 The top right panel of Figure \ref{Fig:Single_Disks_GNs} shows the mean relative echo amplitude $\langle \sqrt{\mu_{\rm EP}} \rangle$ as a function of $r_{\rm orb}$ for different $i$ as labelled. Clearly, $\langle \sqrt{\mu_{\rm EP}} \rangle$ increases monotonically with $i$ from the face-on to the edge-on disk configuration, and it decreases monotonically for larger $r_{\rm orb}$. The former trend arises because a systematically higher fraction of $\gamma$ values occur closer to either $\gamma \sim 0$ or $\gamma \sim \pi$ for lower $i$, where $\sqrt{\mu_{\rm EP}}$ is the highest. The latter trend follows from the $r_{\rm orb}$ dependence of $P_i(\sqrt{\mu_{\rm EP}})$. In particular, $\langle \sqrt{\mu_{\rm EP}} \rangle$ is \mbox{$\sim 2.03 \left(\log(r_{\rm orb} / 10 \, \rS) \right)^{2.1} + 1.92$} times higher for edge-on disks ($\sin{i} = 0$ blue curve) than for face-on disks ($\sin{i} = 1$ green curve), e.g. a factor of \mbox{$\sim \{2,\, 4,\, 11 \}$} times higher for $r_{\rm orb} \sim \{ 10 \, \rS,\, 100 \, \rS,\, 1000 \, \rS \}$.\footnote{Throughout the paper, we use the non-linear least squares method to fit the coefficients of functions to be fitted, and resulting values are given at $95 \%$ confidence level.} Therefore, astrophysical GW echoes with the highest amplitudes are expected to be typically detected from nearly edge-on AGN/BH disks in the inner regions close to the SMBH.

 Similar to the top left panel of Figure \ref{Fig:Single_Disks_GNs}, the bottom left panel shows the distribution of the logarithmic relative GW echo amplitude $P(\log \sqrt{\mu_{\rm EP}})$ for isotropically distributed BBH mergers around a single SMBH or around SMBHs in a mock complete volume-limited survey. $P(\sqrt{\mu_{\rm EP}})$ is determined by marginalizing $P_i(\sqrt{\mu_{\rm EP}})$ over $i$.\footnote{$P(\sqrt{\mu_{\rm EP}})$ was generated from an MC sample of $5 \times 10^6$ elements.} The figure shows that $P(\sqrt{\mu_{\rm EP}})$ has one peak close to \mbox{$\sqrt{\mu_{\rm EP}} \sim 0.121 \, (r_{\rm orb}/ 10 \, \rS)^{-1}$}, which corresponds to BBH mergers at $\gamma \sim \pi/2$. The lower bound, where $P(\sqrt{\mu_{\rm EP}})$ is nonzero, corresponds to Equation \eqref{eq:mu_EP_min}, while the upper bound corresponds to mergers at either $\gamma \sim 0$ or $\gamma \sim \pi$. Furthermore, $P(\sqrt{\mu_{\rm EP}})$ shifts systematically toward lower values with increasing $r_{\rm orb}$ roughly as $r_{\rm orb}^{-1}$ because  $P_i(\sqrt{\mu_{\rm EP}})$ also shifts to lower values with increasing $r_{\rm orb}$ for all $i$.

 Finally, the bottom right panel of Figure \ref{Fig:Single_Disks_GNs} shows the cumulative fraction of BBH mergers located at a radial distance of $r_{\rm orb}$ to the SMBHs, for which the relative GW echo amplitude is larger than the values shown in the legend, $\mathcal{F}(\sqrt{\mu_{\rm EP}})$. Here we assume an isotropic merger population around a single SMBH or around SMBHs in a mock complete volume-limited survey. We find that $\mathcal{F}(\sqrt{\mu_{\rm EP}})$ decreases monotonically with increasing $r_{\rm orb}$, which arises because $P(\sqrt{\mu_{\rm EP}})$ shifts systematically toward lower $\sqrt{\mu_{\rm EP}}$ values for larger $r_{\rm orb}$. The figure shows that $\mathcal{F}(\sqrt{\mu_{\rm EP}})=1$ for \mbox{$r_{\rm orb} \leqslant 20 \, \rS$} for $\sqrt{\mu_{\rm EP}} > 0.02$, meaning that the GW echo amplitude is at least 0.02 of the primary amplitude for all sources with \mbox{$r_{\rm orb} \leqslant 20 \, \rS$} (see Equation \ref{eq:mu_EP_min}). Furthermore, at least $1 \%$ of GW echoes have amplitudes larger than $\{ 0.4, 0.2, 0.1, 0.05, 0.02 \}$ times the primary GW amplitude if \mbox{$r_{\rm orb} \lesssim \{ 60 \, \rS, 210 \, \rS, 500 \, \rS, 1100 \, \rS, 2400 \, \rS \}$}, respectively, and more than $10 \%$ of GW echoes have amplitudes larger than $\{0.2, 0.1, 0.05, 0.02 \}$ times the primary amplitude, respectively, for BBH mergers closer to the SMBH than \mbox{$\sim \{ 25 \, \rS, 60 \, \rS, 135 \, \rS, 300 \, \rS \}$}. This indicates that for the most typical weak primary GW signals in a complete volume-limited survey, there is at least a $1\%$ probability of detecting a GW echo with $0.4$ of the primary's amplitude for BBH mergers within $\sim 60 \, \rS$ of the SMBH, but for very strong primary signals for which an echo is detectable at $0.02$ of the primary's amplitude, there is close to $100\%$ chance to detect a GW echo for sources within $20 \rS$, and still a nonzero ($1\% \lesssim$) chance to detect an echo from distances up to $\sim 2400 \, \rS$.

\subsection{GW echos from AGN/BH disks and isotropic nuclear star clusters in an SNR-limited survey}
\label{subsec:Disk_IsotropDist_ObsBias}
 
 In this section, we investigate the impact of observational bias on the previous results presented for single AGN/BH disks and isotropic distributions of BBH mergers around SMBHs in Section \ref{subsec:SingleDisk_IsotropDist}. Here we examine the distributions in a mock SNR-limited GW observational survey of the primary GWs, where BBH mergers with larger primary GW amplitudes are detectable from a larger volume.

 As we have seen in Section \ref{subsec:SingleDisk_IsotropDist}, primary GWs typically have higher amplitudes compared to GW echoes, and by definition, arrive earlier at the detector due to their lower light-travel time around the SMBH (e.g. \citealt{Kocsis2013}). Thus, assuming that the primary GW signal has been detected above a specific SNR detection limit, we now investigate what fraction of sources may have a detectable GW echo. Here we account for the amplitude dependence of the detection volume of the primary GW signal, which leads to an \textit{observational bias} skewing the echo amplitude distribution of single SMBHs presented above in Section \ref{subsec:SingleDisk_IsotropDist}. The number of detectable mergers scales with the detection volume. Hence, the relative fraction of primary GWs with $\sqrt{ \mu_{\rm P} }$ in an MC sample scales as $N_{\rm P} \propto V_{\rm P} \propto \mu_{\rm P}^{3/2}$. Thus, we assign the detection-volume based weight $w_{\rm P} = \mu_{\rm P}^{3/2}$ to each element in the MC sample to define the detection-volume weighted PDF of any given variable $x$, $P_{\rm det,P}(x)$.\footnote{This method assumes a homogeneous density of BBH mergers in an Euclidean geometry, which gives an adequate first estimate for the observable distribution despite for its simplicity of neglecting the inhomogeneous clustering of mergers on small scales and cosmological effects (see also, \citealt{Abadieetal2010,OShaughnessyetal2010,Dominiketal2015,Belczynskietal2016b,Tagawaetal2020b}, for a similar estimate for various other source populations).} Note that here and throughout the paper we assume a $100\%$ detector duty cycle and a polarization-averaged waveform. Presented results may be affected at some level for a sparse GW detector network when the line of nodes of the antenna beam pattern of the individual instruments is considered.

 In practice, we generate the PDF as follows. We first assign a weight $w_{\rm P}$ to each element in an MC sample of Section \ref{subsec:SingleDisk_IsotropDist} that was generated for sources near an individual SMBH or near a population of SMBHs in a mock complete volume-limited survey. Such an MC sample is a table of GW source parameters ($r_{\rm orb}, \gamma, \phi_{\rm AM}, \Theta_{\rm AM}$). Next, $\sqrt{\mu_{\rm P}}$, $\sqrt{\mu_{\rm E}}$, and  $\sqrt{\mu_{\rm EP}}$ are calculated using Equations \eqref{eq:Magnif_GWs} and \eqref{eq:Magnif} from the GW source parameters. Then we determine the cumulative distribution function (CDF) of the detection-volume weighted elements\footnote{i.e., the CDF of the variable $x$ is given as ${\rm CDF}_{\rm det,P}(x_k) = \sum_k w_k / \sum_g w_g$, where $k$ indices those elements in the MC sample for which the corresponding $x$s are lower than $x_k$, and $g$ runs over the sample. Here, $x$ may represent $\sqrt{\mu_{\rm EP}}$, for example, and $g$ may represent all sources with some fixed $r_{\rm orb}$ and $i$ or cases with fixed $r_{\rm orb}$ but with arbitrary $i$.}, and finally differentiate the CDF with respect to $x$ to get $P_{\rm det,P}(x)$ probability density function.\footnote{An MC sample of $5 \times 10^7$ elements was used to generate $P_{\rm det,P}(x)$.}

\begin{figure*}
    \centering
    \includegraphics[width=80mm]{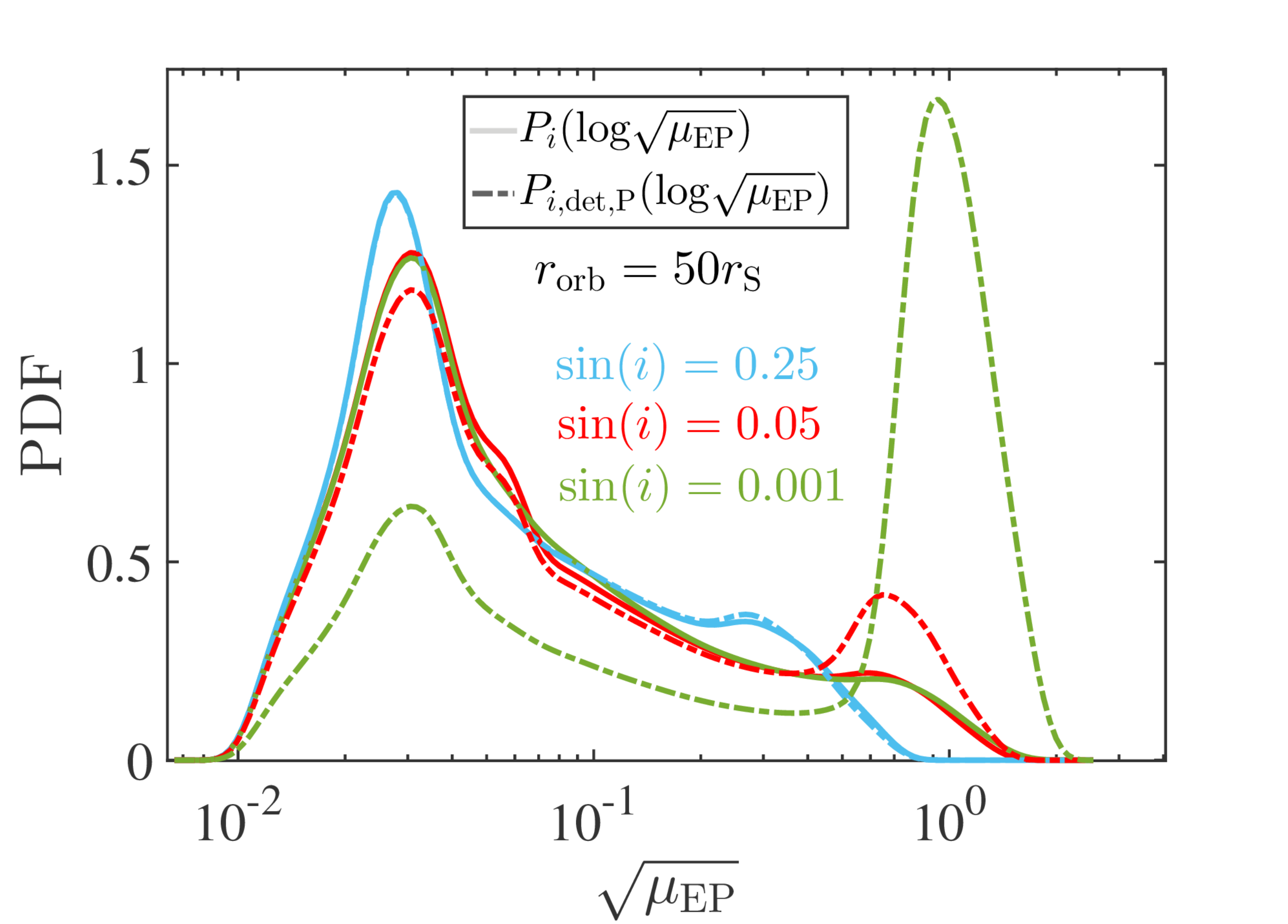}
    \includegraphics[width=80mm]{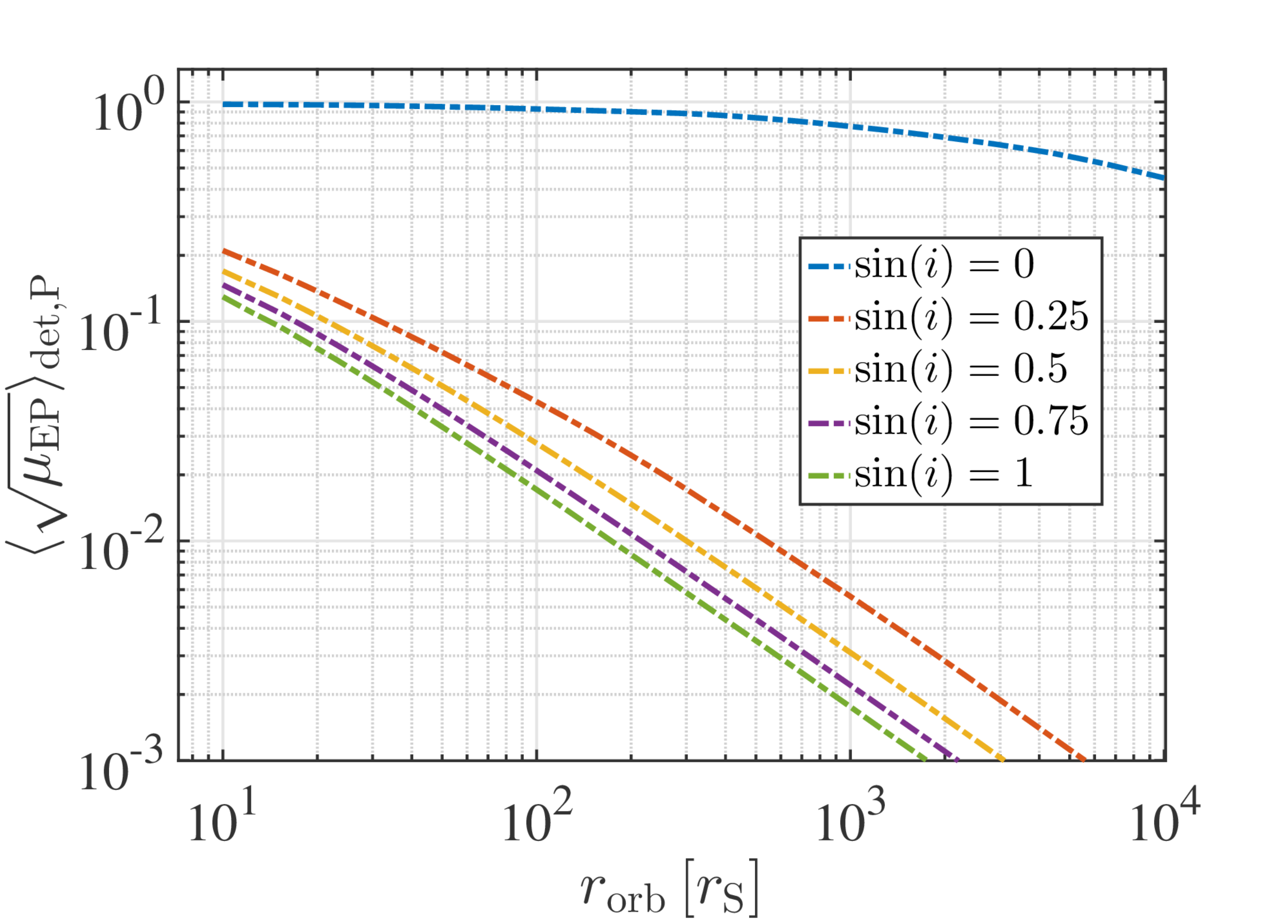}
    \\
    \includegraphics[width=80mm]{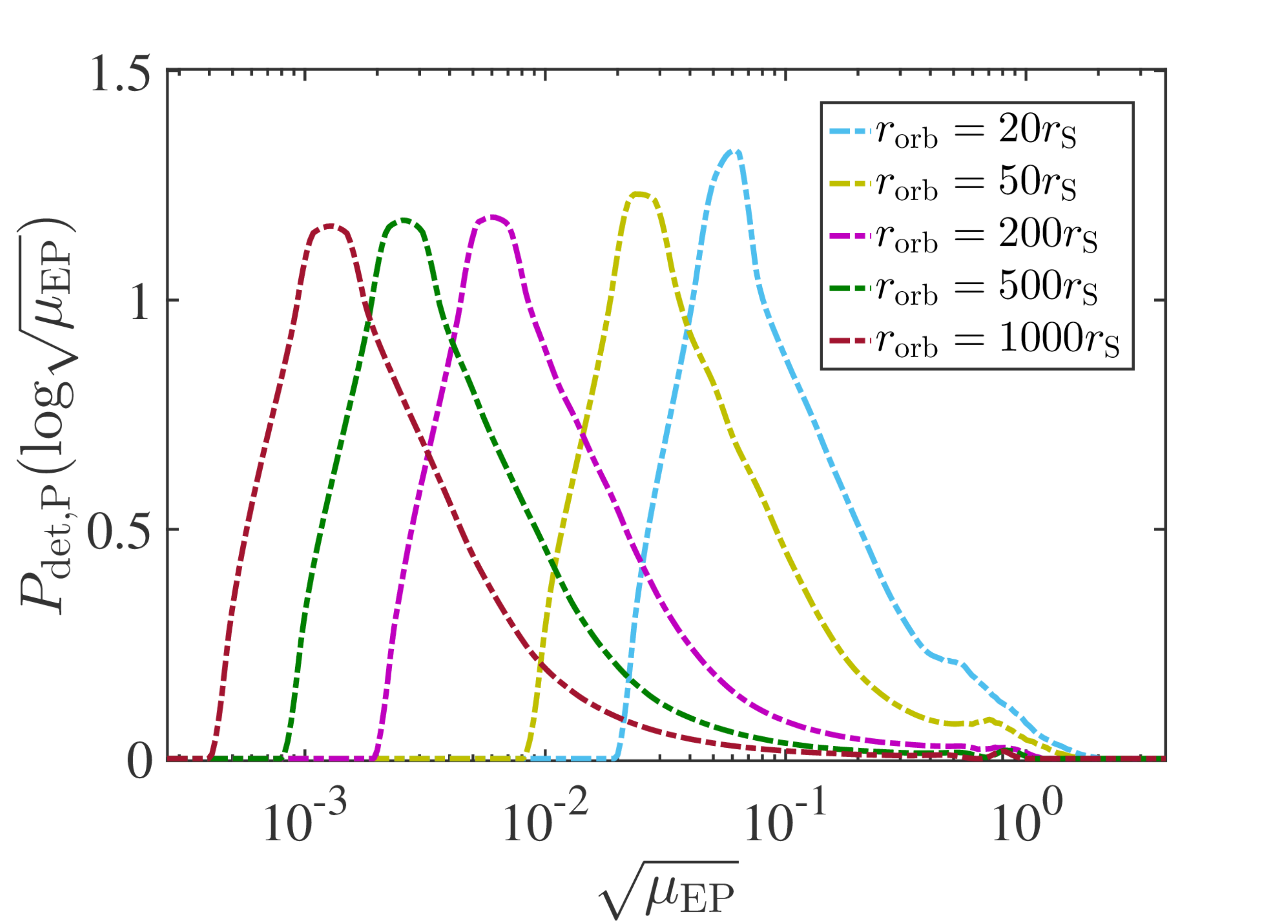}
    \includegraphics[width=80mm]{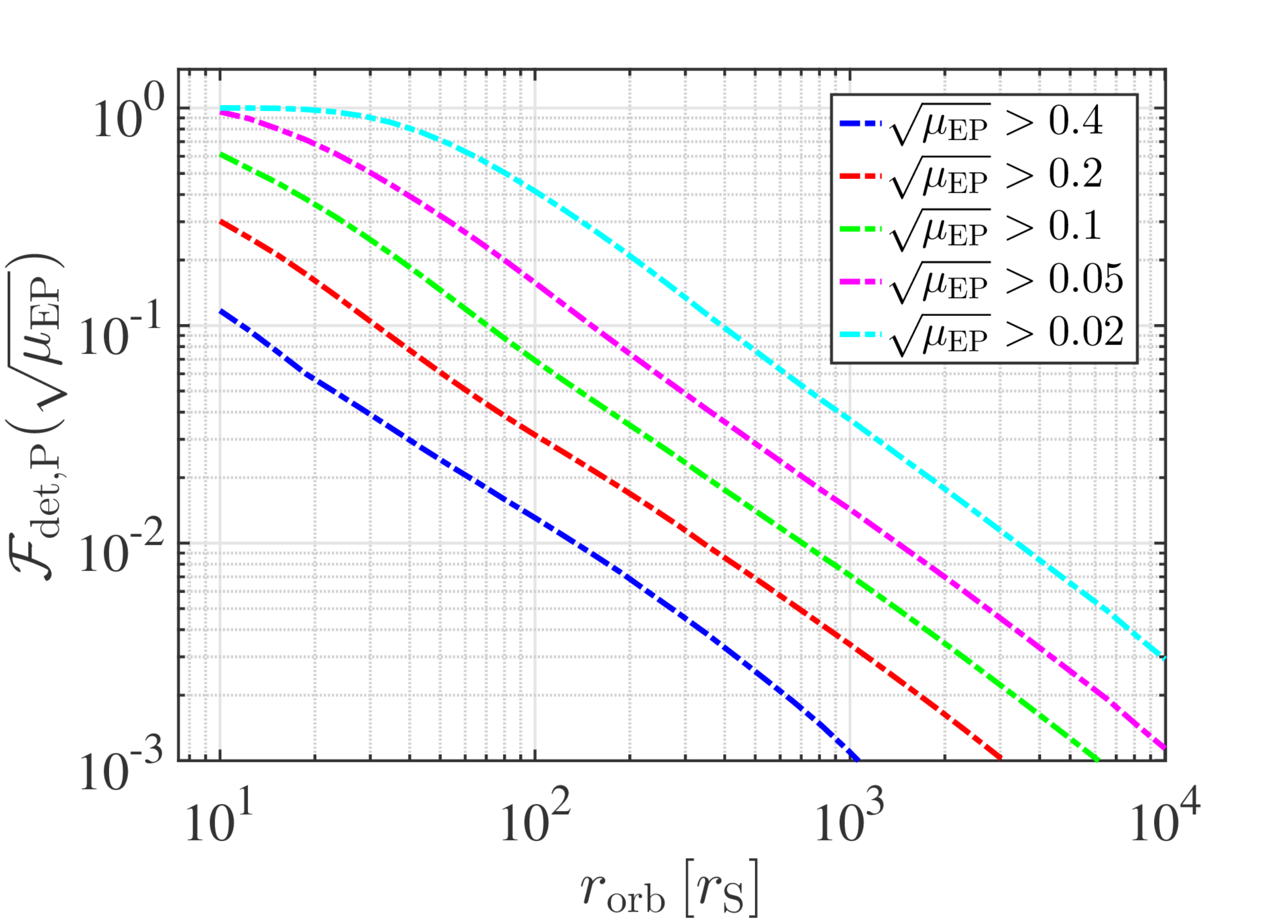}
    \\
    \includegraphics[width=80mm]{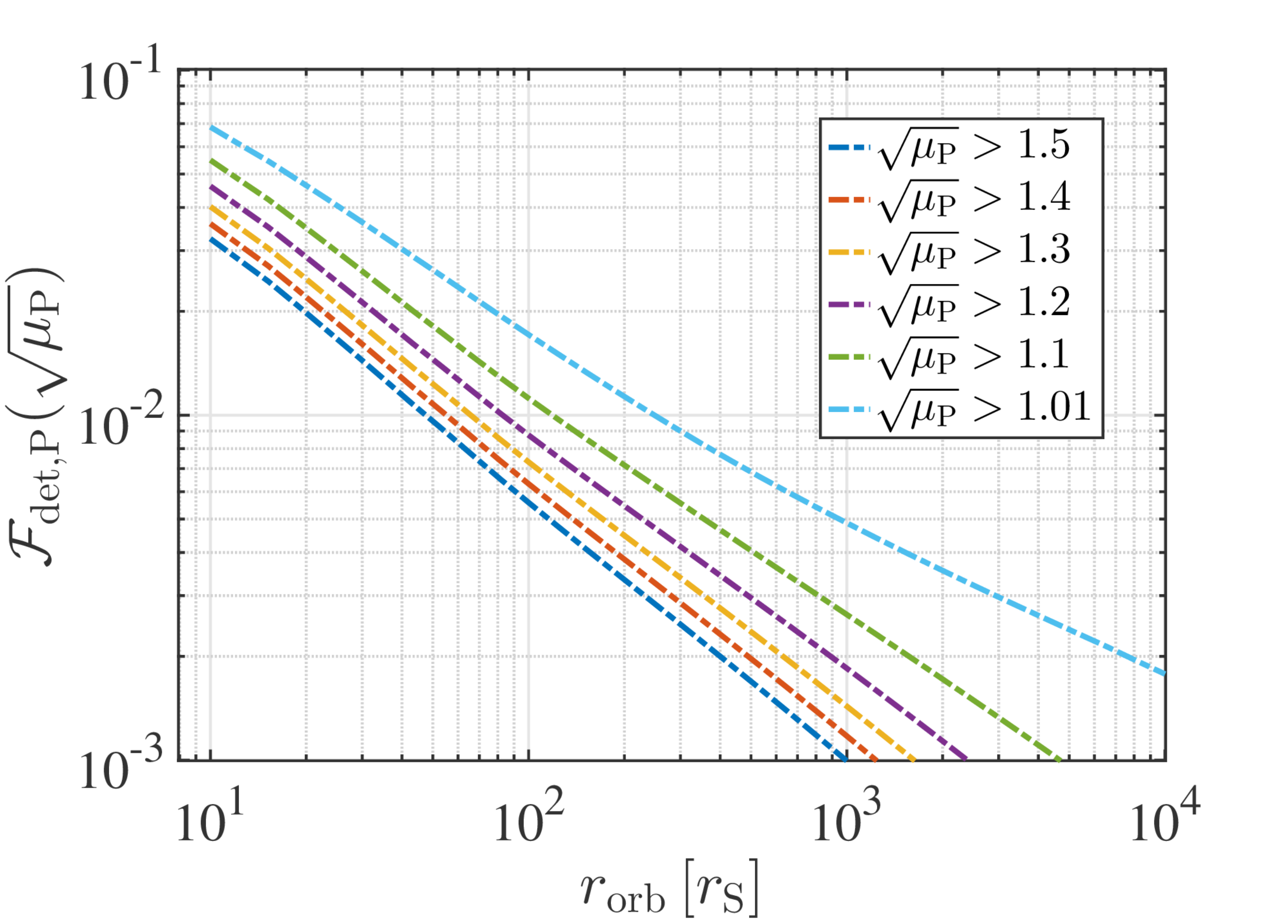}
\caption{The first and second rows are similar to Figure \ref{Fig:Single_Disks_GNs}, showing the properties of the relative GW echo amplitude distribution over the primary GWs, but for a mock SNR-limited GW observational survey accounting for observational bias by scaling the probabilities with the detection volume of the primary GW signal (Section \ref{subsec:Disk_IsotropDist_ObsBias}). The top left panel shows the distributions of $\log \sqrt{\mu_{\rm EP}}$ with and without observational bias using dashed and solid lines, respectively. The inclination angle of the AGN/BH disks is fixed as labelled. The top right panel shows that the mean relative echo amplitude is strongly increased in a mock SNR-limited survey by observational bias for edge-on disks ($\sin{i} = 0$) due to strong lensing. The middle row left panel shows that the distributions of $\sqrt{\mu_{\rm EP}}$ for isotropic BBH merger populations around SMBHs (or an ensemble of AGN/BH disks) acquire a long tail to high echo amplitudes due to strongly lensed sources, but this affects only a tiny fraction of sources at any fixed $r_{\rm orb}$. The right panel in the middle row shows the fraction of sources above a given $\sqrt{\mu_{\rm EP}}$ in a mock SNR-limited survey, as labelled, which is very similar to that of Figure \ref{Fig:Single_Disks_GNs} as observational bias affects only a tiny fraction of sources. \textit{Bottom panel:} The fraction of BBH mergers for which the amplification of primary GWs relative to the unlensed GWs $\sqrt{\mu_{\rm P}}$ is higher than selected values as labelled as a function of $r_{\rm orb}$. Results are given for isotropic distributions of BBH mergers around SMBHs, including the effects of observational bias.
  \label{Fig:ObsBias_Disks_GNs} } 
\end{figure*}

 The top left panel shows the distribution of the GW echo amplitude relative to the primary amplitude for GW detections of BBH mergers in all AGN/BH disks with an inclination $i$ in a mock SNR limited survey, $P_{i, \rm det,P}(\sqrt{\mu_{\rm EP}})$, for various fixed inclination angle $i$ at fixed $r_{\rm orb } = 50 \, \rS$. As seen, observational bias significantly changes the corresponding $P_i(\sqrt{\mu_{\rm EP}})$ distribution for nearly edge-on disks. This is due to the high-magnification strongly lensed mergers in the sample for which $\sqrt{\mu_{\rm EP}} \sim 1$. Observational bias has a gradually less impact for increasing inclination when \mbox{$\sin{i} \gtrsim 0.1$}, or equivalently $i \gtrsim 6^{\circ}$. The reasons are that for BBH mergers with lower $i$, a systematically higher fraction of highly-magnified mergers ($\sqrt{\mu_{\rm EP}} \sim 1$) lies closer to the optical axis (Section \ref{subsec:SingleDisk_IsotropDist}), and the distribution is biased towards $\sqrt{\mu_{\rm EP}} \sim 1$ for small $i$. Note that BBH mergers behind the SMBH close to the optical axis have $\sqrt{\mu_{\rm EP}} \lesssim 1$, which contribute to the peak occurring in $P_{i, \rm det,P}(\sqrt{\mu_{\rm EP}})$ at $\sqrt{\mu_{\rm EP}} \sim 1$. Mergers in front of the SMBH close to the optical axis, i.e. $\alpha \sim \pi$, contribute typically to the $\sqrt{\mu_{\rm EP}} \gtrsim 1$ part of the peak. Finally, the impact of observational bias on $P_i(\sqrt{\mu_{\rm EP}})$ decreases with increasing $r_{\rm orb}$ because a systematically lower fraction of mergers has $\sqrt{\mu_{\rm EP}} \sim 1$ for larger $r_{\rm orb}$ (Section \ref{subsec:SingleDisk_IsotropDist}). The important conclusion is that in an SNR-limited GW observational survey, the amplitudes of GW echoes are comparable to primary amplitudes for nearly edge-on disks or, more generally, for sources along the line-of-sight, but such configurations are expected to be uncommon.

 Next, we determine the detection-volume weighted mean of the GW echo amplitude relative to the primary GW signal $\langle \sqrt{\mu_{\rm EP}} \rangle_{\rm det,P}$ as a function of $r_{\rm orb}$ for the same inclination angles as in the top right panel of Figure \ref{Fig:Single_Disks_GNs}. In comparison, $\langle \sqrt{\mu_{\rm EP}} \rangle_{\rm det,P}$ is significantly higher than $\langle \sqrt{\mu_{\rm EP}} \rangle$ for nearly edge-on disk configurations with small $i$, while it is only marginally higher than $\langle \sqrt{\mu_{\rm EP}} \rangle$ if \mbox{$\sin{i} \gtrsim 0.1$}. For nearly edge-on disks, $\langle \sqrt{\mu_{\rm EP}} \rangle_{\rm det,P} \gtrsim \{ 0.91, 0.45 \}$ if $r_{\rm orb} \lesssim \{ 150 \, \rS , 10^4 \, \rS \}$, respectively, in a mock SNR-limited survey, while $\langle \sqrt{\mu_{\rm EP}} \rangle_{\rm det,P}$ is above $\sim \{ 0.046, 0.004 \}$ for the same $r_{\rm orb}$ limits for single nearly edge-on disks. Furthermore, $\langle \sqrt{\mu_{\rm EP}} \rangle_{\rm det,P}$ decreases much less steeply than in a single AGN/BH disk with increasing $r_{\rm orb}$ for \mbox{$\sin{i} \sim 0$} but decreases similarly as in a single disk for \mbox{$\sin{i} \gtrsim 0.1$}. Note that similar to the case of single AGN/BH disks, nearly edge-on disks produce on average GW echoes with the highest relative amplitude in the inner regions close to the SMBH in an SNR-limited survey, and the relative echo amplitude decreases systematically with both $i$ and $r_{\rm orb}$. In comparison to the case of a single AGN/BH disk (Section \ref{subsec:SingleDisk_IsotropDist}), we find that $\langle \sqrt{\mu_{\rm EP}} \rangle_{\rm det,P}$ is $\sim 21.18 \left(\ \log(r_{\rm orb} / 10 \, \rS) \right)^{4.38} + 26.65$ times higher for edge-on disks than for face-on disks, e.g. this enhancement is \mbox{$\sim \{27, 48, 470 \}$} for $r_{\rm orb} \sim \{ 10 \, \rS, 100 \, \rS, 1000 \, \rS \}$, respectively.

 We repeat the analysis of Section \ref{subsec:SingleDisk_IsotropDist} for an isotropic distribution of BBH mergers around an SMBH by marginalizing $P_{i, \rm det,P}(\sqrt{\mu_{\rm EP}})$ over the inclination angle to obtain the distribution function of the relative GW echo amplitude $P_{\rm det,P}(\sqrt{\mu_{\rm EP}})$ in an SNR-limited survey. Examples are displayed in the middle left panel of Figure \ref{Fig:ObsBias_Disks_GNs} (c.f. Figure \ref{Fig:Single_Disks_GNs}). The fraction of mergers with $\sqrt{\mu_{\rm EP}} \sim 1$ is increased due to observational bias predominantly for $r_{\rm orb} \lesssim 100 \rS$. The reason is that the Einstein angle is smaller for higher $r_{\rm orb}$, hence the probability for GW echoes with $\sqrt{\mu_{\rm EP}} \sim 1$ is suppressed for isotropic merger populations. However, we find that the fraction of mergers with $\sqrt{\mu_{\rm EP}} \gtrsim 1$ is typically smaller than a per cent. This is not surprising since $\sqrt{\mu_{\rm E}}$ drops at a much faster rate at \mbox{$\gamma \sim \pi$} ($\alpha \sim \pi$) than at $\gamma \sim 0$ ($\alpha \sim \alpha_{\rm Einstein}$), while (i) the fraction of mergers in the $\gamma$ distribution are the same at either $\gamma \sim 0$ or $\gamma \sim \pi$ as it is symmetric (Section \ref{subsec:GeometricConvects}) and (ii) the corresponding weights are comparable as $\sqrt{\mu_{\rm EP}} \sim 1$ at either $\gamma \sim 0$ or $\gamma \sim \pi$ (Section \ref{subsec:Magnif_GWs}). Therefore, GW echoes with amplitudes higher than that of primary GWs are expected to constitute a small fraction of the detected BBH mergers (e.g. $\sqrt{\mu_{\rm EP}} \gtrsim 1$ for $\sim 0.65 \%$ of mergers at fixed $r_{\rm orb} = 20 \, \rS$).

 The middle right panel of Figure \ref{Fig:ObsBias_Disks_GNs} shows the cumulative fraction of isotropically distributed BBH mergers around SMBHs in a mock SNR-limited survey $\mathcal{F}_{\rm det, P}(\sqrt{\mu_{\rm EP}})$ that have relative GW echo amplitudes higher than the given values shown in the legend as a function of $r_{\rm orb}$ (c.f. Figure \ref{Fig:Single_Disks_GNs}). Clearly, observational bias increases $\mathcal{F}(\sqrt{\mu_{\rm EP}})$, although the degree of increase is significant only for relatively high $\sqrt{\mu_{\rm EP}}$ limits ($\sqrt{\mu_{\rm EP}} \gtrsim 0.1$) since observational bias preferentially enhances the fraction of BBH mergers with \mbox{$\sqrt{\mu_{\rm EP}}\sim 1$} in $P(\sqrt{\mu_{\rm EP}})$. For instance, the fraction of mergers with $\sqrt{\mu_{\rm EP}} > 0.2$ is $\mathcal{F}_{\rm det, P}(\sqrt{\mu_{\rm EP}}) \sim \{ 30\%,\, 3\%,\, 0.33\% \}$ for $r_{\rm orb} \sim \{ 10 \, \rS, 100 \, \rS, 1000 \, \rS \}$, respectively, in a mock SNR-limited survey, while these fractions drop to $\mathcal{F}(\sqrt{\mu_{\rm EP}}) \sim \{ 25\%,\, 2\%,\, 0.15\%\}$ in a mock complete volume-limited survey for the same radii. Furthermore, a non-negligible fraction of mergers produce GW echoes with amplitudes comparable to those of primary signals. For instance, $\sim 60\%$ ($\sim 10\%$) of mergers have $\sqrt{\mu_{\rm EP}} > 0.1$ ($0.4$) at $r_{\rm orb} \sim 10 \, \rS$ in a mock SNR-limited survey, which drops to $\sim 1\%$ ($\sim 0.1\%$) at $r_{\rm orb} \sim 1000 \, \rS$. Note that similar results stand for a mock complete volume-limited survey. In comparison to the case of a mock complete volume-limited survey (Section \ref{subsec:SingleDisk_IsotropDist}), we find that at least $1 \%$ of all GW echoes have amplitudes larger than $\{ 0.4, 0.2, 0.1, 0.05, 0.02 \}$ times the primary GW amplitude if \mbox{$r_{\rm orb} \lesssim \{ 135 \, \rS, 330 \, \rS, 690 \, \rS, 1350 \, \rS, 3300 \, \rS \}$}, respectively, and more than $10 \%$ of echoes have amplitudes larger than $\{ 0.2, 0.1, 0.05, 0.02 \}$ times the primary GW amplitude if BBHs merge closer to the SMBH than \mbox{$\sim \{ 30 \, \rS, 65 \, \rS, 140 \, \rS, 370 \, \rS \}$}, respectively. Thus, in an SNR-limited survey, there is at least a $1\%$ probability of detecting GW echoes for BBH mergers within $\sim 135 \, \rS$ for the most typical weak primary GW signals, and there is a nonzero ($1\% \lesssim$) chance to detect echoes from distances up to $\sim 3300 \, \rS$ for very strong primary signals. These limits are respectively $\sim 60 \, \rS$ and $\sim 2400 \, \rS$ for a complete volume-limited survey (Section \ref{subsec:SingleDisk_IsotropDist}). Finally, we note that the fraction of detectable echos $\mathcal{F}_{\rm det, P}(\sqrt{\mu_{\rm EP}})$ decreases with $r_{\rm orb}$ as expected from the $r_{\rm orb}$ dependence of the relative echo amplification distribution $P_{\rm det,P}(\sqrt{\mu_{\rm EP}})$.

 For further astrophysical applications, we also investigate how the SMBH may amplify the primary GW signal relative to the unlensed GW signal for isotropically distributed BBH mergers around SMBHs in a mock SNR-limited survey irrespective of the GW echo. For this purpose, we assume detection for the primary GW signal and generate the $P_{\rm det,P}(\sqrt{\mu_{\rm P}})$ distribution as before. Similarly, we also determine the fraction of mergers $\mathcal{F}_{\rm det,P}(\sqrt{\mu_{\rm P}})$ for which the amplification of primary GWs relative to the unlensed GWs is higher than a given limit as a function of $r_{\rm orb}$. Results are displayed in the bottom panel of \mbox{Figure \ref{Fig:ObsBias_Disks_GNs}}. $\mathcal{F}_{\rm det,P}(\sqrt{\mu_{\rm P}})$ decreases with $r_{\rm orb}$ because the fraction of strongly lensed mergers by the SMBH, which are located near $\alpha_{\rm Einstein}$, decreases with $r_{\rm orb}$ as well (Section \ref{subsec:Magnif_GWs}). We find that only a small fraction of BBH mergers have a prominent $\sqrt{\mu_{\rm P}}$, even in the best-case scenario, when binaries merge closest to the SMBH. For instance, $\sim \{ 7 \%, 4 \%, 3 \% \}$ of mergers have $\sqrt{\mu_{\rm P}} > \{1.01, 1.3, 1.5 \}$, respectively, for $r_{\rm orb} = 10 \, \rS$ and these merger fractions drop to $\sim \{ 1.7 \%, 0.7 \%, 0.5 \% \}$ and $\sim \{ 0.5 \%, 0.14 \%, 0.1 \% \}$ for $r_{\rm orb} = 100 \, \rS$ and $1000 \, \rS$, respectively. This is caused mainly by the quick drop of $\sqrt{\mu_{\rm P}}$ close to $\alpha_{\rm Einstein}$ (Section \ref{subsec:Magnif_GWs}) and partly because a small fraction of mergers are located close to the optical axis whose image is near $\alpha_{\rm Einstein}$. Therefore, we conclude that strongly lensed GW signals are not expected to be common in an SNR-limited survey of isotropically distributed BBH mergers around SMBHs. In particular, one in 30 (200) BBH mergers at a distance of $10 \, \rS$ ($100 \, \rS$) of an SMBH may be amplified by $50\%$.

 Note that since highly magnified primary GWs are located near the Einstein radius most such mergers originate in near edge-on disk configurations ($\sin{i} \sim 0$)  (Section \ref{subsec:GeometricConvects}). Furthermore, this subset produces GW echoes with amplitudes comparable to that of primary GWs $\sqrt{\mu_{\rm EP}} \sim 1$. This explains why nearly edge-on AGN/BH disk may produce GW echoes with $\sqrt{\mu_{\rm EP}} \sim 1$ in an SNR-limited survey, yet the overall fraction of highly amplified primary GWs is low for an isotropic population as disks with $\sin{i} \sim 0$ constitute a small portion of possible configurations in the $[-1,1]$ range.

\subsection{Time delay between the primary GW and the GW echo in an SNR-limited survey}
\label{subsec:RelPos_TimeDelay}
 
 In this section, we determine the time-delay distribution between the primary GW and the GW echo for isotropic distributions of BBH mergers around SMBHs in a mock SNR-limited GW observational survey as a function of SMBH-binary distance. We examine detections for both GW echoes and primary GWs since the detection volume is different for the two cases, which will skew the time-delay distribution in an observational sample. First, we determine the distribution of time delay assuming that the GW echo was detected irrespective of the primary GW signal. Second, we examine the time-delay distribution in cases where the primary GW signal has a high SNR, such that the echo may be fainter than the primary by a factor of either $\sqrt{\mu_{\rm EP}} > 0.02$ or $> 0.2$.

\begin{figure*}
    \centering
    \includegraphics[width=80mm]{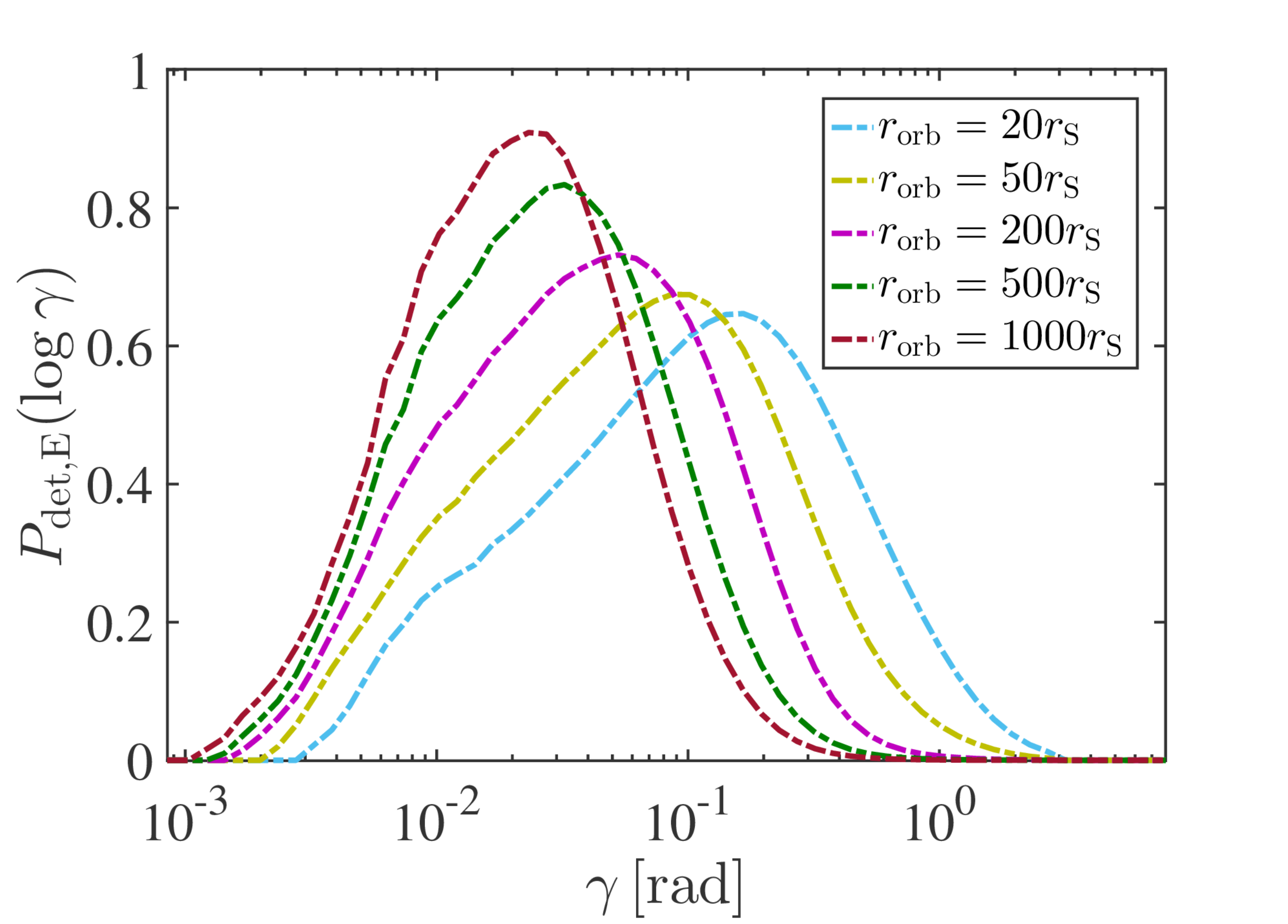}
    \includegraphics[width=80mm]{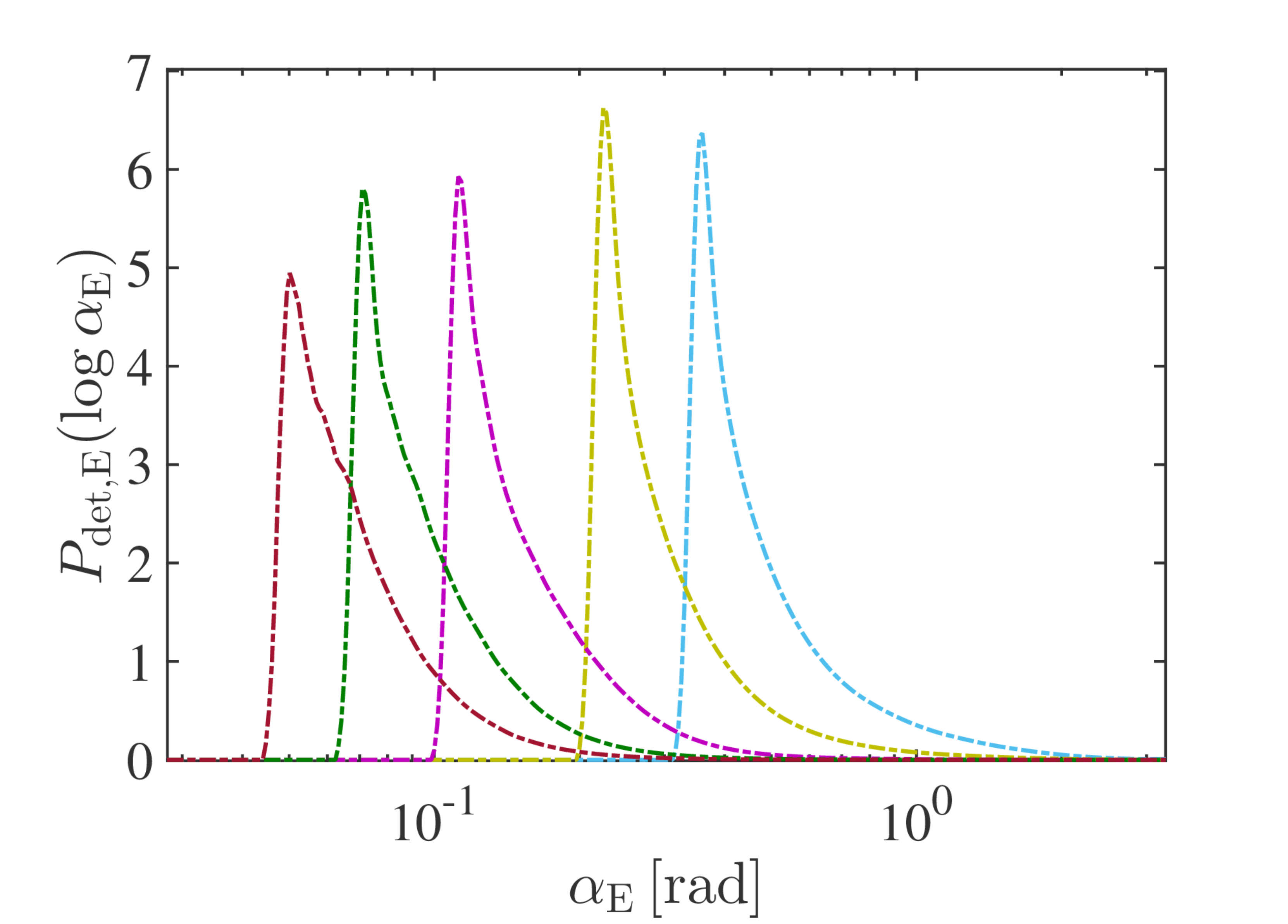}
    \\
    \includegraphics[width=80mm]{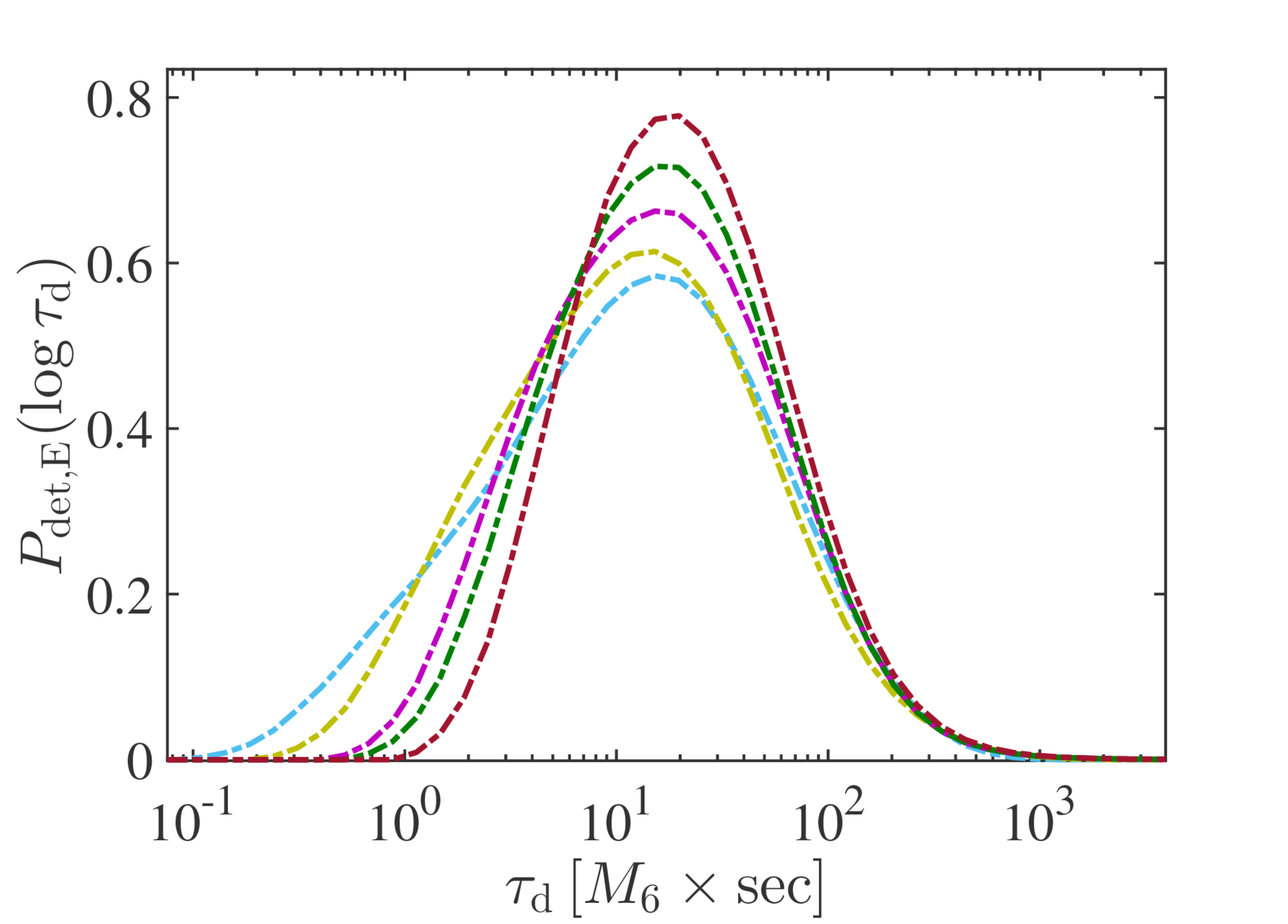}
    \includegraphics[width=80mm]{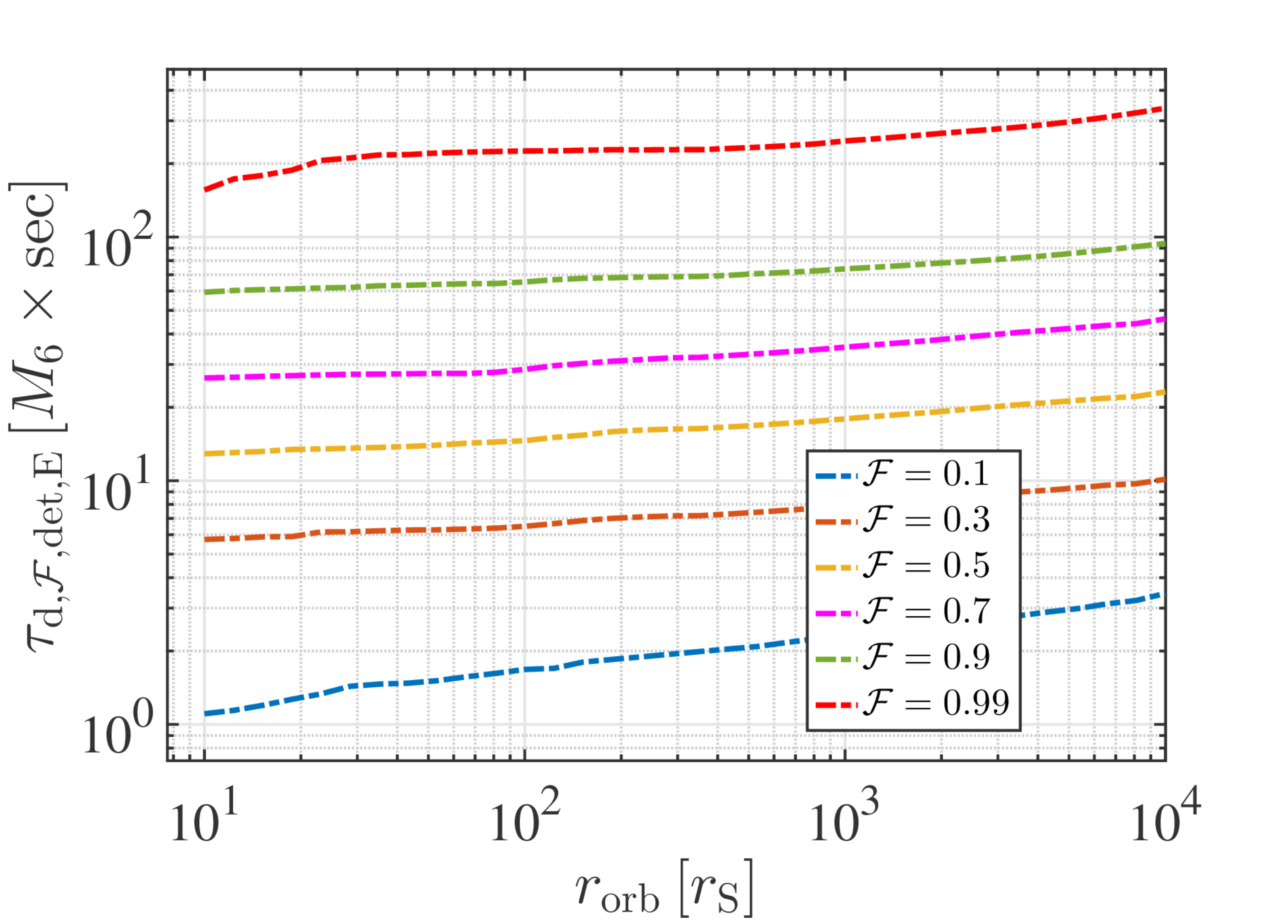}
\caption{\textit{Left panels:} Distribution of the relative angular position $\gamma$ (\textit{top left}) and the deflection angle of the GW echo $\alpha_{\rm E}$ (\textit{top right}) on a logarithmic scale and the distribution of corresponding time delay $\log \tau_{\rm d}$ (\textit{bottom left}) between primary GWs and GW echoes for isotropic distributions of BBH mergers around SMBHs at a fixed distance $r_{\rm orb}$ shown in the legend in a mock SNR-limited GW observational survey of detectable GW echoes. The bottom right panel shows the corresponding contours of fixed cumulative distribution levels $\mathcal{F}$ of $\tau_{\rm d}$ shown in the legend as a function of $r_{\rm orb}$. In all panels, observational bias was calculated using the detection volume of the GW echo. Note that $\tau_{\rm d}$ is shown for a redshifted SMBH mass of $M_{\rm SMBH,z} = (1+z)M_{\rm SMBH}= 10^6 \, \Msun$. For other values of $M_{\rm SMBH,z}$ the units must be scaled by the factor $M_6 = M_{\rm SMBH,z} / 10^6 \, \Msun$ (Section \ref{subsec:RelPos_TimeDelay}).
  \label{Fig:DistPos_DistDelay_ObsBias} } 
\end{figure*}

 We start by generating a merger sample. For fixed SMBH-binary distance $r_{\rm orb}$, we first generate an isotropic prior MC sample around the SMBH with an isotropic binary orientation $(\gamma, \phi_{\rm AM}, \Theta_{\rm AM})$ such that the prior distribution of $\cos{\gamma}$ is uniform between $[-1,1]$ (Section \ref{subsec:GeometricConvects}). We also determine the corresponding deflection angle $\alpha_{\rm E}$ for each $\gamma$ at the considered $r_{\rm orb}$ as discussed in Section \ref{subsec:Lens_Delay}. Then, we apply the methodology introduced in Section \ref{subsec:Disk_IsotropDist_ObsBias} to obtain the detection-volume weighted distribution $P_{\rm det,E}(\gamma)$, the angular position distribution of merging BBHs relative to the SMBH and the line-of-sight in a mock SNR-limited survey, and the corresponding distribution of deflection angle for the GW echo $P_{\rm det,E}(\alpha_{\rm E})$ by assuming detection for the GW echo. Here we compute the detection-volume weight according to the echo amplitude as $w_{\rm E} = \mu_{\rm E}^{3/2}$. This assumption also means a detection for the primary signal as typically $\sqrt{\mu_{\rm EP}} \lesssim 1$ (Section \ref{subsec:Disk_IsotropDist_ObsBias}). Note that in this type of survey, the vast majority of mergers with observable GW echoes are expected to be in the high-magnification strong-lensing regime, and hence the primary GWs and GW echoes have similar amplitudes. This implies that both the primary GWs and the GW echoes are typically near the SNR detection limit in an observational sample.

 Examples for $P_{\rm det,E}(\log \gamma)$ and $P_{\rm det,E}(\log \alpha_{\rm E})$ are displayed in the top left and top right panels of Figure \ref{Fig:DistPos_DistDelay_ObsBias}, respectively, for various $r_{\rm orb}$ as labelled.\footnote{We use an MC sample of $2.5 \times 10^7$ elements to generate $P_{\rm det,E}(\gamma)$.} We find that (i) the vast majority of BBH mergers with detectable GW echoes are strongly lensed (i.e. $\sqrt{\mu_{\rm EP}} \sim 1$) as most of them have $\alpha_{\rm E} \gtrsim \alpha_{\rm Einstein}$ and accordingly are located behind the SMBH relatively close to the optical axis, (ii) and a marginal fraction of mergers are located in front of the SMBH ($\lesssim 1 \% $ retrolensing among all events at $r_{\rm orb} \gtrsim 10 \, \rS$). Note that $P_{\rm det,E}(\gamma)$ systematically shifts toward lower $\gamma$ with increasing $r_{\rm orb}$ because mergers with the largest detectable volumes preferentially form near $\alpha_{\rm Einstein}$ that systematically shifts toward lower values for larger SMBH-binary distances, while the corresponding $P_{\rm det,E}(\alpha_{\rm E})$ shifts toward lower $r_{\rm orb}$ since $\alpha_{\rm Einstein} \propto 1 / \sqrt{ r_{\rm orb} }$. For further investigations, we define $\gamma_{\mathcal{F},{\rm det,E}}$ using $P_{\rm det,E}(\gamma)$ as the half opening angle of the cone in which a cumulative $\mathcal{F}$ fraction of BBH mergers are located at a distance $r_{\rm orb}$ from the SMBH. We find that (i) $\gamma_{\mathcal{F},{\rm det,E}}$ decreases with increasing $r_{\rm orb}$, and (ii) the mergers that take place behind the SMBH are located close to the optical axis only for relatively large $r_{\rm orb}$ (not shown). For instance, $50 \%$ and $90 \%$ of detected echoes correspond to mergers located within a half opening angle of $\gamma_{\mathcal{F},{\rm det,E}} \sim \{ 0.15, 0.05, 0.02 \}$ radian and $\sim \{ 1.4, 0.6, 0.2 \}$ radian, respectively, for $r_{\rm orb} = \{ 10 \, \rS, 100 \, \rS, 1000 \, \rS \}$. We define similarly $\alpha_{\rm E, \mathcal{F}, det,E}$ and find that the vast majority of mergers have deflection angles close to $\alpha_{\rm Einstein}$. In particular, $\{ 30 \%, 50 \%, 70\%, 90\%\}$ of mergers are located typically within $\sim \{ 1.1, 1.3, 1.6, 2.3 \} \times \alpha_{\rm Einstein}$.

 Next, we determine the distribution of time delay $P_{\rm det,E}(\tau_{\rm d})$ between primary GW signals and GW echoes for BBH mergers with detectable GW echoes in isotropic populations around SMBHs in a mock SNR-limited survey. To do so, we first assign a time delay $\tau_{\rm d}$ to a given relative position $(r_{\rm orb}, \gamma)$ and redshifted SMBH mass $M_{\rm SMBH,z}$ as introduced in Section \ref{subsec:Lens_Delay}, then proceed as prescribed in Section \ref{subsec:Disk_IsotropDist_ObsBias} to generate $P_{\rm det,E}(\tau_{\rm d})$. The distributions corresponding to $P_{\rm det,E}(\log \gamma)$ in the bottom left panel of Figure \ref{Fig:DistPos_DistDelay_ObsBias} are displayed in the bottom right panel for $M_{\rm SMBH,z} = 10^6 \, \Msun$. Results for other $M_{\rm SMBH,z}$ can be given by simply scaling the obtained results with the factor \mbox{$M_6 = M_{\rm SMBH,z} / 10^6 \, \Msun$} as $\tau_{\rm d} \propto \rS$ (Section \ref{subsec:Lens_Delay}). As seen, the peak of the $P_{\rm det,E}(\log \tau_{\rm d})$ distribution is approximately independent of $r_{\rm orb}$, but the distribution shifts slowly to higher $\tau_{\rm d}$ with increasing $r_{\rm orb}$. This feature is due to the fact that $\tau_{\rm d}$ increases at a faster rate than how $P_{\rm det,E}(\gamma)$ shifts toward smaller angles for increasing $r_{\rm orb}$.

\begin{figure*}
    \centering
    \includegraphics[width=75mm]{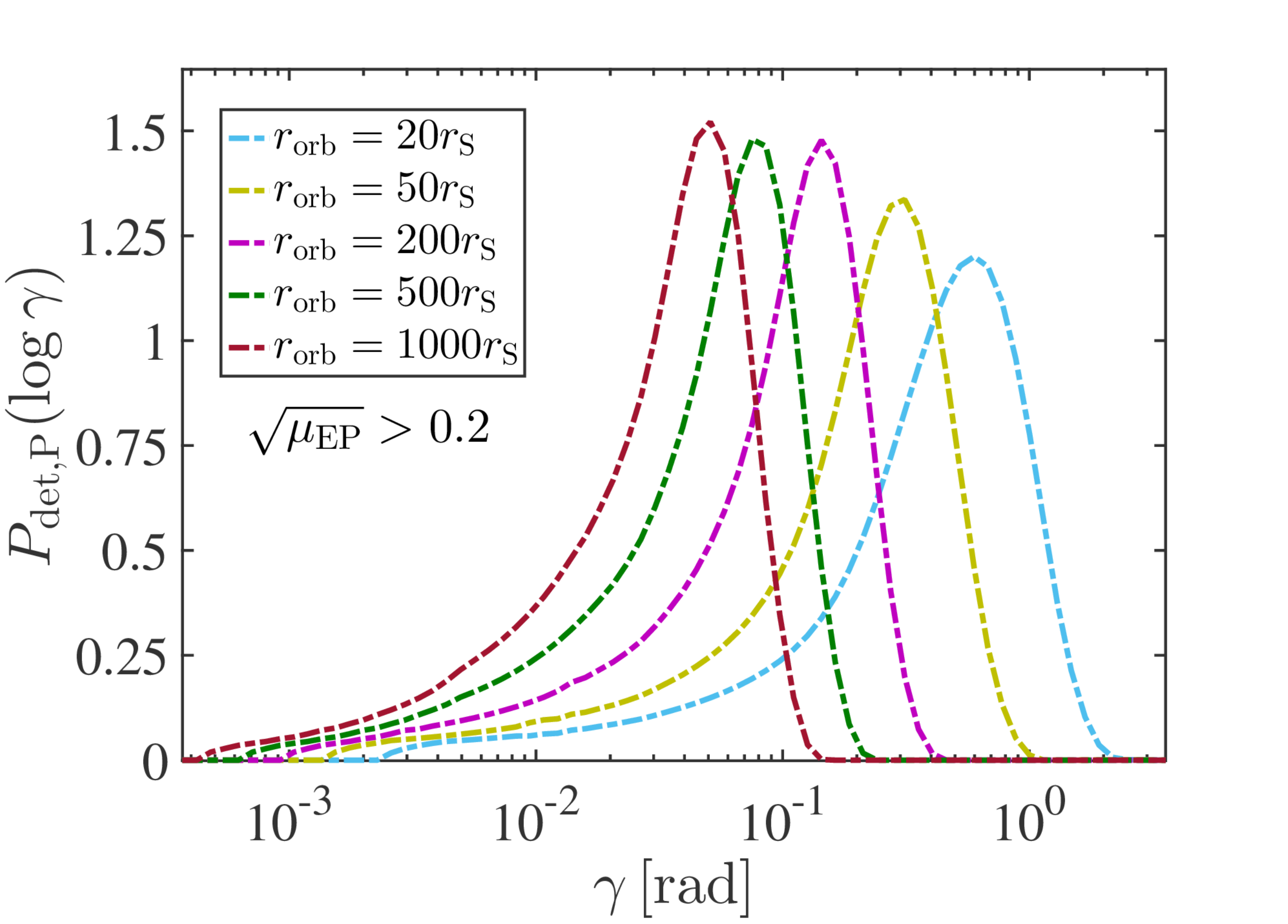}
    \includegraphics[width=75mm]{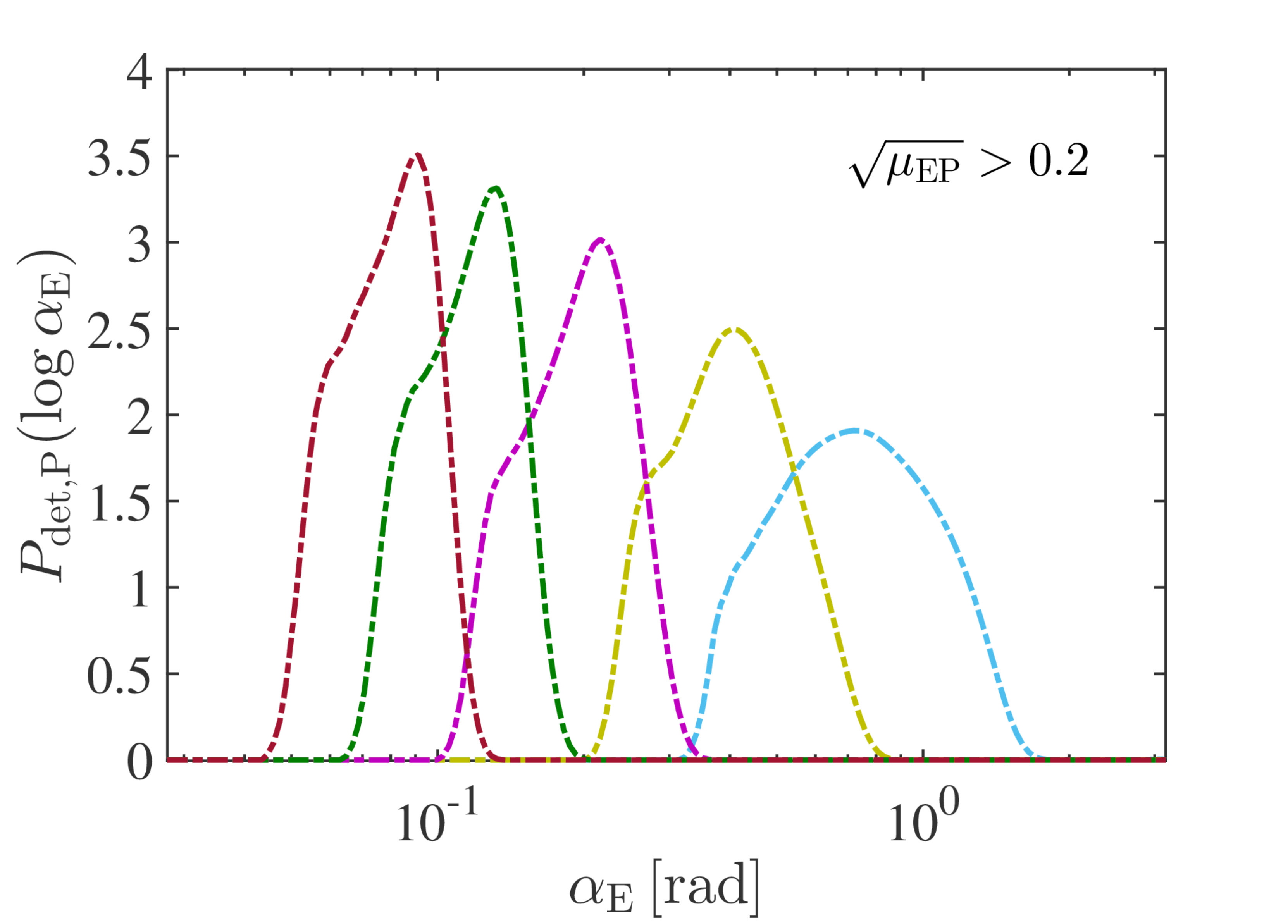}
    \\
    \includegraphics[width=75mm]{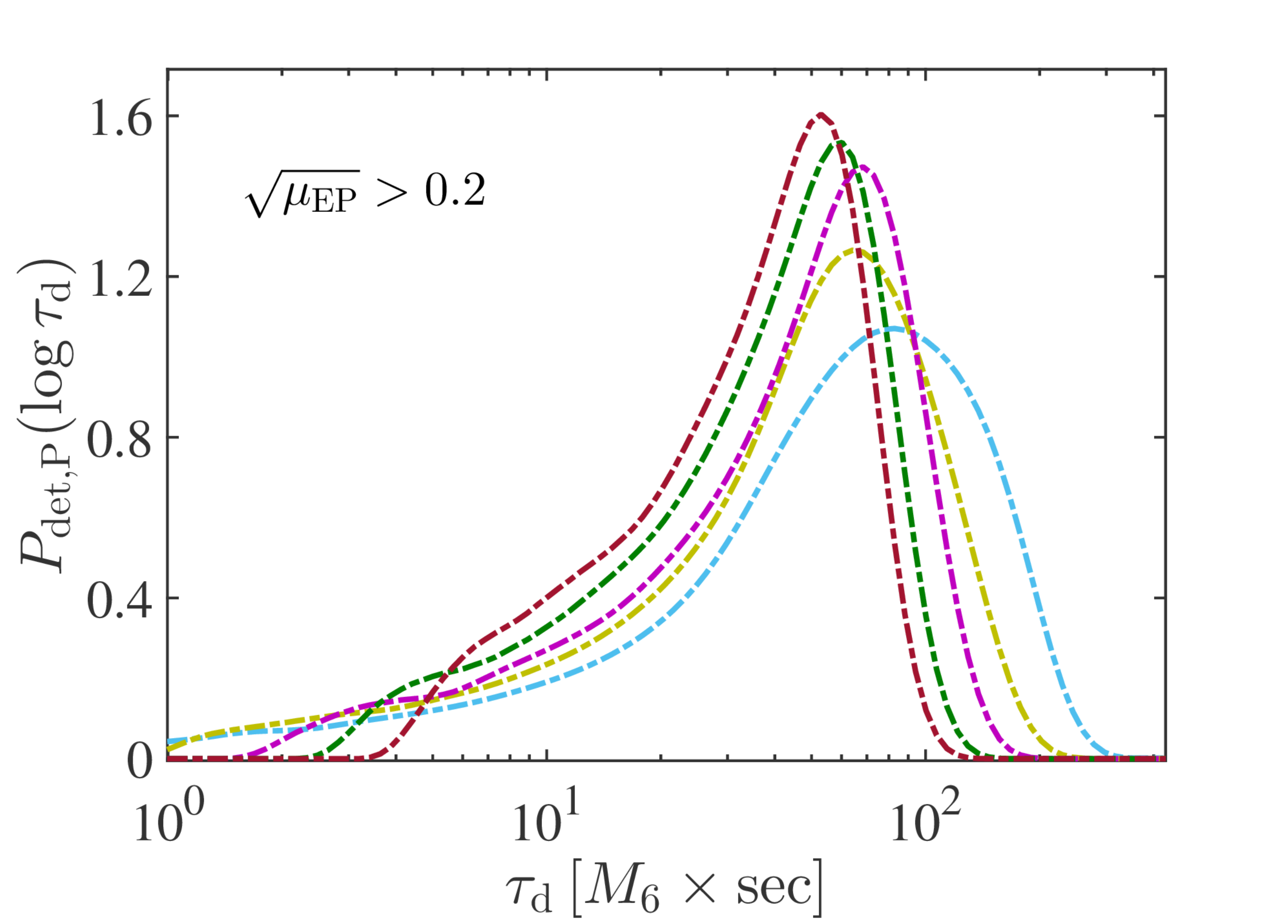}
    \includegraphics[width=75mm]{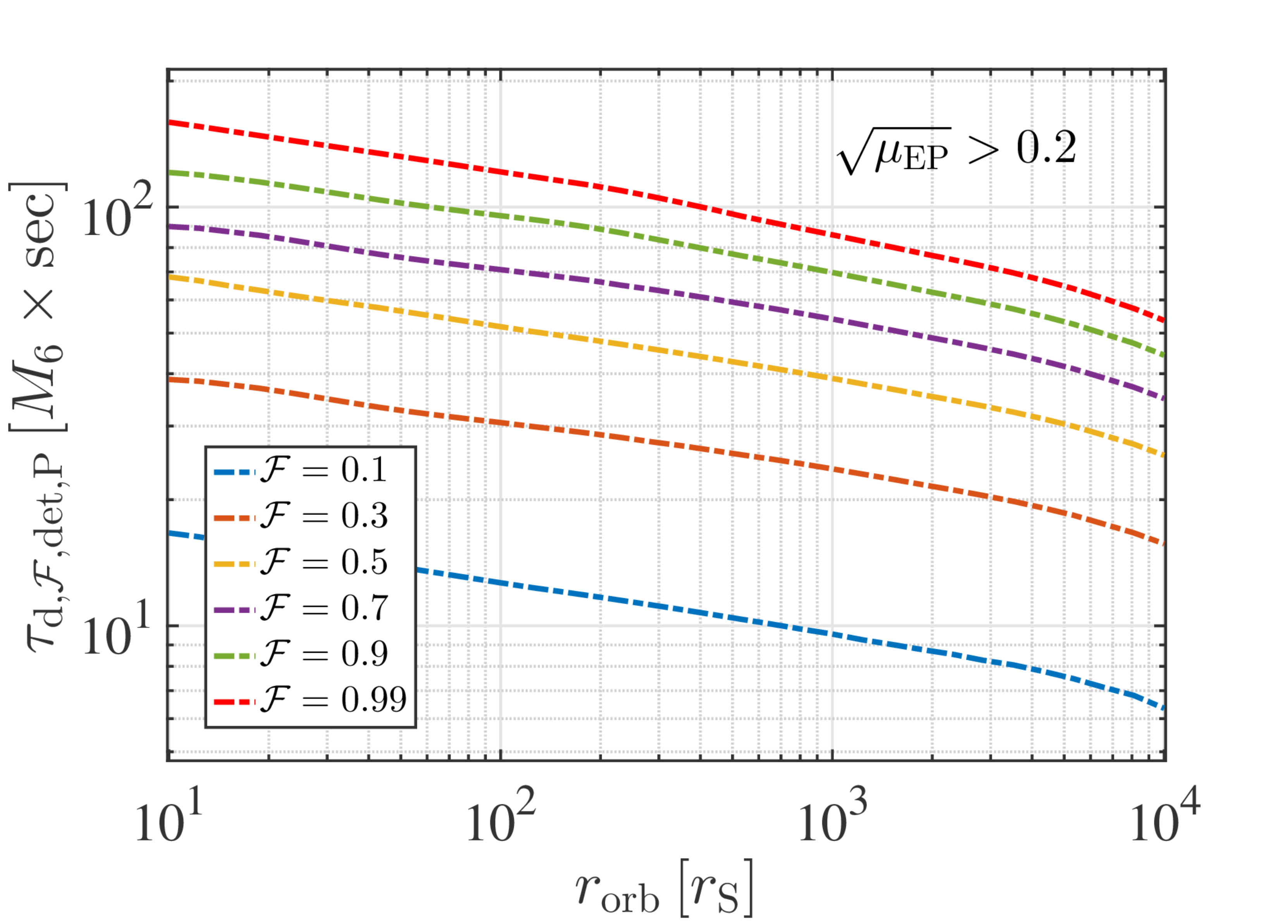}
    \\
    \includegraphics[width=75mm]{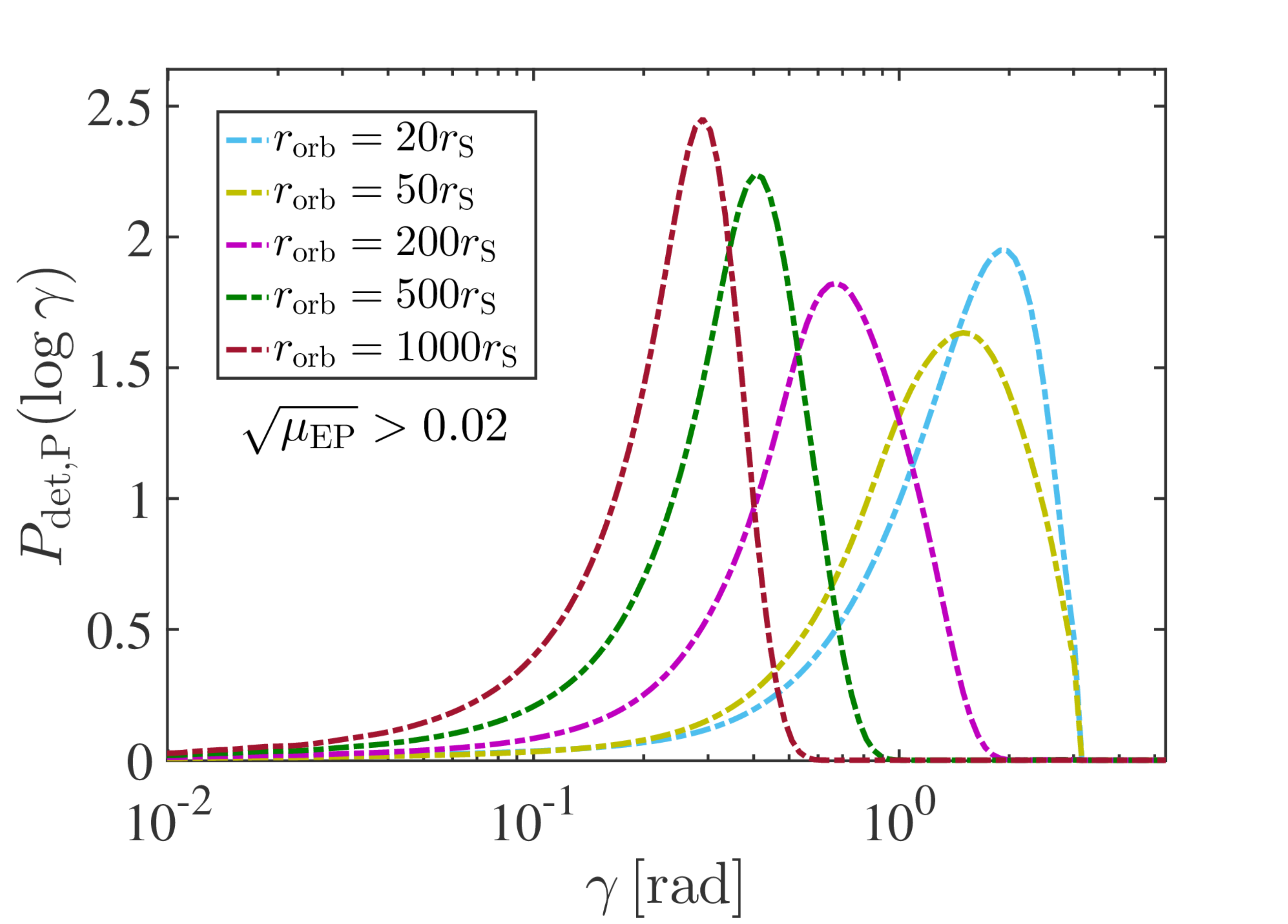}
    \includegraphics[width=75mm]{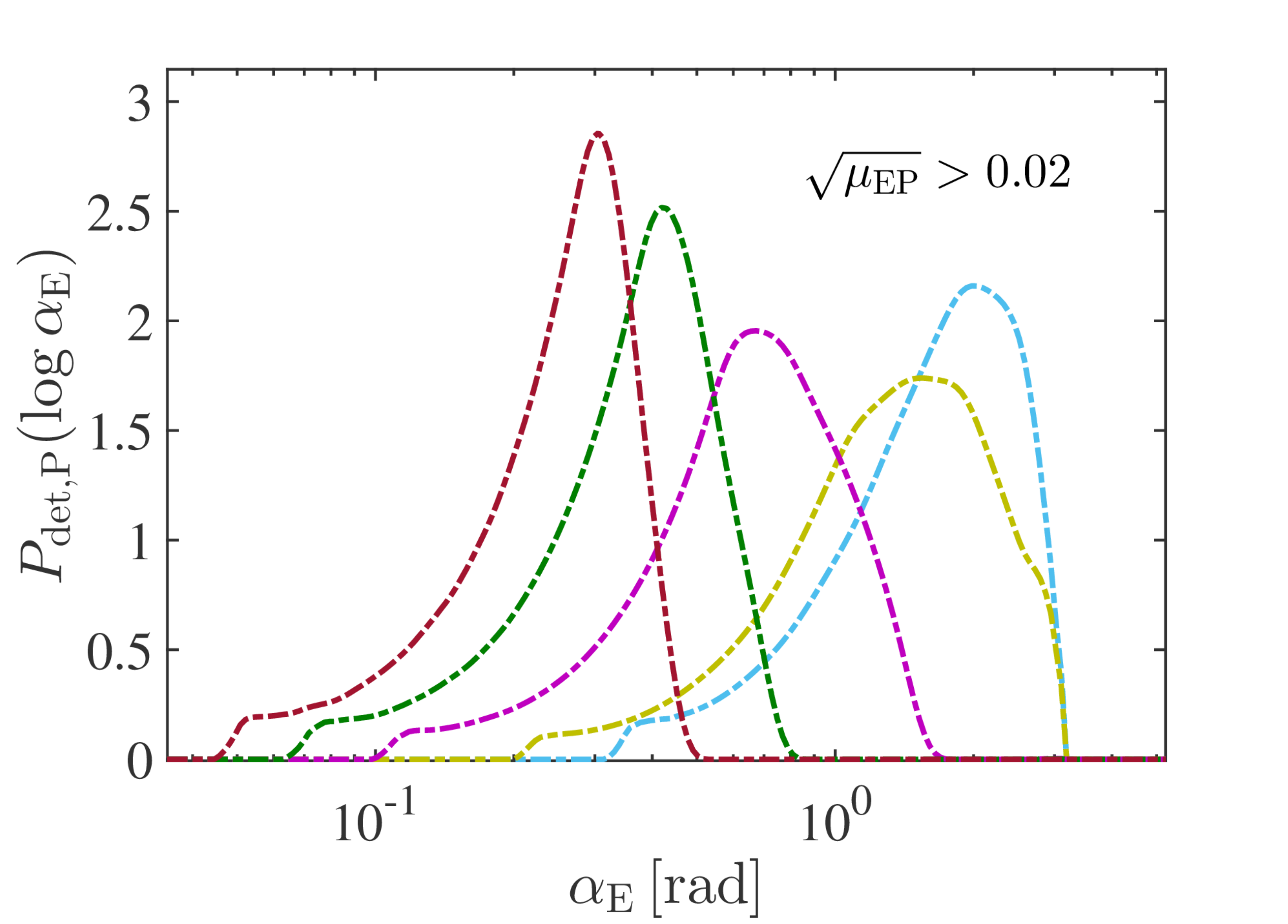}
    \\
    \includegraphics[width=75mm]{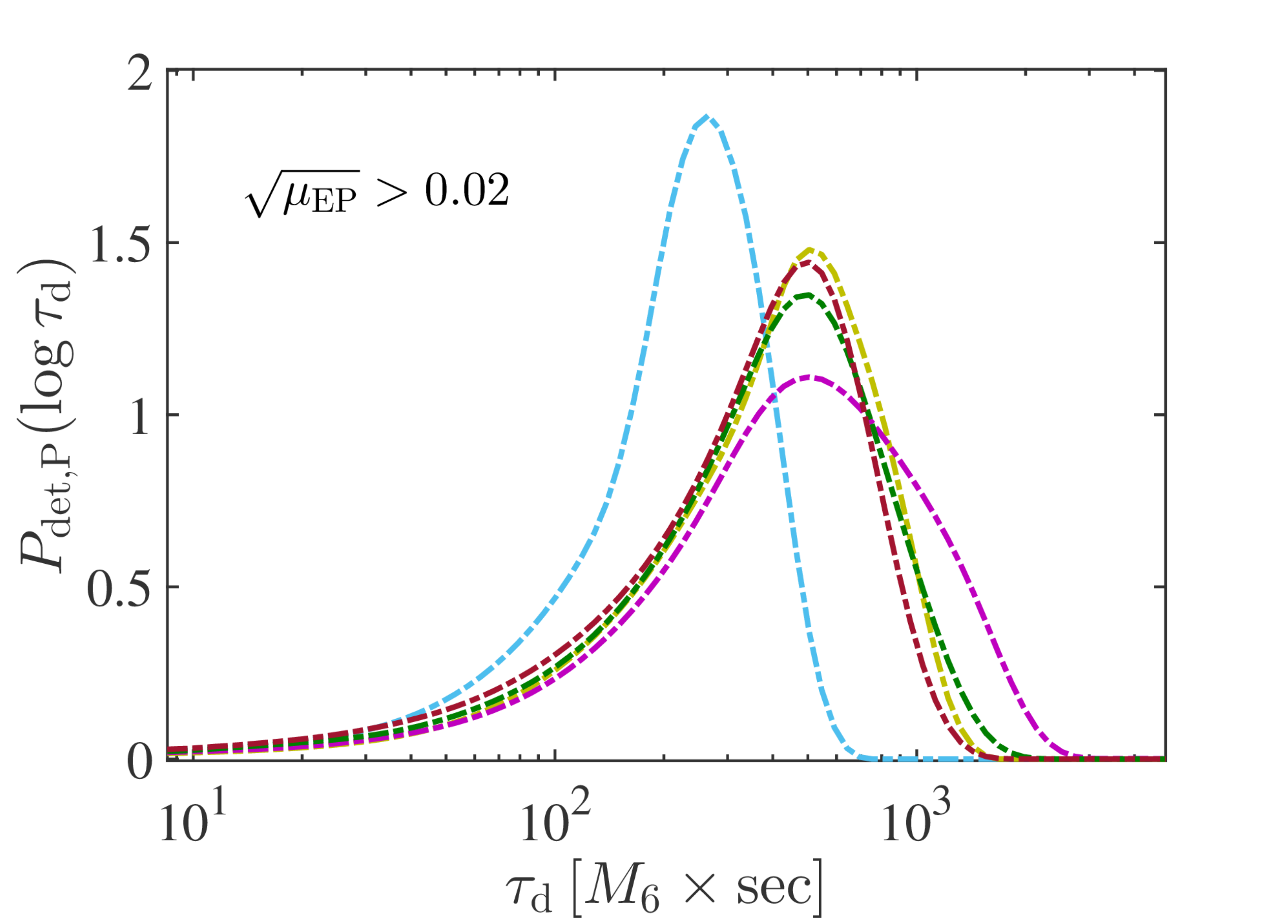}
    \includegraphics[width=75mm]{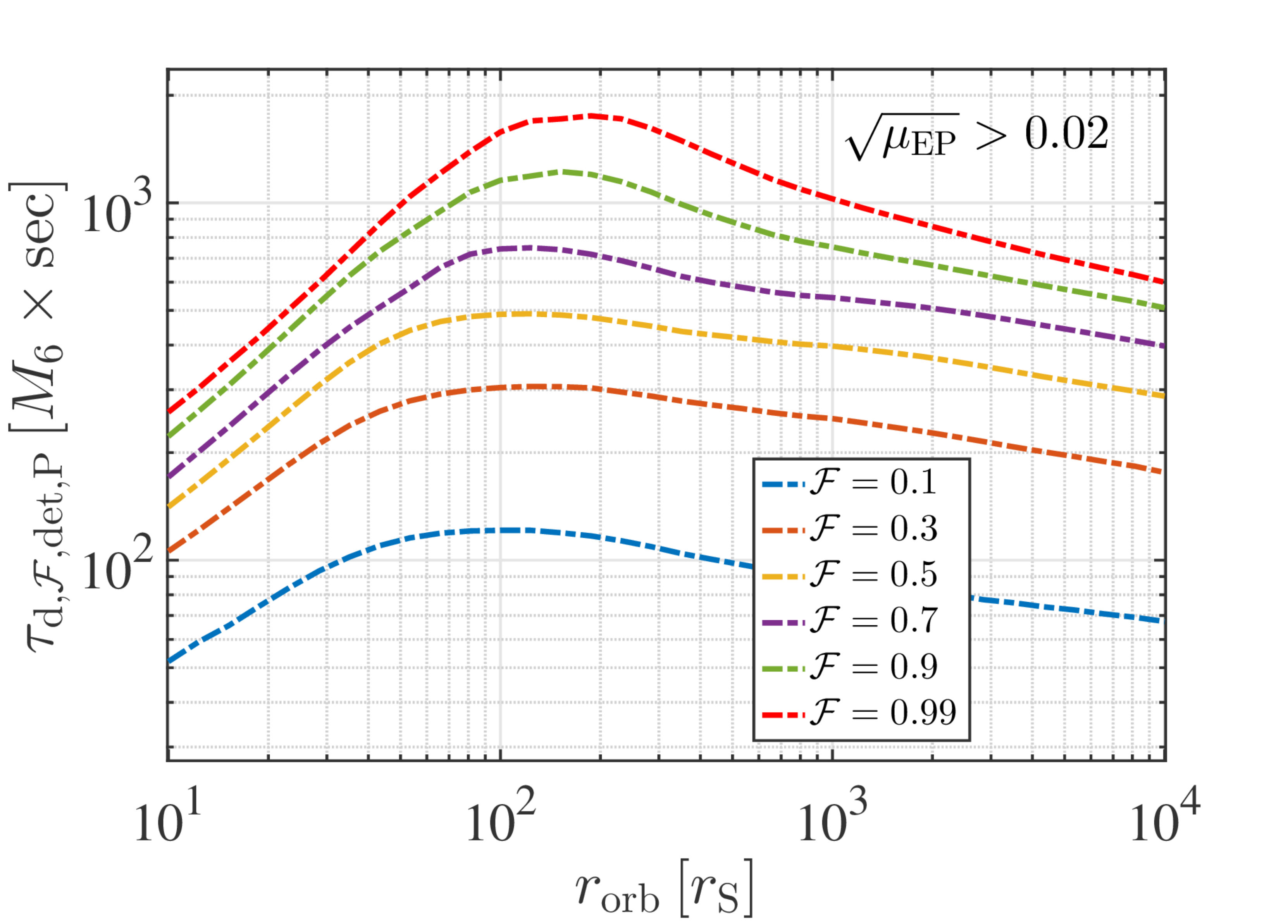}
\caption{Similar to Figure \ref{Fig:DistPos_DistDelay_ObsBias}, showing the distribution of the relative angular position $\gamma$, the deflection angle of echo $\alpha_{\rm E}$, and the corresponding time delay $\tau_{\rm d}$ between the primary GW and its GW echo for isotropic distributions of BBH mergers around SMBHs in a mock SNR-limited survey for fixed $r_{\rm orb}$ as shown in the legend. The figure also shows the corresponding lines of constant cumulative distribution levels $\mathcal{F}$ of $\tau_{\rm d}$ labelled in the legend as a function of $r_{\rm orb}$. Observational bias is calculated using the detection volume of the primary GW (cf. Figure \ref{Fig:DistPos_DistDelay_ObsBias}, which used the detection volume of the GW echo). The top four and bottom four panels show the distributions assuming that the primary GW has a moderately high or very high SNR, respectively, such that the GW echo may be detected if its amplitude is at least $0.2$ or $0.02$ times the primary amplitude, respectively; see also Figures \ref{Fig:Single_Disks_GNs} and \ref{Fig:ObsBias_Disks_GNs}.
 \label{Fig:DistPos_DistDelay_ObsBiasPrimary} } 
\end{figure*}

\begin{figure}
    \centering
    \includegraphics[width=90mm]{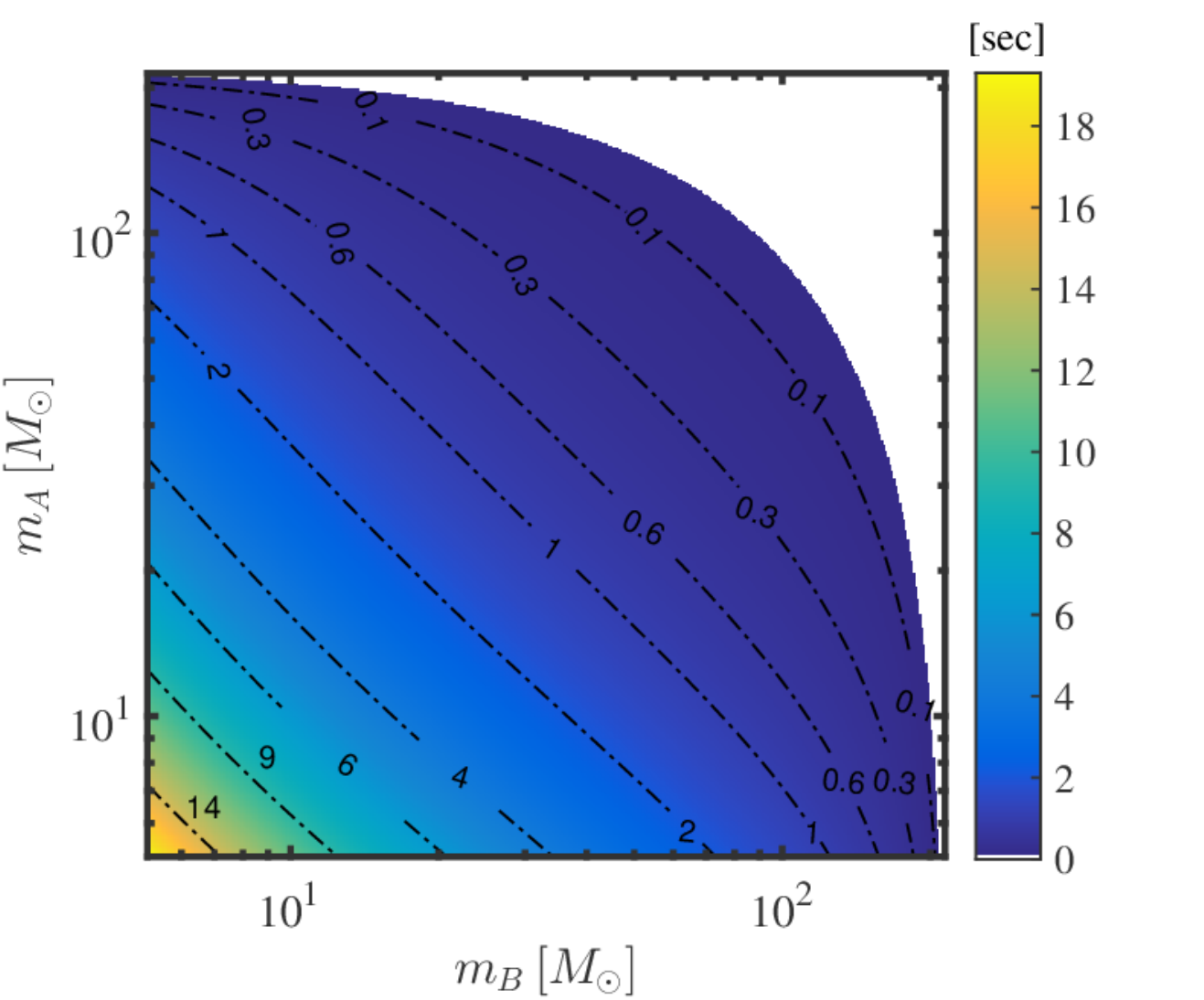}
\caption{The duration of time that quasi-circular BBH mergers' GW signals spend in the advanced GW detectors' sensitive frequency band above $20 \, \Hz$ as a function of component masses using the 1.5PN calculation; see Appendix \ref{sec:SignalTimeDuration} for details.   \label{Fig:tau_obs_circ}} 
\end{figure}

 Finally, the bottom right panel of Figure \ref{Fig:DistPos_DistDelay_ObsBias} shows the time delay at which a cumulative $\mathcal{F}$ fraction of isotropically distributed BBH mergers with detectable echoes are located in the vicinity of SMBHs in a mock SNR-limited survey, which we denote by $\tau_{\rm d,\mathcal{F}, det,E}$. We find that $\tau_{\rm d,\mathcal{F}, det,E}$ increases slowly with $r_{\rm orb}$. For instance, $99\%$ of the systems have a delay-time $\tau_{\rm d,0.99, det,E}$, which increases slowly between $20 \, \rS$ and $1000 \, \rS$ from $\sim 200 \, {\rm sec}$ up to $\sim 250 \, {\rm sec}$. Since a non-negligible fraction of GW echoes with amplitudes comparable to the primary amplitudes may be produced by merging BBHs out to a few hundred $\rS$ (Section \ref{subsec:Disk_IsotropDist_ObsBias}), the maximum time delay between the primary GW and its GW echo may be roughly estimated as ${\rm max}(\tau_{\rm d}) \sim 250 \, {\rm sec} \times M_6$. The median time delay between the primary signal and the echo $\tau_{\rm d,0.5, det,E}$ represents the typical time delay.\footnote{Note that the median of $P_{\rm det,E}(\tau_{\rm d})$ estimates to within $\sim 10 \%$ the peak location in $P_{\rm det,E}(\log \tau_{\rm d})$.} As seen, the median also increases slowly with $r_{\rm orb}$, e.g. $\tau_{\rm d,0.5, det,E}$ is $\sim 13 \, {\rm sec}$ and $\sim 18 \, {\rm sec}$ for \mbox{$r_{\rm orb} = 20 \, \rS$} and $1000 \, \rS$, respectively. Similar to the estimate of ${\rm max}(\tau_{\rm d})$, the typical time delay between the primary signal and its echo is set to be \mbox{$\sim 18 \, {\rm sec} \times M_6$}.

 Up to this point, the distributions describing GW echos were obtained by marginalizing over the properties of primary GWs. Let us now consider the distributions for a subset of GW echoes, which arrive after a moderately strong or a very strong primary GW signal that has been detected such that the detection of the GW echo is made possible for e.g. $\sqrt{\mu_{\rm EP}} > 0.2$ or $\sqrt{\mu_{\rm EP}} > 0.02$, respectively (c.f. Figures \ref{Fig:Single_Disks_GNs} and \ref{Fig:ObsBias_Disks_GNs}). We generate such distribution by applying the detection-volume weights according to the primary GW signal and by also discarding the samples that do not satisfy the above-mentioned conditions on $\sqrt{\mu_{\rm EP}}$. We denote these distributions with the subscript ${\rm det,P}$ and additionally specify $\sqrt{\mu_{\rm EP}} > 0.2$ or $\sqrt{\mu_{\rm EP}} > 0.02$ in the figures. The results are shown in Figure \ref{Fig:DistPos_DistDelay_ObsBiasPrimary}. The top four panels show the case of $\sqrt{\mu_{\rm EP}} > 0.2$ for primary GWs with moderately high SNRs, and the bottom four panels assume $\sqrt{\mu_{\rm EP}} > 0.02$ for primary GWs with very high SNRs. The figure shows a systematic shift of the source angular positions and the deflection angles to higher values (cf. Figure \ref{Fig:DistPos_DistDelay_ObsBias}). The $\alpha_{\rm E}$ distribution is no longer strongly peaked near the Einstein angle, instead it is shifted to higher values by a factor of $\sim 2-3$ for $\sqrt{\mu_{\rm EP}} > 0.2$ and by a factor of $6 - 7$ for $\sqrt{\mu_{\rm EP}} > 0.02$ over the radial distance range of $r_{\rm orb} = 20 \, \rS - 1000 \, \rS$. Here, higher factors correspond to larger $r_{\rm orb}$. In comparison to GW echo detections, we find that $\{ 30 \%, 50 \%, 70\%, 90\%\}$ of mergers are located typically within $\sim \{ 1.5, 1.9, 2.2, 2.6 \} \times \alpha_{\rm Einstein}$ and within $\sim \{ 4.5, 5.8, 7.2, 9.1 \} \times \alpha_{\rm Einstein}$ for echoes with $\sqrt{\mu_{\rm EP}} > 0.2$ and $\sqrt{\mu_{\rm EP}} > 0.02$, respectively, while the corresponding limits for GW echo detections range between $\sim 1.1 - 1.6$. These differences are not surprising since if the primary has a very high (moderately high) SNR such that the detection of GW echo requires e.g. $\sqrt{\mu_{\rm EP}} > 0.02$ (e.g. $> 0.2$), the source does not need to lie close to the optical axis, but also could be detected even for scattering configurations with large deflection angles.

 Furthermore, Figure \ref{Fig:DistPos_DistDelay_ObsBiasPrimary} shows that the corresponding time-delay distributions of GW echoes for primary GWs with moderately high and very high SNRs are systematically shifted to higher time-delay values compared to those distributions presented in Figure \ref{Fig:DistPos_DistDelay_ObsBias} for GW echo detections marginalized over all primary GWs. Again, this is due to the fact that the latter case is dominated by primary GWs with low SNRs for which both the primary and the echo are strongly lensed, while for primary GWs with moderately high and very high SNRs, even GW echoes with high deflection angles are detectable. Therefore, among primary GWs with moderately high and very high SNRs, the time delays between primary signals preceding echoes may be much larger, in many cases peaking near $r_{\rm orb}/c$ for small $r_{\rm orb}$. This leads to a systematic linear increase of $\tau_{\rm d}$ with $r_{\rm orb}$ for small $r_{\rm orb}$. For large $r_{\rm orb}$, the delay-time decreases weakly with the SMBH-binary distance. Note that these trends are different from the GW echo distribution marginalized over all primary GWs (Figure \ref{Fig:DistPos_DistDelay_ObsBias}), which showed a nearly constant, weakly increasing trend with $r_{\rm orb}$. Similar to our considerations on the characteristic time delays for GW echo detections, we set the typical time delay to be the median of $P_{\rm det,P}(\tau_{\rm d})$ and the maximum time delay to be the $99\%$ quantile of $P_{\rm det,P}(\tau_{\rm d})$ in the mock observational sample. The typical time delay for primary GWs with moderately high SNRs ($\sqrt{\mu_{\rm EP}} > 0.2$ in examples) is $\tau_{\rm d} \sim (40 \, {\rm sec} - 60 \, {\rm sec} ) \times M_6$ and it is $\sim ( 260 \, {\rm sec} - 500 \, {\rm sec} ) \times M_6$ for primary GWs with very high SNRs (e.g. when $\sqrt{\mu_{\rm EP}} > 0.02$) over the radial distance range of $r_{\rm orb} = 20 \, \rS - 1000 \, \rS$. Similarly, the maximum time delay is ${\rm max}(\tau_{\rm d}) \sim (85 \, {\rm sec} - 144 \, {\rm sec}) \times M_6$ for primary GWs with $\sqrt{\mu_{\rm EP}} > 0.2$ and it is $ \sim (500 \, {\rm sec} - 1750 \, {\rm sec}) \times M_6$ for $\sqrt{\mu_{\rm EP}} > 0.02$ over the same radial distance range.

 To put these time-delay values in context, we determine the typical time duration of signals in the sensitive region of the aLIGO/AdV/KAGRA frequency band in Appendix \ref{sec:SignalTimeDuration}. Figure \ref{Fig:tau_obs_circ} shows the result for quasi-circular BBHs; the signals spend less than $10$ seconds above $20 \, \Hz$ for BBHs with component masses above $10 \, \Msun$. The signal duration is longer in the band for highly eccentric BBHs, which may reach a few minutes for the lightest binaries (i.e. $5 \, \Msun - 5 \, \Msun$ BBHs) for also a lower frequency cutoff of $20 \, \Hz$, but higher-mass binaries typically spend much less time in the band. These timescales are comparable to the typical time delays for SMBHs with masses $\lesssim 10^{6} \, \Msun$ for most detectable GW echoes (i.e. those which correspond to high-magnification events). However, they are generally lower than the typical time-delay values for echoes with moderately high or very high SNR primary GWs. Thus, we conclude that the GW echo may often overlap with the primary signal for low-mass BBHs merging in the vicinity of low-mass SMBHs, particularly for GW echo detections marginalized over the primary GWs (Figure \ref{Fig:DistPos_DistDelay_ObsBias} where both the echoes and the primary signal amplitudes are typically near the SNR detection limit). Note that interference effects between the primary GW signal and its GW echo may be important in these scenarios. However, the primary and echo signals are well separated in time and are distinct for higher-mass BBHs or massive SMBHs. They are also typically distinct in the case when the primary GW has at least a moderately high SNR.

%% file: Sections/Summary_Conclusions.tex
\section{Summary and Conclusions} 
\label{sec:Conclusions}
 
 In this paper, we studied the expected properties of astrophysical GW echoes produced by merging BBHs around SMBHs in the sensitive frequency band of aLIGO/AdV/KAGRA. The astrophysical GW echo is generated by scattering of the GWs by the SMBH. It follows a very similar time-frequency evolution and is observed from the same direction as the primary GW signal but arrives with a time delay. The time-frequency evolution of the GW echo may differ from the primary GW for a large deflection angle due to the Doppler shift, which is different if looking at the source from different directions. In addition, the GW polarization of the primary GW and the GW echo are generally different as the emitted signal is intrinsically anisotropic.

 We investigated the expected properties of GW echoes for general configurations not limited to highly-magnified small-deflection angle cases, accounting for the intrinsic anisotropy of the emitted GWs. We carried out an MC-based study to explore the characteristics of the expected magnification distribution and hence the relative amplitude of the GW echo compared to the primary GW signal in mock SNR-limited GW observational surveys and volume-limited surveys focusing on AGN/BH disks and isotropic populations of merging BBHs around SMBHs. We also studied the impact of disk inclination on the resulting distributions. Finally, we explored the time delay distribution between the primary GW signal and its astrophysical GW echo in mock SNR-limited surveys and the corresponding distribution of the angular position of the merging BBH relative to the SMBH and the line-of-sight.

 Our conclusions are summarized as follows:
\begin{enumerate}
   \item We find that the inclination of the AGN/BH disk $i$ has a significant impact on the amplitude of the GW echo relative to the primary GW, denoted by $\sqrt{\mu_{\rm EP}}$. Echoes with the most prominent $\sqrt{\mu_{\rm EP}}$ are expected from the inner regions of edge-on disks with $i \lesssim 6^{\circ}$, and $\sqrt{\mu_{\rm EP}}$ systematically decreases with both $i$ and $r_{\rm orb}$. This holds for BBH mergers in single AGN/BH disks (Section \ref{subsec:SingleDisk_IsotropDist}) and BBH mergers with detectable primary GWs in AGN/BH disks in SNR-limited surveys (Section \ref{subsec:Disk_IsotropDist_ObsBias}).\footnote{Here we assume detection for the primary GW signal above a certain SNR detection threshold and multiply the single GN probability by the detection volume $\propto \mu_{\rm EP}^{3/2}$.} Nearly edge-on disks may produce \mbox{$\sim 27 - 470$} times higher $\sqrt{\mu_{\rm EP}}$ on average than face-on disks for $10 \, \rS \lesssim r_{\rm orb} \lesssim 1000 \, \rS$ in SNR-limited surveys, where higher fractions correspond to larger SMBH-binary distances (Figure \ref{Fig:ObsBias_Disks_GNs}). This enhancement factor reduces to $\sim 2 - 11$ for the same $r_{\rm orb}$ range in the case of single disks (Figure \ref{Fig:Single_Disks_GNs}).
   
   \item The GW echo amplitude is comparable to the primary amplitude for strongly lensed signals, which represents a significant fraction of sources in SNR-limited surveys for edge-on disks, but $\sqrt{\mu_{\rm EP}}$ is typically much less than unity for disks with moderate inclinations (Figure \ref{Fig:ObsBias_Disks_GNs}). We find that the mean $\sqrt{\mu_{\rm EP}}$ is expected to be $\gtrsim \{0.91, 0.45 \}$ respectively for $r_{\rm orb} \lesssim \{200 \, \rS, 10^4 \, \rS \}$ for nearly edge-on disks in an SNR-limited survey for the primary GW (Figure \ref{Fig:ObsBias_Disks_GNs}), while it is $\gtrsim \{ 0.046, 0.004 \}$ for the same $r_{\rm orb}$ range for nearly edge-on disks (Figure \ref{Fig:Single_Disks_GNs}). Note that nearly edge-on disk configurations are expected to be uncommon.
   
   \item A non-negligible fraction of BBH mergers with detectable primary GWs produces GW echoes with amplitudes comparable to that of primary signals in isotropic populations in close proximity to SMBHs (e.g. either isotropic GNs or an ensemble of AGN/BH disks) both in complete volume-limited surveys\footnote{The detection of primary GW signals is complete within some maximum detection distance.} (Figure \ref{Fig:Single_Disks_GNs}) and in SNR-limited surveys (Figure \ref{Fig:ObsBias_Disks_GNs}). For instance, we find that $\sim 60\%$ ($\sim 10\%$) of mergers have $\sqrt{\mu_{\rm EP}} > 0.1$ ($0.4$) at $r_{\rm orb} \sim 10 \, \rS$ in mock SNR-limited surveys, which drops to $\sim 1\%$ ($\sim 0.1\%$) at $r_{\rm orb} \sim 1000 \, \rS$.
   
   \item There is a nonzero ($>1\%$) chance to detect GW echoes for BBH mergers within $\sim 135 \, \rS$ ($\sim 60 \, \rS$) of the SMBH for the most typical weak primary GW signals, and there is a nonzero ($1\% \lesssim$) chance to detect echoes from distances up to $\sim 3300 \, \rS$ ($\sim 2400 \, \rS$) for very strong primary signals in SNR-limited surveys (complete volume-limited surveys) (Sections  \ref{subsec:SingleDisk_IsotropDist} and \ref{subsec:Disk_IsotropDist_ObsBias}).
   
   \item Strongly lensed GWs by the SMBH are expected to be rare for BBH mergers with detectable primary GWs in isotropic populations around SMBHs in SNR-limited surveys (Section \ref{subsec:Disk_IsotropDist_ObsBias}). We find that the primary GWs have amplitudes larger than $\{1.01, 1.3, 1.5\}$ times the amplitude of unlensed GWs for $\sim \{ 7 \%, 4 \%, 3 \% \}$ of mergers for fixed $r_{\rm orb} = 10 \, \rS$ and these fractions drop significantly for higher $r_{\rm orb}$ (Figure \ref{Fig:ObsBias_Disks_GNs}). However, these percentages increase by a factor of $8$ for the combined SNR of the primary signal and its echo because the GW echo is also strongly lensed.
   
   \item In an SNR-limited survey of BBH mergers in isotropic populations around SMBHs, the vast majority of mergers with observable GW echoes are expected to be in the high-magnification strong-lensing regime (Figure \ref{Fig:DistPos_DistDelay_ObsBias}), where primary GWs and GW echoes have similar amplitudes. Although these mergers, which take place behind the SMBH close to the optical axis, are intrinsically rare, their GW echos can be detected from a much larger volume than high deflection angle weak echos. However, for primary GWs with high SNR, the GW echo may be much weaker in comparison and still be detected, implying that the GW echo may be in the large deflection angle regime, especially if the source is relatively close to the SMBH, within a few hundred $\rS$ (Figure \ref{Fig:DistPos_DistDelay_ObsBiasPrimary}). 
   
   \item In an SNR-limited survey of GW echoes irrespective of the primary GWs for isotropically distributed BBH mergers around SMBHs, the typical time delay between the primary GW and the GW echo can be roughly estimated as $\sim 18 \, {\rm sec} \times M_6$, where $M_6 = (1+z) M_{\rm SMBH} / 10^6 \, \Msun$ (Figure \ref{Fig:DistPos_DistDelay_ObsBias}). In practice, primary GWs and GW echos typically have comparable amplitudes and are near the SNR detection limit. For moderately high or very high SNR primary GWs in an SNR-limited survey of primary GWs, much weaker GW echoes may be detected, which generally arrive with a much longer time delay after the primary, up to $r_{\rm orb}/c$ in many cases (Figure \ref{Fig:DistPos_DistDelay_ObsBiasPrimary}). For instance, the typical time delay is $\sim (40 \, {\rm sec} - 60 \, {\rm sec} ) \times M_6$ and $\sim ( 260 \, {\rm sec} - 500 \, {\rm sec} ) \times M_6$ respectively, if the amplitude of detected echo is at least $0.2$ and $0.02$ times the primary amplitude. These timescales characterize BBH mergers out to few hundred $\rS$ since a non-negligible fraction of GW echoes have amplitudes comparable to primary amplitudes in this distance range (Section \ref{subsec:RelPos_TimeDelay}).
   
   \item The signal duration above $20 \,\rm Hz$ for quasi-circular BBHs with component masses $\gtrsim 10 \, \Msun$ and highly eccentric BBHs with relatively high component masses (Appendix \ref{sec:SignalTimeDuration}) are comparable to the typical time delays for SMBHs with masses $\lesssim 10^{6} \, \Msun$ for most detectable GW echoes (Section \ref{subsec:RelPos_TimeDelay}). However, they are generally lower than the typical time-delay values for echoes with moderately high or very high SNR primary GWs. Accordingly, the primary signal and its echo may interfere for very low-mass BH and NS mergers around low-mass SMBHs with \mbox{$M_{\rm SMBH} \lesssim 10^{6} \, \Msun$}, particularly for GW echo detections marginalized over the primary GWs.
   
   \item We find that the distribution of GW amplitude magnification $P(\sqrt{\mu})$ at each deflection angle $\alpha$ and SMBH-binary distance $r_{\rm orb}$ is broadened due to the anisotropic emission of the GW source profile for merging binaries. However, this broadening is found to be negligible in the strong lensing scenario ($\alpha \sim \alpha_{\rm Einstein}$) for relatively distant mergers from the SMBH (Appendix \ref{Sec:Anistropy_Properties}). For instance, the fraction of the upper or the lower bound of $P(\sqrt{\mu})$ relative to its median is $\sim \{ 10 \%, 5 \%, 1 \% \}$ for mergers at $r_{\rm orb} \sim \{300 \, \rS, 10^3 \, \rS, 2.2 \times 10^4 \rS \}$.
   
   \item We identified a strong lower bound for the GW echo amplitude relative to the primary GW $\sqrt{\mu_{\rm EP}}$ emitted by merging BBHs in the vicinity of an SMBH as a function of distance from the SMBH $r_{\rm orb}$,
   \begin{equation}\label{eq:muEPmin}
      \sqrt{\mu_{\rm EP}} \gtrsim 0.0411 \left(\frac{r_{\rm orb}}{10 \, \rS}\right)^{-1} - 0.0021 \left(\frac{r_{\rm orb}}{10 \, \rS}\right)^{-2}
   \end{equation}
   (Appendix \ref{Sec:Anistropy_Properties}). Its inverse gives the minimum distance of a merging binary from an SMBH upon detection of a primary GW together with the corresponding GW echo with $\sqrt{\mu_{\rm EP}}$,
   \begin{equation}
      r_{\rm orb} \lesssim 4.11 \, \rS \left( \frac{\sqrt{\mu_{\rm EP}} }{ 0.1 } \right)^{-1} - 0.516 \, \rS \, .
   \end{equation}
   This theoretical bound may be used to rule out the possibility that a BBH merged closer to an SMBH than the estimated minimum radial distance (Figure \ref{Fig:RelmagnifMin_vs_rorb}). In addition, this limit may also offer a conclusive way to rule out that a given BBH merger could have originated near an SMBH.
\end{enumerate}

 \citet{Chen2019} suggested that the high-mass BBH mergers observed by LIGO-VIRGO are in reality have lower masses because Doppler and gravitational redshift could increase the `apparent' BH mass and distance by up to a factor of $\sim 1.9 - 3.4$ if the mergers lie within $r_{\rm orb} = 10 \, \rS$ near an SMBH. However, in this case, the minimum GW echo amplitude is larger than $0.04$ of the primary GW (Equation \ref{eq:muEPmin}). Thus, if targeted searches rule out the detection of a GW echo from the same direction at this level\footnote{Note, however, that we neglected the observing direction-dependence of Doppler beaming, which may vary the GW amplitude by up to $\sim 25 \%$ (Appendix \ref{Sec:Additional_Effects}).} within $4\,M_6\,$minutes after the primary GW signal (i.e. within $\sim 25 \, {\rm sec} - 29 \, {\rm days}$ for SMBH masses between $10^5 \, \Msun - 10^{10} \, \Msun$), this argues against the possibility that the source mass and distance have been systematically overestimated for these sources.

 If detected with a consistent value of magnification and time delay, the astrophysical GW echo serves as a smoking gun signature to identify the origin of a GW source from GNs hosting an SMBH. Furthermore, it may be used to test theories describing the astrophysical formation scenarios of BBH mergers in the central regions of AGN/BH disks and GNs with an SMBH in their centre. Moreover, the primary signal and the echo may have a different amplitude and polarization but the same frequency when accounting for the time delay, given that the corresponding detected waves are originally emitted by the source in different directions and at different times. The relative amplitude and time delay between the primary signal and the echo may be used to constrain the geometry of the system and the distance to the SMBH. Note that the recently performed LIGO-Virgo searches can already constrain multiple images by looking for events with overlapping parameters without including the time delay (which was fixed to the scale of galaxy lensing) \citep{Abottetal2021a}. For very low-mass SMBHs and binary mergers, the time-dependent GW diffraction pattern between the primary GW signal and the GW echo may help to constrain the SMBH mass and spin \citep{Kocsis2013}. Detections of astrophysical GW echoes may also be useful to confirm the prediction of general relativity for the large-angle bending of GWs around BHs or to offer new possibilities for testing strong gravity.

%% file: Sections/Acknowledgments.tex
\section*{Acknowledgment}

 We thank the anonymous referee for constructive comments that helped improve the quality of the paper. We are grateful to M\'aria Kolozsv\'ari for help with logistics and administration related to the research. L\'aszl\'o Gond\'an is supported by the \mbox{{\'U}NKP-18-3} and \mbox{{\'U}NKP-21-4} New National Excellence Programmes of the Ministry for Innovation and Technology from the source of the National Research, Development and Innovation Fund. This work received funding from the European Research Council (ERC) under the European Union's Horizon 2020 Programme for Research and Innovation ERC-2014-STG under grant agreement No. 638435 (GalNUC).

%% file: Sections/Data_Availability.tex
\section*{Data Availability}
 
 The data underlying this article will be shared on reasonable request to the corresponding author.

%% file: Appendices/Properties_AmplDist.tex
\section{Properties of the amplification distribution}
\label{Sec:Anistropy_Properties}
 
 In this section, we carry out investigations regarding $P(\sqrt{\mu})$ and the minimum value of the relative echo amplitude ${\rm min} \, \sqrt{\mu_{\rm EP}}$. We also note that we work under the assumption of detectable primary GW and GW echo pairs.

 First, we note that ${\rm max}(\sqrt{\mu}) / {\rm med}(\sqrt{\mu}) = {\rm med}(\sqrt{\mu}) / {\rm min}(\sqrt{\mu})$ is due to the isotropy of $\mathbf{e}_{\rm AM}$. Furthermore, $P(\sqrt{\mu})$ forms a sharp peak as a function of deflection angle $\alpha$ at $\sqrt{ \mu_{\rm IS}}$ in the $\alpha \rightarrow 0$ and $\alpha \rightarrow \pi$ limits, respectively. Finally, ${\rm max}(\sqrt{\mu}) / {\rm med}(\sqrt{\mu})$ as a function of $\alpha$ can be best fitted numerically with the quartic polynomial
\begin{equation}  \label{eq:Var_MagnifFrac}
  \frac{ {\rm max}(\sqrt{\mu}) }{ {\rm med}(\sqrt{\mu}) } \simeq 0.165 \alpha^4 -1.035 \alpha^3 + 1.295 \alpha^2 + 1.036 \alpha + 1 \, ,
\end{equation}
 which approximates the fraction in question to within $4 \%$ for any $r_{\rm orb} \gtrsim 10 \, \rS$. ${\rm max}(\sqrt{\mu}) / {\rm med}(\sqrt{\mu})$ is symmetric about $\alpha = \pi/2$ and gradually decreases with $\alpha$ to unity in the $\alpha \rightarrow 0$ and $\alpha \rightarrow \pi$ limits. Finally, based on Equation \eqref{eq:Var_MagnifFrac}, the broadening of $\sqrt{\mu}$ in the strong lensing scenario ($\alpha \sim \alpha_{\rm Einstein}$) due to the anisotropic GW emission pattern of merging binaries at a fixed $r_{\rm orb}$ can be given as
\begin{align}  \label{eq:Var_MagnifFrac_StrLens}
  \nonumber
  \frac{ {\rm max}(\sqrt{\mu}) }{ {\rm med}(\sqrt{\mu}) } \bigg|_{ \alpha_{\rm Einstein} } 
  &\simeq 0.66 \left( \frac{ r_{\rm orb} }{ \rS } \right)^{-2} - 2.93 \left( \frac{ r_{\rm orb} }{ \rS } \right)^{-3/2}
  \\
  &+ 2.59 \left( \frac{ r_{\rm orb} }{ \rS } \right)^{-1} + 1.47 \left( \frac{ r_{\rm orb} }{ \rS } \right)^{-1/2} + 1 \, .
\end{align}
 Clearly, the broadening of $\sqrt{\mu}$ is negligible at $\sim \alpha_{\rm Einstein}$ for relatively distant merger events from the SMBH. For instance, the broadening is \mbox{$\sim \{ 10 \%, 5 \%, 1 \% \}$} for mergers at $\sim \{300 \, \rS, 10^3 \, \rS, 2.2 \times 10^4 \rS \}$.

\begin{figure}
   \centering
   \includegraphics[width=80mm]{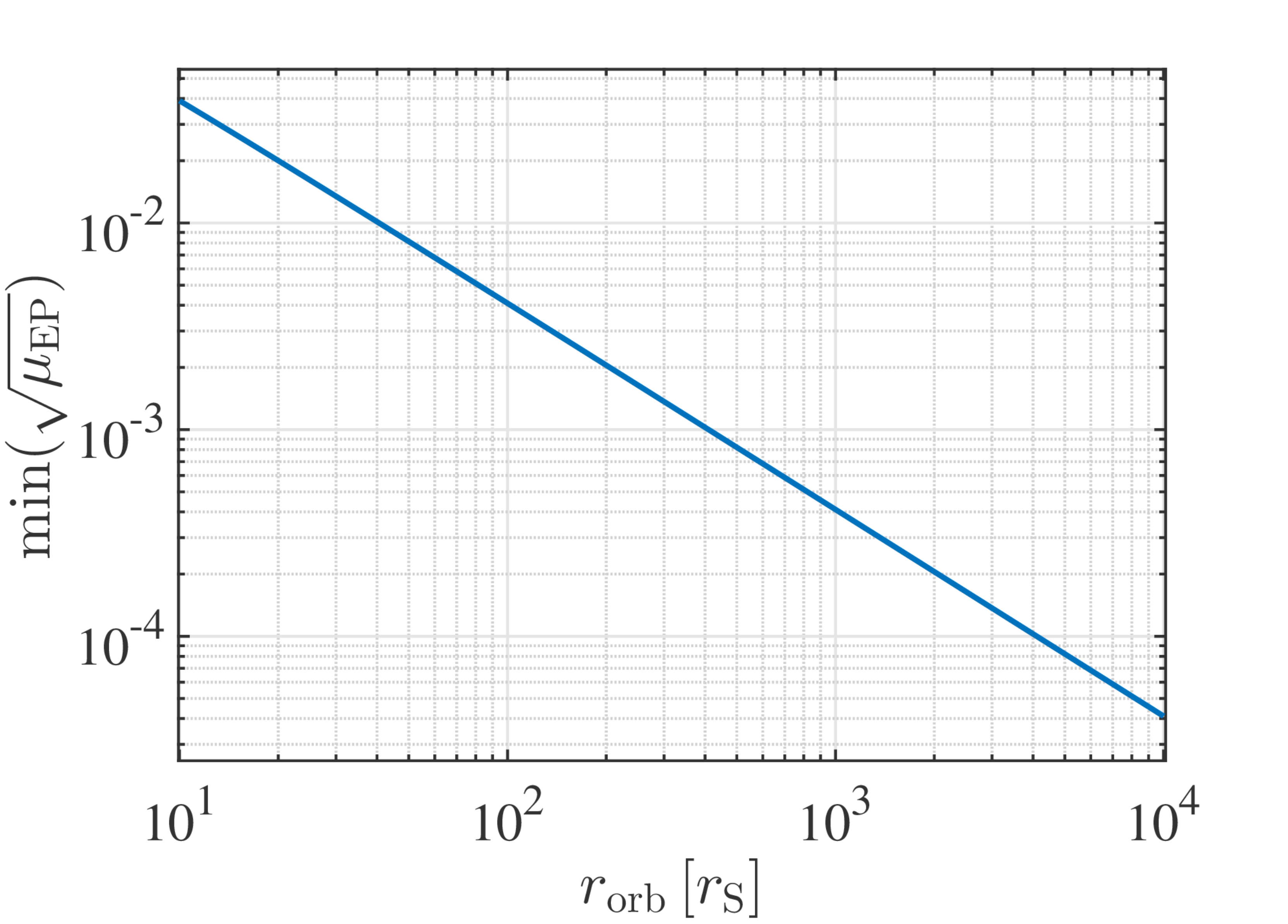}
   \\
   \includegraphics[width=80mm]{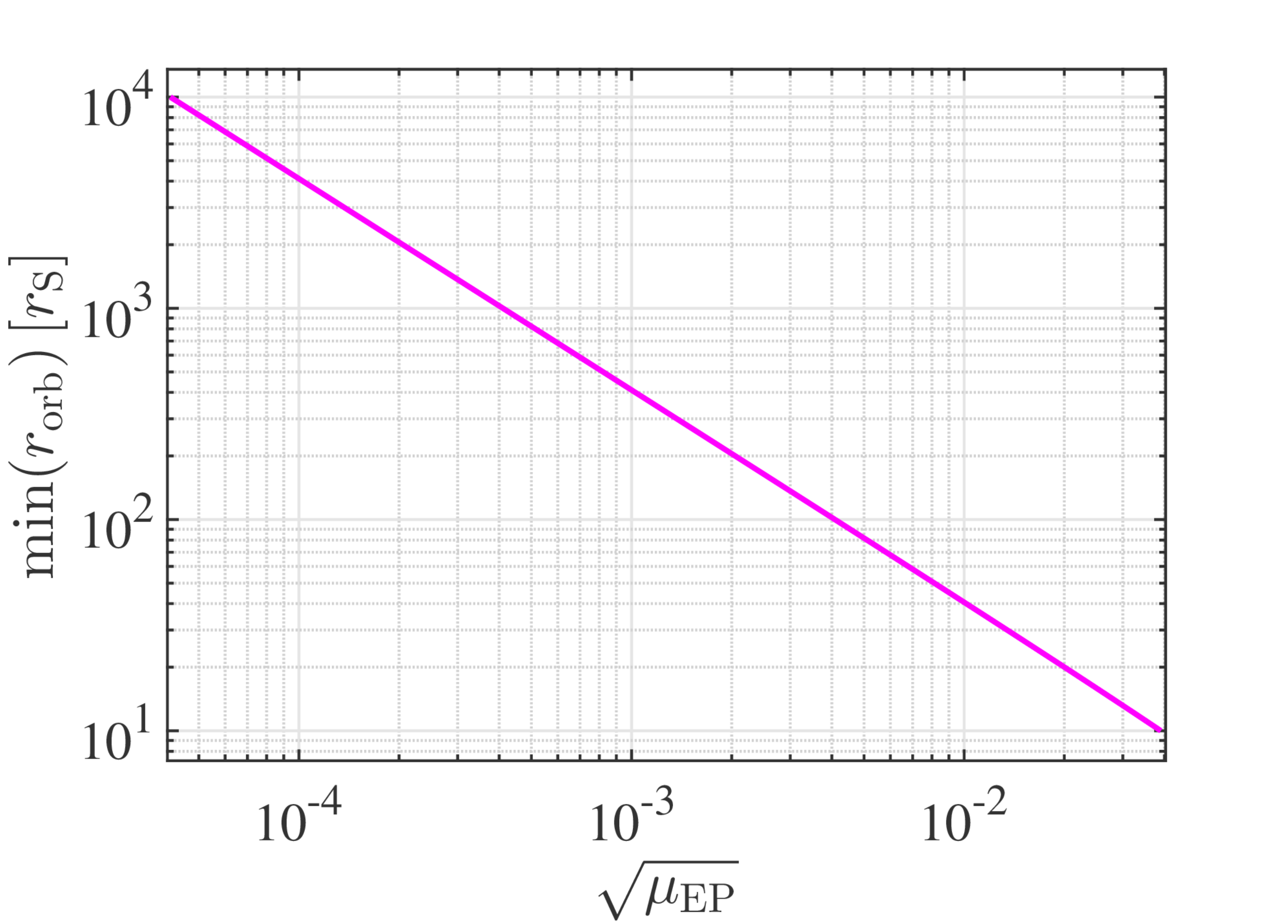}
\caption{\textit{Top panel:} The minimum echo amplitude relative to the amplitude of the primary GW signal ${\rm min}(\sqrt{\mu_{\rm EP}})$ assuming that the merging binary is within a radial distance of $r_{\rm orb}$ from the SMBH. \textit{Bottom panel:} The minimum distance of the merging binary from an SMBH for a given GW echo detection amplitude $\sqrt{\mu_{\rm EP}}$ relative to the primary GW.
 \label{Fig:RelmagnifMin_vs_rorb}} 
\end{figure}

 We also note that $\sqrt{\mu_{\rm E}}$ and $\sqrt{\mu_{\rm EP}}$ have minima at a certain $\alpha_{\rm magn,min}$ (or equivalently $\gamma_{\rm magn,min}$) for fixed $r_{\rm orb}$.\footnote{Note that both $\alpha_{\rm magn,min}$ and $\gamma_{\rm magn,min}$ do not depend significantly on $r_{\rm orb}$. For instance, $\{ \alpha_{\rm magn,min}, \gamma_{\rm magn,min} \} \sim \{ 117^{\circ}, 104^{\circ}\}$ for \mbox{$r_{\rm orb} = 10 \, \rS$}, while these angles shift to $\sim \{ 121^{\circ}, 121^{\circ}\}$ for \mbox{$r_{\rm orb} = 10^4 \, \rS$}.} Both minima may be best fitted by
\begin{align}  
  \label{eq:mu_EP_min}
  & {\rm min}(\sqrt{\mu_{\rm EP}}) \approx 0.0411 \left(\frac{r_{\rm orb}}{10 \, \rS}\right)^{-1} - 0.0021 \left(\frac{r_{\rm orb}}{10 \, \rS}\right)^{-2} \, , 
  \\
  \label{eq:mu_EU_min}
  & {\rm min}( \sqrt{\mu_{\rm E}}) \approx 0.0398 \left(\frac{r_{\rm orb}}{10 \, \rS}\right)^{-1} - 0.0078 \left(\frac{r_{\rm orb}}{10 \, \rS}\right)^{-2} \, , 
\end{align}
 which estimate ${\rm min}(\sqrt{\mu_{\rm EP}})$ and ${\rm min}( \sqrt{\mu_{\rm E}})$ to within $\sim 3 \%$ and $\sim 5 \%$, respectively. The top panel of Figure \ref{Fig:RelmagnifMin_vs_rorb} shows ${\rm min}(\sqrt{\mu_{\rm EP}})$ as a function of $r_{\rm orb}$. Upon detection of a GW echo with $\sqrt{\mu_{\rm EP}}$, this defines the minimum radial distance ${\rm min}(r_{\rm orb})$ of the merging binary from the SMBH lens, \footnote{We expressed $r_{\rm orb}$ as a function of $\sqrt{\mu_{\rm EP}}$ from Equation \eqref{eq:mu_EP_min}, which may be best fitted by a reciprocal power function. We also used the non-linear least squares method to fit the coefficients, and resulting values are given at $95 \%$ confidence level.}
 \begin{equation}  \label{eq:rorb_min}
  {\rm min}(r_{\rm orb}) \approx 4.11 \, \rS \left( \frac{\sqrt{\mu_{\rm EP}} }{ 0.1 } \right)^{-1} - 0.516 \, \rS \, .
\end{equation}
 The resulting bound is displayed in the bottom panel of Figure \ref{Fig:RelmagnifMin_vs_rorb}. This theoretical bound may be used to rule out the possibility that any binary has been merged near an SMBH at a smaller radial distance from the SMBH than ${\rm min}(r_{\rm orb})$ for a measured relative echo amplitude of $\sqrt{\mu_{\rm EP}}$.

 Finally, similar to the $\gamma$ distribution (Section \ref{subsec:GeometricConvects}), the support of the corresponding $\alpha$, $\sqrt{\mu_{\rm E}}$, and $\sqrt{\mu_{\rm EP}}$ distributions also shrinks as $i$ increases from 0 to $\pi/2$. More precisely for the $\sqrt{\mu_{\rm E}}$ and $\sqrt{\mu_{\rm EP}}$ distributions, the lower bound increases with $i$ only if \mbox{$\gamma_{\rm max} < \gamma_{\rm magn,min}$} or equivalently if \mbox{$i \gtrsim \pi - \gamma_{\rm magn,min}$} and decreases with $i$ otherwise as both $\sqrt{\mu_{\rm E}}$ and $\sqrt{\mu_{\rm EP}}$ attain minima at $\gamma_{\rm magn,min}$.

%% file: Appendices/SignalTimeDuration.tex
\section{Signal time duration in the frequency band of advanced GW detectors}
\label{sec:SignalTimeDuration}

 We estimate the time $\tau_{\rm band}$ that the GW signals may spend in the advanced GW detectors' sensitive frequency band for BBH mergers originating in AGN/BH disks and isotropic distributions around SMBHs. We consider two scenarios regarding the eccentricity with which inspiraling BBHs may enter the band: (i) quasi-circular BBHs that are expected to form in AGN disks and KL-induced BBH mergers in GNs (e.g. \citealt{AntoniniPerets2012,Hamersetal2018,RandallXianyu2018b,Fragioneetal2019,Samsing2020,Tagawa2021a}) and (ii) eccentric BBHs that are expected to originate mainly from the single-single GW capture channel in GNs \citep{OLearyetal2009,Gondanetal2018b,GondanKocsis2021} and also in part from AGN disks and KL-induced BBH mergers in GNs.

 The time duration that the GW signals of isolated quasi-circular BBHs at cosmological redshift $z$ spend in the aLIGO/AdV/KAGRA band may be estimated using the leading-order estimate as
\begin{align}  \label{eq:tau_band}
  \tau_{\rm band} &= 5 (8\pi f_{\rm L})^{-8/3} \mathcal{M}_z^{-5/3} \left[ 1 - \left( \frac{ f_{\rm L} }{ f_{\rm ISCO} } \right)^{8/3} \right] \nonumber
  \\  
  & \approx 18.9\,{\rm sec} \, \left( \frac{ M_{\rm tot,z} }{ 10 \, \Msun } \right)^{-5/3} \left( \frac{ \eta }{ 0.25 } \right)^{-1} \left( \frac{ f_{\rm L }}{ 20\,\Hz } \right)^{-8/3} \, ,
\end{align}
 where $\mathcal{M}_z = M_{{\rm tot},z} \eta^{3/5}$ is the redshifted chirp mass. Here, $M_{\rm tot}$ and $M_{{\rm tot},z} = (1+z)M_{{\rm tot}}$ are the intrinsic and redshifted total binary mass, respectively, and $\eta = m_1m_2/(m_1+m_2)^2$ is the symmetric mass ratio. We set the lower bound of the detection band to be \mbox{$f_{\rm L} = 20 \, \Hz$} based on recent analyses in LIGO-Virgo-KAGRA searches \citep{Abottetal2020a}, while the upper bound is the frequency at the innermost stable circular orbit (ISCO) $f_{\rm ISCO} = (6^{3/2} \pi M_{\rm tot,z})^{-1}$. Note that higher Post-Newtonian (PN) corrections beyond leading order affect $\tau_{\rm band}$ by more than $\sim 10 \%$ only in cases where $\tau_{\rm band} \lesssim 3.4\, {\rm sec}$. Figure \ref{Fig:tau_obs_circ} shows $\tau_{\rm band}$ as a function of component masses using the 1.5PN calculation. \footnote{We estimate $\tau_{\rm band}$ more accurately for quasi-circular BBHs by integrating the frequency evolution equation including PN terms up to 1.5PN order neglecting spin effects (e.g. \citealt{Cutleretal1994}). Generally, $f_{\rm ISCO}$ sets an upper limit for $M_{\rm tot}$ by the condition $f_{\rm ISCO} > f_{\rm L}$ (i.e. $\tau_{\rm band} > 0$), that is $M_{\rm tot} \lesssim 220 \, \Msun$ for $f_{\rm L} = 20 \, \Hz$.} The white region in the figure corresponds to binaries with \mbox{$f_{\rm ISCO} < 20 \, \Hz$}. Clearly, quasi-circular BBHs typically spend a few seconds in the aLIGO/AdV/KAGRA band above $f_{\rm L} = 20 \, \Hz$.

 In the case of highly eccentric BBHs, the GW signal may spend up to hours in the band in the repeated burst phase\footnote{To estimate $\tau_{\rm band}$, we first determine the dimensionless pericenter distance $\rho_{\rm p, 20 Hz}$ with which the GW frequency ($f_{\rm GW}$; \citealt{Wen2003}) enters the $20 \, \Hz$ frequency band, then determine $\tau_{\rm band}$ by calculating the merger time between $\rho_{\rm p, 20 Hz}$ and the dimensionless pericenter distance characterizing the last stable orbit by applying the Peters' formula improved by \citet{Zwicketal2020}, which accounts for the eccentricity evolution and 1PN order effects. We neglect higher PN order terms because the merger timescale is weakly affected by these terms \citep{Zwicketal2021}.}. However, the vast majority of the cumulative SNR of the inspiral signal arrives typically within the final few minutes (e.g. $\sim 50 \%$ within the final minute) of the merger \citep{KocsisLevin2012,Gondanetal2018b}, even for $f_{\rm L} = 10 \, \Hz$. Generally, $\tau_{\rm band}$ may reach a few minutes for the lightest BBHs for $f_{\rm L} = 20 \, \Hz$ based on \citet{Gondanetal2018b}.

 Note that additional effects such as gravitational and Doppler shift can be neglected when estimating $\tau_{\rm band}$; see Appendix \ref{Sec:Additional_Effects} for details.

%% file: Appendices/Additional_Effects.tex
\section{Additional Effects}
\label{Sec:Additional_Effects}
 
 In addition to the deflection of GW rays considered in detail in this paper, the SMBH may also influence the properties of the GW source in other ways. These include the gravitational redshift and Doppler shift (e.g. \citealt{Chen2019,Chen2021}) and time delay effects (the Shapiro and the Roemer delay, the geometric delay).\footnote{Other effects such as peculiar acceleration of binaries around SMBHs and the influence of gas affect more seriously the low-mass binaries in the LISA band (e.g. \citealt{Chen2021}, and references therein).} Following \citet{Chen2021}, we estimate the impact of these effects on the physical parameters of GWs investigated in this paper, such as the observed GW amplitude, $\tau_{\rm band}$, and $\tau_{\rm d}$. We denote the leading order estimate of the observed GW amplitude of a merging binary without these additional effects but accounting for cosmological redshift $z_{\rm cos}$ with $h_{\rm cos} \sim (\mathcal{M}_{\rm cos}/D_{\rm L,cos})( \pi f_{\rm cos} \mathcal{M}_{\rm cos})^{2/3}$. For clarity, we denote the cosmological redshift by $z_{\rm cos}$ in this section. Furthermore, \mbox{$f_{\rm cos} = f/(1+z_{\rm cos})$} is the cosmological redshifted (observed) GW frequency, $D_{\rm L,cos} = (1 + z_{\rm cos}) D_{\rm C}$ is the luminosity distance to the binary, and finally $f$ and $D_{\rm C}$ are the intrinsic (emitted) GW frequency and the comoving distance, respectively.

 Close to an SMBH, gravitational redshift also distorts a GW signal causing additional time-dilation by the same factor for the primary GW and the GW echo. The redshifted (observed) frequency including both the cosmological and gravitational redshift is \mbox{$f_z = f_{\rm cos}/(1+z_{\rm grav})$}. Similarly, the redshifted chirp mass is $\mathcal{M}_{z} = (1+z_{\rm grav})\mathcal{M}_{\rm cos} $ (e.g. \citealt{Chen2019}), where the gravitational redshift is
\begin{equation}
   1 + z_{\rm grav} = \left( 1 - \frac{ \rS }{r_{\rm orb}} \right)^{-1/2} \, .    
\end{equation}
 This implies that the observed GW amplitude and the signal time duration in the band are further distorted as \mbox{$h = (1+z_{\rm grav}) h_{\rm cos}$} and $(1+z_{\rm grav}) \tau_{\rm band}$, respectively; see Section \ref{subsec:Lens_Delay} for details. As seen, gravitational redshift has a small impact on $h$ which equals $h_{\rm cos}$ to within $\lesssim 5.5\%$ for $r_{\rm orb} \geqslant 10 \, \rS$, which is marginal compared to the factor of $\sqrt{3}$ variation caused by the intrinsic anisotropy of the GW emission pattern (Section \ref{subsec:Lensing_Geom}). Consequently, we neglected this effect in our investigation. Similarly, the effect of gravitational redshift on $\tau_{\rm band}$ is also negligible.

 The Doppler shift may cause an additional observation direction-dependent time-delay and amplification (Doppler beaming) if the source is moving with respect to the lens and the observer. The effect of Doppler shift can be characterized by the relativistic Doppler factor \mbox{$1 + z_{\rm dop} = \gamma (1 + \mathbf{v}\cdot \mathbf{n})$}, where \mbox{$\gamma = 1/\sqrt{1 - v^2}$} is the Lorentz factor, and $\mathbf{v}$ is the velocity of the source. Finally, $\mathbf{n}$ is a unit vector tangent to GW rays at the source, i.e. $\mathbf{e}_\mathrm{P}$ and $\mathbf{e}_\mathrm{E}$ in Figure \ref{Fig:GeomConvMagnif} for the primary signal and its echo, respectively. Thus, the observed GW amplitude and the signal time duration in the band are further distorted as $(1+z_{\rm dop}) h_{\rm cos}$\footnote{See e.g. \citet{DOrazioLoeb2020} for a detailed derivation.} and $(1+z_{\rm dop}) \tau_{\rm band}$, respectively.\footnote{See \citet{Gong2021,Yu2021} for more detailed estimates for sources close to the horizon.} To estimate the maximum effect of Doppler shift on both quantities, we note that to date BBH mergers are expected to merge closest to an SMBH in migration traps \citep{Bellovaryetal2016,Secundaetal2019,Secundaetal2020a,Secundaetal2020b}, which may exist in slim disks even near the ISCO \citep{PengChen2021}. Accordingly, the motion of BBHs merging closest to an SMBH happen approximately on a circular orbit ($v = \sqrt{ \rS / 2 r_{\rm orb}}$) implying that to leading order $1 + z_{\rm dop}$ can be approximated as
\begin{equation}
  1 + z_{\rm dop} \approx \left( 1 + \cos{\zeta} \sqrt{ \frac{ \rS }{ 2 r_{\rm orb} } } \right) \left( 1 + \frac{ \rS }{ 4 r_{\rm orb} } \right) \, ,
\end{equation}
 where $\cos{\zeta}$ is the cosine of the angle between $\mathbf{v}$ and $\mathbf{n}$. We find that Doppler shift causes a maximum of $\sim 25\%$ variation in $h_{\rm cos}$ at $r_{\rm orb} = 10 \, \rS$, and it drops relatively quickly with increasing $r_{\rm orb}$ ($\sim \{ 12\%, 7\% \}$ for $r_{\rm orb} = \{40 \, \rS, 100 \, \rS \}$). Note that this effect is still negligible compared to the $\sqrt{3}$ factor of variation due to the anisotropic GW emission pattern. Furthermore, the same results apply for both $\tau_{\rm band}$ and $\tau_{\rm d}$. From these, we conclude that the effect of the Doppler shift can also be neglected in our investigations.

 Time-delay effects such as the Shapiro and the Roemer delay together with the geometric delay occur in the weak field regime in the first PN order when determining the coordinate time $t$ between two spatial coordinates in a curved background geometry, e.g. near a Schwarzschild BH  (e.g. \citealt{Shapiro1964,BlandfordTeukolsky1976,LaiRafikov2005}). Since we used a fully general relativistic treatment to determine $t$ and thereby $\tau_{\rm d}$ in Section \ref{subsec:Lens_Delay}, these effects are already incorporated in our investigations.

%% file: Appendices/Regimes.tex
\section{Regimes for BBHs merging around SMBHs} 
\label{sec:LensingRegimes}
 
 In this section, we check the validity of our assumptions on the geometric optics limit and the stationary-lens regime for BBH mergers around SMBHs in Appendices \ref{subsec:OpticsRegimes} and \ref{subsec:LensingRegimes}, respectively.

\subsection{Geometric optics versus wave optics regime}
\label{subsec:OpticsRegimes}
 
 GWs or EM waves passing near a BH are deflected or lensed (e.g., \citealt{Schneideretal1992,TakahashiNakamura2003}). The geometric optics and wave optics regimes are delineated by the parameter
\begin{equation}  \label{eq:xi_geomopt}
   \chi = 8 \pi M_{\rm SMBH,z} f_{\rm GW} = 2.4 \times 10^3 \frac{M_{\rm SMBH,z}}{10^6 \, \Msun} \frac{f_{\rm GW}}{20 \, \Hz}\, 
\end{equation}
 \citep{TakahashiNakamura2003}, where $f_{\rm GW} = f_{\rm GW,source}/(1+z)$ is the observer-frame frequency, and the normalization factor $20 \, \Hz$ accounts for the cutoff frequency upon detection of a binary merger in recent exploratory analyses in LIGO-Virgo-KAGRA searches \citep{Abottetal2020a}. The wave-optics treatment asymptotes to the case of geometrical optics in the $\chi \gg 1$ limit (i.e. in the short-wavelength limit), in which case GWs travel on null geodesics of the curved background geometry (i.e. near an SMBH) in exact analogy to light waves \citep{Isaacson1968}.

 SMBH masses range from $\sim 10^5 \, \Msun$ to $\sim 10^{10} \, \Msun$. Accordingly, Equation (\ref{eq:xi_geomopt}) implies that GW sources observed by aLIGO/AdV/KAGRA have $\chi \gtrsim 10^2$ if lensed by SMBHs. Specifically for AGNs, the vast majority of SMBHs have masses between $\sim 10^6 \, \Msun$ and $\sim 10^8 \, \Msun$ \citep{GreeneHo2007}, indicating that $\chi \gtrsim 10^3$. From these results, we conclude that these sources are typically in the short-wavelength limit. Thereby, GWs propagate in approximately the same way near SMBHs as light rays.

 Note that interference effects may become significant only in special configurations for sources that happen to be close to the line-of-sight behind or in front of the SMBH \citep{Kocsis2013}. We conservatively neglected this effect in our investigations.

\subsection{Lensing regimes}
\label{subsec:LensingRegimes}
 
 Three different lensing regimes can be delineated based upon the motion of the GW source with respect to the SMBH lens over its observable lifetime $\tau_{\rm band}$ for circular orbits around the SMBH \citep{DOrazioLoeb2020}:
\begin{itemize}
   \item the repeating-lens regime: $1 / f_{\rm orb} \lesssim \tau_{\rm band}$,
   
   \item the slowly-moving lens regime: $\tau_{\rm band} \gtrsim \tau_{\rm lens}$, $1 / f_{\rm orb} \gtrsim \tau_{\rm band}$,
   
   \item the stationary-lens regime: $1 / f_{\rm orb} \gtrsim \tau_{\rm band}$ and $\tau_{\rm band} \lesssim \tau_{\rm lens}$.
\end{itemize}
 Here, $f_{\rm orb}$ is the orbital period of the GW source around the SMBH in the observer's frame. Furthermore, $\tau_{\rm lens}$ is the time for the source to cross the Einstein radius of the lens $\tau_{\rm lens} = r_{\rm Eins}^{\rm max} / v_{\rm orb}$, where \mbox{$r_{\rm Eins}^{\rm max} = \sqrt{4 M_{\rm SMBH,z} r_{\rm orb} \sin{i}}$} is the Einstein radius when the source is directly behind the lens, $v_{\rm orb}$ is the orbital velocity of the source around the lens, and $i$ is the inclination of the binary plane around the SMBH to the line-of-sight \citep{DOrazioDiStefano2018}. Note that redshift effects cancel out for the ratio of timescales. Since both $1 / f_{\rm orb}$ and $\tau_{\rm lens}$ increase for larger distances, while $\tau_{\rm band}$ is fixed, merger events far from the SMBH are well in the stationary-lens regime in LIGO-Virgo-KAGRA searches. Here, we examine sources in close proximity to the SMBHs to determine where the stationary-lens approximation breaks down. We estimate the fraction of merging BBHs in the stationary-lens regime for two separate cases due to the distinct $M_{\rm SMBH}$ and $r_{\rm orb}$ ranges, one in which BBHs merge in AGN disks and one in which they merge in GNs.

 First, the vast majority of SMBH masses of AGN disks lie between $\sim 10^6 \, \Msun$ and $\sim 10^8 \, \Msun$ \citep{GreeneHo2007}. Here we focus on the hypothetical migration traps in slim disks as BBH mergers may merge closest to the SMBH in this channel (even near the ISCO; \citealt{PengChen2021}) among other merger channels in AGNs (see the Introduction). In this paper, we examine $r_{\rm orb} \geqslant 10\,\rS$ and assume circular orbits around the SMBH. As the eccentricity of the binary thermalizes due to binary-single interactions in AGN disks (especially in slim disks; \citealt{Samsing2020,Tagawa2021a}), we assume $\tau_{\rm band} \lesssim 2 \, {\rm min}$ as an upper limit, which corresponds to the lightest BBH mergers on highly eccentric orbits, and $\tau_{\rm band}$ is significantly lower for BBH mergers with either lower eccentricities or higher masses (Appendix \ref{sec:SignalTimeDuration}, and references therein). Most BBH mergers in migration traps may be higher mass BHs \citep{Secundaetal2020a}. The developed MC routine implements the following steps to estimate the fraction of BBH mergers in the stationary-lens regime. (i) We first draw $10^6 \, \Msun \leqslant \Msmbh \leqslant 10^8 \, \Msun$ and $10\,\rS \leqslant r_{\rm orb} \leqslant 10^4\,\rS$ and $\sin{i}$ values from a uniform distribution between $[-1,1]$ (Section \ref{subsec:GeometricConvects}). (ii) Next, we calculate $1/f_{\rm orb}$, and calculate the corresponding $\tau_{\rm lens}$ value to each $i$ in the MC sample. (iii) Finally, we determine the fraction of BBHs in the stationary-lens regime using the relations introduced between the timescales $\{ 1/f_{\rm orb}, \tau_{\rm band}, \tau_{\rm lens} \}$. We find that $\sim 90\%$ of mergers lies in the stationary-lens regime even for the lowest $r_{\rm orb}$ and $M_{\rm SMBH}$ values and for $\tau_{\rm band} = 2 \, {\rm min}$, and this fraction increases steeply for higher $M_{\rm SMBH}$ and $r_{\rm orb}$ and for lower $\tau_{\rm band}$. Here we note that the mergers outside of the stationary-lens regime are located at $\gamma \sim \{ 0, \pi \}$. We conclude that the stationary-lens approximation is typically applicable for this study.

 Now we consider GN hosts with $10^5 \, \Msun \leqslant M_{\rm SMBH} \leqslant 10^{10} \, \Msun$. To estimate the lower bound of the fraction of sources in the stationary-lens regime, we focus on the single-single GW capture channel in which BBHs form closest to the SMBH, typically outside of a few hundred $\rS$ for standard models of mass segregation (e.g. \citealt{OLearyetal2009,Gondanetal2018a,GondanKocsis2021}). Note that BBH mergers in the single-single capture channel may form closer to the SMBH than a few hundred $\rS$ for strong mass segregation. However, in this case, high-mass BBHs dominate the merger rates, and among them, the most massive ones are located closest to the SMBH, for which $\tau_{\rm band}$ is much lower. Repeating the above analysis, we find that more than $90\%$ of binaries are in the stationary-lens regime for $\{ M_{\rm SMBH}, r_{\rm orb}, \tau_{\rm band} \} = \{ 10^5 \, \Msun, 100 \, \rS, 2 \, {\rm min} \}$, and this fraction becomes much higher for higher $M_{\rm SMBH}$ or $r_{\rm orb}$ or for lower $\tau_{\rm band}$. We conclude that the stationary-lens approximation is typically justified for BBH mergers in GNs.
 